\documentclass[3p]{elsarticle}

\usepackage[bookmarks=false]{hyperref}

\journal{Journal of \LaTeX\ Templates}

\usepackage{geometry}                		% See geometry.pdf to learn the layout options. There are lots.
\usepackage{graphicx}
\usepackage{tikz}
\usepackage{bm}

\usepackage{amsmath}
\usepackage{amssymb}

\usepackage{amsfonts}
\usepackage{combelow}
\usepackage{color}

\def\abs#1{{\left|#1\right|}}

\begin{document}

\begin{frontmatter}

%\title{Elsevier \LaTeX\ template\tnoteref{mytitlenote}}

%\title{Lattice Boltzmann model for liquid-vapour thermal flows}

\title{{Two-dimensional off-lattice Boltzmann model for van
der Waals fluids\\ with variable temperature}}


\author[acadaddress,uvtaddress]{Sergiu Busuioc}
\corref{mycorrespondingauthor}
\cortext[mycorrespondingauthor]{Corresponding author}\ead{sergiu.busuioc@e-uvt.ro}

\author[acadaddress,uvtaddress]{Victor E. Ambru\cb{s}}
\ead{victor.ambrus@e-uvt.ro}

\author[acadaddress,uvtaddress]{Tonino Biciu\cb{s}c\u{a}}
\ead{biciusca.tonino@gmail.com}

\author[acadaddress]{Victor Sofonea}
\ead{sofonea@acad-tim.tm.edu.ro}

%\author[mysecondaryaddress]{Global Customer Service
%\corref{mycorrespondingauthor}}

\address[acadaddress]{Center for Fundamental and Advanced Technical
Research, Romanian Academy\\
Bd. Mihai Viteazul 24, 300223 Timi\cb{s}oara, Romania}

\address[uvtaddress]{Department of Physics, West University of
Timi\cb{s}oara, Bd. Vasile P\^{a}rvan 4, 300223 Timi\cb{s}oara, Romania}


\begin{abstract}
We develop a two-dimensional Lattice Boltzmann model for liquid-vapour systems with variable temperature. Our model is based on a
single particle distribution function expanded with respect to the full-range Hermite polynomials.  
In order to ensure the recovery of the hydrodynamic equations for thermal flows, we use a fourth order expansion together with a set
of momentum vectors with $25$ elements whose Cartesian projections are the roots of the
Hermite polynomial of order $Q=5$. Since these vectors are
{\emph{off-lattice}}, a fifth-order projection scheme is used
to evolve the corresponding set of distribution functions.
A fourth order scheme employing a $49$
point stencil is used to compute the gradient operators in the force term that ensures the liquid-vapour
phase separation and diffuse reflection boundary conditions are
used on the walls. 
We demonstrate at least fourth order convergence
with respect to the lattice spacing in the contexts of shear and longitudinal 
wave propagation through the van der Waals fluid. 
For the planar interface, fourth order convergence can be seen at small enough 
lattice spacings, while the effect of the spurious velocity on the temperature 
profile is found to be smaller than $1.0\%$, even when $T_w  \simeq 0.7\,T_c$.
We further validate our scheme by considering the Laplace pressure test. 
Galilean invariance is shown to be preserved up to second order with 
respect to the background velocity.
We further investigate the liquid-vapour phase separation 
between two parallel walls kept at a constant temperature $T_w$
smaller than the critical temperature $T_c$ and discuss the 
main features of this process.
\end{abstract}


\end{frontmatter}

\section{Introduction}\label{sec:intro}

Lattice Boltzmann (LB) models with variable temperature are known
since at least two decades
\cite{massaioliEPL1993,alexanderPRE1993,qianJSciComput1993}
and their development is still in progress today.
Basically, the thermal LB models belong to one of the following families
\cite{GuoShuLBapplications,liPECS2016}:
\emph{multi-speed models}
\cite{chenPRE1994,watariPRE2003,watariPRE2004,sofoneaPRE2005,
	philippiPRE2006,shanJFM2006,gonnellaPRE2007,
	sbragagliaJFM2009,chenEPL2010,
	frapolliPRE2014},
\emph{double distribution function models}
\cite{heJCPH1998,guoPRE2007,zhangEPL2007,karlinPRE2013,
	yasuokaPRE2016,pareschiPRE2016}
and \emph{hybrid models}
\cite{zhang03pre,lallemandIJMPC2003,lallemandPRE2003,gonnellaPRE2010,
	safariPRE2013,safariPRE2014,liJTHT2014,liIJHMT2015}.
Such models are currently applied to investigate physical and
engineering processes involving heat transfer with or without
phase change, as well as micro- and nano-scale flow phenomena.
The diversity of these applications are confirmed by the rich literature
related to LB models and by two series of regular conferences
\cite{icmmes,dsfd}.

The purpose of this paper is to explore the  capabilities of a minimal LB
model with variable temperature used to simulate the behaviour of a
two-dimensional ($2D$)
phase-separating fluid which obeys the Van der Waals equation of state.
The model is constructed using the Gauss-Hermite quadrature of order
$Q = 5$ and has $K = Q \times Q = 25$ velocity vectors. 
Since the roots of the fifth order Hermite 
polynomial are irrational numbers, the ensuing velocity set
is off-lattice. For this reason,
finite difference techniques must be employed
in order to obtain the numerical solution of the
evolution equations in the LB model.
In this paper, we employ the third order total variation 
diminishing (TVD) Runge-Kutta (RK-3) time stepping procedure, together 
with the fifth-order weighted essentially non-oscillatory (WENO-5) 
scheme for the advection. 

This paper is structured as follows. Section~\ref{sec:model} introduces 
the off-lattice Boltzmann model, with a brief discussion 
of our non-dimensionalization convention, as well as of the 
 momentum space discretization procedure, the
expansion of the equilibrium distribution and the implementation of 
the forcing term which ensures the recovery of the van der Waals equation of state.
The finite difference schemes RK-3 and WENO-5 are introduced in Sec.~\ref{sec:scheme}, 
together with a discussion of the
 implementation of the boundary 
conditions. Sec.~\ref{sec:planar} presents
our validation tests (interface width, phase diagram)
for the case of a planar interface (when the system is assumed to be homogeneous 
along the direction parallel to the walls).
The Laplace pressure test is performed for circular gas bubbles in 
Sec.~\ref{sec:laplace}. We discuss the transport coefficients 
appearing in our model and its Galilean invariance in 
Sec.~\ref{sec:galilei}, in the context of the damping of shear and 
longitudinal waves in a periodic one-dimensional domain, as well as for a droplet 
in a constant velocity background flow.
Section~\ref{sec:pspp} presents our simulation 
results for the phase separation process induced in a van der Waals fluid 
at the critical point 
enclosed between parallel plates which are cooled suddenly.
Our conclusions are summarized in Sec.~\ref{sec:conc}.

\section{Description of the model}\label{sec:model}

\subsection{Non-dimensionalized quantities}\label{sec:model:nondim}

In LB models, all quantities of interest are expressed in non-dimesional form.
For convenience, in this subsection the tilde ($\widetilde{\hspace{0.5em}}$)
symbol over a letter which denotes a physical (measurable) quantity makes
the difference between its dimensional form $\widetilde{A}$ and the
non-dimensionalized form $A$. These two forms are
related through $\widetilde{A} = A\, \widetilde{A}_{R}$, where
$\widetilde{A}_{R}$ is the corresponding reference quantity.
The non-dimensionalization procedure of the Boltzmann 
equation \cite{sofoneaPRE2005} amounts to defining four basic reference quantities, namely 
the particle number density $\widetilde{n}_{R}$,
the length $\widetilde{\ell}_{R}$,
the mass $\widetilde{m}_{R}$ and the energy
$\widetilde{e}_{R} = \widetilde{k}_{B} \widetilde{T}_{R}$, where
$\widetilde{k}_{B}$ is the Boltzmann constant and $\widetilde{T}_{R}$ is
the reference temperature. In our LB model, the reference length 
$\widetilde{\ell}_{R}$ is some characteristic length of the fluid system, which
may be, e.g., the width of the flow channel. For 
the single component van der Waals fluid considered in this paper,
the mass of its particles is the natural choice for
$\widetilde{m}_{R}$,
% and the temperature value $\widetilde{T}_{c}$
% at its critical point is a convenient choice for the reference temperature
% ($\widetilde{T}_{R}=\widetilde{T}_{c}$).
while the properties of the fluid at the critical point 
are the natural choice for the reference temperature and the
reference density,
i.e.~$\widetilde{T}_R = \widetilde{T}_c$ and 
$\widetilde{n}_R = \widetilde{n}_c$.
The values of $\widetilde{T}_c$ and $\widetilde{n}_c$ defining the 
critical point can be considered free parameters of the van der Waals 
model.

The reference values for other physical quantities in the LB model are
derived from the above-mentioned basic quantities. In particular, we
get the reference speed $\widetilde{c}_{R} = \sqrt{\widetilde{k}_{B}
\widetilde{T}_{R} / {\widetilde{m}_R}}$ and the reference time
$\widetilde{t}_{R} = \widetilde{\ell}_{R} / \widetilde{c}_{R}$
\cite{sofoneaPRE2005,sofoneaPRE2004}. We choose
the reference pressure to be the 
pressure of the ideal gas at $\widetilde{n} = \widetilde{n}_c$ 
and $\widetilde{T} = \widetilde{T}_c$, namely 
$\widetilde{p}_R = \widetilde{n}_R\,\widetilde{k}_{B}\widetilde{T}_{R}$.
% at the temperature $\widetilde{T}_{R}=\widetilde{T}_{c}$, with
% the particle number density $\widetilde{n} = n\, \widetilde{n}_{R}$,
The reference density can also be written in terms of the molar volume 
$\widetilde{V}_m(\widetilde{T}_R)$ at the reference temperature as follows:
\begin{equation*}
\widetilde{n}_{R}=\,\frac{N_{A}}{\,\widetilde{V}_{m}(\widetilde{T}_{R})\,}\,=\,
\frac{N_{A}}{\,V_{m;R}\,(\widetilde{\ell}_{R})^{3}\,} \qquad {\mathrm{and}} \qquad
n\,=\,\frac{\,V_{m;R}\,}{\,V_{m;T}\,}\, .
\end{equation*}
Here $N_{A}$ is the Avogadro number, while $V_{m;R}$ and $V_{m;T}$ are the nondimensionalized molar volumes at temperatures
$\widetilde{T}_{R}$ and $\widetilde{T}$, respectively.
With this choice, the non-dimensionalized form of the van der Waals
equation reads \cite{gonnellaPRE2007,sofoneaPRE2004}
\begin{equation}
p^w=\frac{3n T}{3-n}-\frac{9}{8}n^2 \, .
\label{vdw}
\end{equation}

Finally, we note that in our model, the lattice spacing is 
$\widetilde{\delta s} = {\widetilde{\ell}}_{R}/{\mathfrak{N}}$ 
and hence its non-dimensionalized value is 
$\delta s = 1/{\mathfrak{N}}$, where ${\mathfrak{N}}$ 
is an integer number.

\subsection{Discretization of the momentum space,
evolution equations and the equilibrium distribution functions}
\label{sec:model:disc}

In LB models, the non-dimensionalized values of the fluid particle number 
density $n\equiv n({\bm{x}},t)$, velocity ${\bm{u}}\equiv \bm{u}({\bm{x}},t)$
and temperature $T\equiv T({\bm{x}},t)$, defined in the nodes ${\bm{x}}$
of a lattice ${\mathcal{L}}$, are retrieved at time $t$ through the calculation
of the moments (up to second order) of the single particle distribution function
$f\equiv f({\bm{x}},{\bm{p}},t)$ defined in the point $({\bm{x}},{\bm{p}})$ of
the phase space \cite{GuoShuLBapplications,shanJFM2006,deville}.
Current multispeed LB models use
the Cartesian coordinate system in the $D$-dimensional momentum space 
and the moments of
$f({\bm{x}},{\bm{p}},t)$ are computed using a convenient quadrature.
For this purpose,  $f({\bm{x}},{\bm{p}},t)$ is
approximated by its expansion $f^{N}({\bm{x}},{\bm{p}},t)$
up to a certain order $N$
with respect to a set of orthogonal polynomials, e.g., the full-range Hermite
polynomials defined on each Cartesian axis of the momentum space
\cite{GuoShuLBapplications,shanJFM2006,deville,
fedeIJMPF2015,ambrus16jcp,ambrus16jocs}. 
%hildebrand,asteg,nist,shizgal15,
The resulting quadrature points form a discrete vector set 
$\{{\bm{p}}_{\bm{\kappa}}\equiv (p_{k_{1}},p_{k_{2}},\ldots p_{k_{D}})\}$
in the momentum space
(${\bm{\kappa}} \equiv \{k_{1},k_{2},\ldots k_{D}\}$ is a set of integer indices
and $p_{k_{\alpha}}$, $1\leq \alpha \leq D$ is the projection of the vector
${\bm{p}}_{\bm{\kappa}}$ on the Cartesian axis $\alpha$). As a result of
the application of the Gauss-Hermite quadrature method,
in the LB model the fluid system is described by the set of functions
$f_{{\bm{\kappa}}} \equiv f_{{\bm{\kappa}}}({\bm{x}},t)=
f^{N}({\bm{x}},{\bm{p}}_{\bm{\kappa}},t)$,
defined in the nodes ${\bm{x}}$ of a lattice ${\mathcal{L}}$.

In the $D$-dimensional LB model where the full-range Gauss-Hermite
quadrature of order $Q$ is used on each Cartesian axis,
we have $1\leq k_{\alpha} \leq Q$ for all $\alpha$, $1\leq \alpha \leq D$,
and hence the
momentum set $\{{\bm{p}}_{{\bm{\kappa}}}\}$ has $K=Q^{D}$ elements.
The order $Q$ of
the quadrature should satisfy the condition $Q\geq N+1$ in order to retrieve {\emph{all}} the moments of $f({\bm{x}},{\bm{p}},t)$ up to
order $N$
\cite{shanJFM2006,fedeIJMPF2015,ambrus16jcp,piaudIJMPC2014}.
Although the number $K$ of the quadrature points can be reduced
by very elaborated pruning techniques by sacrificing some higher order 
moments of the distribution function
\cite{shanJFM2006,chikatamarlaPRE2009,shimPRE2011},
we will not consider such models in this paper.

When the Bhatnagar-Gross-Krook (BGK) collision term is used
in a LB model with variable temperature,
the moments of the distribution function
$f({\bm{x}},{\bm{p}},t)$ up to order $N=4$ are needed in order to get the
evolution equations of the macroscopic fields at the
Navier - Stokes - Fourier level
\cite{GuoShuLBapplications,watariPRE2003,shanJFM2006,
ambrusPRE2012}. 
Thus, the minimum number of the momentum vectors in the
two-dimensional ($2D$) thermal LB model based on the full-range
Gauss-Hermite quadrature that ensures all the moments of
$f({\bm{x}},{\bm{p}},t)$ up to order $N=4$ is $K=(N+1)^{2}=25$.
In this $ 2D$ LB model,
the full-range Gauss-Hermite quadrature of order $Q=5$
is used on each Cartesian axis. The quadrature points, namely
${\bm{p}}_{\bm{\kappa}}\equiv (p_{k_{1}}, p_{k_{2}})$,
$1\leq k_{1}, k_{2} \leq Q$, are constructed using
the direct product rule \cite{shanJFM2006,ambrus16jcp,ambrus16jocs}.
For each $\alpha \in \{1, 2\}$, the Cartesian projections $p_{k_{\alpha}}$
belong to the set $\{{\cal{R}}_{q}\}$, $1\leq q \leq Q=5$, of
the roots of the full-range Hermite polynomial $H_{5}(p)$
\cite{shanJFM2006,ambrus16jcp,shizgal15}.
For convenience, Table \ref{trots} shows the elements of this set,
as well as their associated weights ${\cal{W}}_{q}$
given by \cite{ambrus16jcp,ambrus16jocs,hildebrand,asteg,nist}
\begin{equation}
{\cal{W}}_{q} = \frac{Q !}{\,[H_{Q+1}({\cal{R}}_{q})]^{2}\,}.
\end{equation}
To avoid confusion, we recall that
the full range Hermite polynomials $H_{\ell}(p)$ used in this paper
are the so-called {\emph{probabilistic} Hermite polynomials,
which are orthogonal with respect to the weight function
\begin{equation}
\omega(p) = \frac{1}{\,\sqrt{2\pi}\,} e^{-p^{2}/2},
\label{eweight}
\end{equation}
and their orthogonality relation reads \cite{hildebrand}
\begin{equation}
 \int_{-\infty}^{+\infty} dp \,\omega(p) H_{\ell}(p) H_{\ell'}(p) =
 \ell ! \, \delta_{\ell,\ell'}.
\end{equation}

\begin{table}[h]
\caption{The roots ${\cal{R}}_{q}$ of the full-range Hermite 
polynomial of order $\ell=5$ and their associated weights ${\cal{W}}_{q}$ \cite{shanJFM2006}.\label{trots}}
\begin{center}
\begin{tabular}{ccccc}
\hline
$q$ &\qquad & ${\cal{R}}_{q}$ & \qquad & ${\cal{W}}_{q}$
\rule[-2mm]{0mm}{6mm}\\
\hline
 1  && $- \sqrt{5 + \sqrt{10}}$  && $(7 - 2\sqrt{10})/60$
 \rule{0mm}{5mm} 
 \\ 
 2  && $- \sqrt{5  - \sqrt{10}}$  && $(7 + 2\sqrt{10})/60$
  \rule{0mm}{5mm}
   \\
 3  && $0$                               && $8/15$
  \rule{0mm}{5mm} \\
 4  && $ \sqrt{5  - \sqrt{10}}$  &&  $(7 + 2\sqrt{10})/60$
  \rule{0mm}{5mm}   \\
 5  && $ \sqrt{5 + \sqrt{10}}$  &&  $(7 - 2\sqrt{10})/60$
  \rule[-2mm]{0mm}{6mm}  \\
  \hline
\end{tabular}
\end{center}
\end{table}

As usual in the current LB models involving the BGK collision term
\cite{shanJFM2006}, the non-dimensionalized form of the evolution
equation of the functions $f_{\bm{\kappa}}$ for a single-component
fluid is:
\begin{equation}\label{evolution}
 \partial_{t} f_{\bm{\kappa}} +
 \frac{1}{\,m\,}{{\bm{p}}_{\bm{\kappa}}}\cdot \nabla f_{\bm{\kappa}} +
 \bm{F} \cdot  (\nabla_{\bm{p}} f)_{\bm{\kappa}}
 = -\frac{1}{\tau}[f_{\bm{\kappa}} - f^{eq}_{\bm{\kappa}}],
\end{equation}
where ${\bm{F}}$ is the force acting on a particle of mass $m$
and $\tau$  is the relaxation time. 
Even though $m = 1$ according to the nondimensionalization conventions 
discussed in Subsec.~\ref{sec:model:nondim}, we keep $m$ 
explicit in all equations in order to avoid confusion.
For simplicity, in this paper we assume that the relaxation
time $\tau$ is constant.
The Cartesian components
$(\partial_{p_{\alpha}} f)_{\bm{\kappa}}$, $\alpha \in \{1,2\}$,
of the elements in the discrete vector set
$\{(\nabla_{\bm{p}} f)_{\bm{\kappa}}\}$
that replace the momentum gradient $\nabla_{\bm{p}} f$
in the Boltzmann equation,
will be detailed in Subsec.~\ref{sec:model:force}.

The equilibrium functions $f^{eq}_{\bm{\kappa}}\equiv
f^{eq}({\bm{x}},{\bm{p}}_{\bm{\kappa}},t)$ are given by
\cite{ambrus16jcp,ambrus16jocs}: 
\begin{subequations}\label{eq:feq}
\begin{equation}
f^{eq}_{\bm{\kappa}}  =  n \prod_{\alpha=1}^{D} g_{k_{\alpha}},
\end{equation}
where
\begin{equation}
 g_{k_{\alpha}}  \equiv  g_{k_{\alpha}}(u_{\alpha}, T) =
  w_{k_{\alpha}}\,\sum_{\ell=0}^{N}
 H_{\ell}(p_{k_{\alpha}}) \sum_{s=0}^{\lfloor \ell/2 \rfloor}
\frac{ \,( mT -1 )^{s} ( mu_{\alpha} )^{\ell-2s}\,}{\,2^{s} s! (\ell-2s)!\,}
\label{g_alpha},
\end{equation}
\end{subequations}
and $\lfloor \ell/2 \rfloor$ is the integer part of $\ell/2$.
In the above, it is understood that $g_{k_{\alpha}} \equiv g_{k_{\alpha}}(\bm{x}, t)$ 
since $u_\alpha$ and $T$ depend on $\bm{x}$ and $t$.
To each $p_{k_{\alpha}}\in \{{\cal{R}}_{q}\}$,
$k_{\alpha} = 1,\,2,\ldots Q$, $\alpha\in \{1,2\}$,
there is an associated weight $w_{k_{\alpha}}\in\{{\cal{W}}_{q}\}$,
given by Eq.~(\ref{eweight}) and we will use the notation $w_{\bm{\kappa}} \equiv w_{k_{1} k_{2}} =
w_{k_{1}}w_{k_{2}}$.

\subsection{Force term}\label{sec:model:force}

The following expression of the force 
$\bm{F}\equiv {\bm{F}}({\bm{x}},t)$, which appears in
Eq.(\ref{evolution}), was used in order to simulate the evolution of a
van der Waals fluid
\cite{GuoShuLBapplications,sofoneaPRE2004,HuangSukopLuMultiphaseLB,hePRE1998,luoPRL1998,biciuscaCRM2015}:
\begin{equation}
 \bm{F} = \frac{1}{\,n\,}\nabla(p^{i}-p^{w}) + \sigma\nabla(\Delta n).
 \label{eq:F_def}
\end{equation}
In this expression, the parameter $\sigma$ controls the surface tension, 
$p^i=nT$ is the non-dimensionalized ideal gas pressure and
the non-dimensionalised form of the van der Waals pressure $p^w$
is given in Eq.~(\ref{vdw}). 
 The spatial gradients appearing in Eq.~\eqref{eq:F_def} 
are computed using 49-point stencils which are given in 
Refs.~\cite{leclaireJSC2014,patraNMPDE2006}. 
For the reader's convenience,
these stencils are summarized in Sec.~\ref{sec:scheme:stencil}.
% spatial gradients in Eq.~\eqref{eq:F_def} is reviewed 
% A 25-point stencil
% is used to compute the two Cartesian components of the force ${\bm{F}}$
% in each lattice node ${\bm{x}}$ at time $t$
% \cite{biciuscaCRM2015,leclaireJSC2014,patraNMPDE2006}.

To account for the
Cartesian components 
$(\partial_{p_{\alpha}} f)_{\bm{\kappa}}$, $\alpha \in \{1,2\}$ of
$\{(\nabla_{\bm{p}} f)_{\bm{\kappa}}\}$, which appear in Eq.(\ref{evolution}), we first 
expand the single particle distribution function $f({\bm{x}},{\bm{p}},t)$ 
with respect to the full-range Hermite polynomials $H_{\ell}(p_{\alpha})$
defined on the Cartesian axis $\alpha \in \{1,\,2\}$ of the momentum space, to get
\cite{ambrus16jocs}:
\begin{equation}
f({\bm{x}},{\bm{p}},t) = \frac{1}{\,\sqrt{2\pi}\,} e^{-p_{\alpha}^{2}/2}
\sum_{\ell=0}^{\infty} \frac{1}{\,\ell !\,}
\mathcal{F}_{\alpha,\ell}
({\bm{x}},
p_{\overline{\alpha}},
t) H_{\ell}(p_{\alpha}),
\label{fexpanded}
\end{equation}
where
\begin{align}
\mathcal{F}_{\alpha,\ell}
({\bm{x}},
p_{\overline{\alpha}},
t)  & = 
\int_{-\infty}^{\infty} f({\bm{x}},{\bm{p}},t) H_{\ell}(p_{\alpha}) dp_{\alpha},\\
\overline{\alpha}
& =  \left\{
\begin{array}{c}
2 \, , \, \alpha = 1\\ 1 \, , \, \alpha = 2 \rule{0mm}{4mm}
\end{array}\right. \rule{0mm}{9mm}
\end{align}
Using the recurrence property of the Hermite polynomials $\partial_x[e^{-x^2/2}H_\ell(x)]=-e^{-x^2/2} H_{\ell + 1}(x)$
\cite{shanJFM2006,ambrus16jocs} to compute the derivative
with respect to $p_{\alpha}$
of $f({\bm{x}},{\bm{p}},t)$ given in Eq.~(\ref{fexpanded}), we get
\begin{equation}
\partial_{p_{\alpha}} f ({\bm{x}},{\bm{p}},t) = - \,
\frac{1}{\,\sqrt{2\pi}\,} e^{-p_{\alpha}^{2}/2}
\sum_{\ell=0}^{\infty} \frac{1}{\,\ell !\,}
\mathcal{F}_{\alpha,\ell}
({\bm{x}},
p_{\overline{\alpha}},
t) H_{\ell+1}(p_{\alpha}).
\end{equation}

After truncation of this expresion up to order $N$, the application
of the discretisation procedure in the momentum space gives:
\begin{align}
(\partial_{p_{\alpha}} f)_{\bm{\kappa}} \equiv
(\partial_{p_{\alpha}} f)_{\bm{\kappa}} ({\bm{x}},t)
 & = 
-\, w_{k_{\alpha}} \sum_{\ell=0}^{N-1} \frac{1}{\,\ell !\,}
\mathcal{F}_{\alpha,\ell;k_{\overline{\alpha}}}
({\bm{x}},t) H_{\ell+1}(p_{k_{\alpha}})
\label{derf}\,,  \\
\mathcal{F}_{\alpha,\ell;k_{\overline{\alpha}}}
({\bm{x}},t) & = 
\sum_{k_{\alpha}=1}^{Q} f_{\bm{\kappa}} ({\bm{x}},t) H_{\ell}(p_{k_{\alpha}}).
\end{align}
Note that the sum in Eq.(\ref{derf}) above runs up to $\ell = N-1$ since
$H_{N+1}(p_{k_{\alpha}}) = 0$ for all $k_{\alpha} = 1,\,2,\ldots Q=N+1$,
$\alpha\in \{1,2\}$.

\subsection{Macroscopic equations}\label{sec:model:macro}
Multiplying the Boltzmann equation \eqref{evolution} with the collision invariants $1$, $\bm{p}$ and $\bm{p}^2/2m$ and 
integrating over the momentum space yields the following macroscopic equations:
\begin{subequations}\label{eq:macro}
\begin{gather}
 \partial_t n + \nabla \cdot (n\bm{u}) = 0,\label{eq:macro_n}\\
 \rho \left[\partial_t u_\alpha + (\bm{u} \cdot \nabla) u_\alpha\right] =
 n F_\alpha - \partial_\alpha p^i - \partial_\beta \Pi_{\alpha\beta},\label{eq:macro_u}\\
 n \left[\partial_t T + (\bm{u} \cdot \nabla) T\right] + \partial_\alpha q_\alpha + p^i 
 (\nabla \cdot \bm{u}) + \Pi_{\alpha\beta} \partial_\alpha u_\beta = 0,\label{eq:macro_e}
\end{gather}
\end{subequations}
where $\Pi_{\alpha\beta}$ is the viscous part of the stress tensor and $q_\alpha$ is the heat flux. This quantities are defined 
in terms of the peculiar momentum $\bm{\xi} = \bm{p} - m\bm{u}$ as follows:
\begin{equation}
 \Pi_{\alpha\beta} + p^i \delta_{\alpha\beta} = \int d^2p \, f \frac{\xi_\alpha \xi_\beta}{m}, \qquad
 q_\alpha = \int d^2p \, f \frac{\bm{\xi}^2}{2m} \frac{\xi_\alpha}{m}.\label{eq:sigmaq_def}
\end{equation}
The force \eqref{eq:F_def} has the effect of replacing the ideal gas pressure $p^i$ in the momentum equation \eqref{eq:macro_u} 
with the van der Waals pressure $p^w$, while also adding a surface tension term:
\begin{equation}
 \rho \left[\partial_t u_\alpha + (\bm{u} \cdot \nabla) u_\alpha\right] =
 n \sigma\nabla(\Delta n) - \partial_\alpha p^w - \partial_\beta \Pi_{\alpha\beta}.
 \label{eq:macro_u_waals}
\end{equation}
The above modification to the momentum equation is sufficient to induce spontaneous phase separation when the 
temperature $T$ of the fluid is smaller than the critical temperature $T_c$.

Furthermore, a Chapman-Enskog analysis shows that, at first order, the viscous stress tensor and the heat flux are 
given by \cite{ambrusPRE2012}:
\begin{equation}
 \Pi_{\alpha\beta} = -\eta (\partial_\alpha u_\beta + \partial_\beta u_\alpha - 
 \delta_{\alpha\beta} \nabla \cdot \bm{u}), \qquad 
 q_\alpha = -\kappa_T \nabla_\alpha T,\label{eq:CE}
\end{equation}
where the dynamic viscosity $\eta$ and heat conductivity $\kappa_T$ have the following expressions:
\begin{equation}
 \eta = \tau n T, \qquad 
 \kappa_T = \frac{2}{m} \tau n T.
\end{equation}
The ensuing Prandtl number ${\rm Pr} = c_p \eta / \kappa_T$ ($c_p = 2/m$ is the specific heat
for a two-dimensional monatomic gas) is fixed in the BGK model at:
\begin{equation}
 {\rm Pr} = 1,\label{eq:Pr}
\end{equation}
while the hard sphere model predicts that ${\rm Pr} = 2/3$.

Given the above mentions there are two remarks we want to highlight: first, 
since the phase separation mechanism is induced
through the use of a body force, the pressure appearing in the energy equation 
\eqref{eq:macro_e} is still the ideal pressure, instead of the van der Waals pressure \cite{luo00pre}; and 
second, the value of ${\rm Pr}$ \eqref{eq:Pr} is fixed at $1$.
The energy equation could be altered such that the ideal fluid pressure is replaced by the van der Waals pressure
by employing the modified Boltzmann (i.e. Enskog) equation \cite{he02jsp}.
Furthermore, there are various methods to alter the value of ${\rm Pr}$, of which we mention the Shakhov \cite{ambrusPRE2012} and the MRT 
\cite{mcnamara95,chen10epl} models.
These possible enhancements are the subject of forthcoming work. 
In this paper, we are interested to perform a first exploration of the capabilities of the single particle distribution function
LB model based on Gauss-Hermite quadratures to simulate liquid-vapour thermal flows and, for simplicity, we only considered
the simple form of both the body force term and of the collision term in Eq. \eqref{evolution}.

\section{Numerical scheme and boundary conditions}\label{sec:scheme}

\subsection{Time stepping}\label{sec:scheme:RK3}

In this paper, the time stepping is implemented using the
explicit third order total variation diminishing (TVD) Runge-Kutta (RK-3) 
time marching procedure \cite{gottlieb98,henrick05,shu88,trangenstein07}
associated to
the fifth-order weighted essentially non-oscillatory (WENO-5) scheme 
\cite{gan11,jiang96} for computing the advection. 
In order to implement the time stepping algorithm, it is convenient 
to cast the Boltzmann equation (\ref{evolution}) in the following form:
\begin{equation}
 \partial_t f_{\bm{\kappa}} = L[f_{\bm{\kappa}}], \qquad 
 L[f_{\bm{\kappa}}] =
 -  \frac{1}{\,m\,}{{\bm{p}}_{\bm{\kappa}}}\cdot \nabla f_{\bm{\kappa}} -
 \bm{F} \cdot  (\nabla_{\bm{p}} f)_{\bm{\kappa}}
  -\frac{1}{\tau}[f_{\bm{\kappa}} - f^{eq}_{\bm{\kappa}}] .
  \label{ecast}
  \end{equation}
%Following the discretisation of the time variable using equal 
%time steps $\delta t$, the distribution function at time step $\ell$ 
%is $f_{\ell} \equiv f(t_{\ell})$, when the time coordinate 
%has the value $t_{\ell} = \ell \delta t$, taken with respect to the 
%initial time $t_0 = 0$. For simplicity, the dependence of the distribution 
%function on the spatial coordinates and on the momentum degrees of freedom
%was omitted.
The third-order TVD Runge-Kutta integrator  gives the following 
algorithm for computing the values of the distribution functions
$f_{\bm{\kappa}}$ at time $t+ \delta t$:
\begin{align}
 f_{\bm{\kappa}}^{(1)}({\bm{x}},t) =& f_{\bm{\kappa}}({\bm{x}},t) + \delta t \, L[f_{\bm{\kappa}}({\bm{x}},t)], \nonumber\\
 f_{\bm{\kappa}}^{(2)}({\bm{x}},t) =& \frac{3}{4} f_{\bm{\kappa}}({\bm{x}},t) + \frac{1}{4} f_{\bm{\kappa}}^{(1)}({\bm{x}},t) + 
 \frac{1}{4} \delta t\, L[f_{\bm{\kappa}}^{(1)}({\bm{x}},t)],\nonumber\\
 f_{\bm{\kappa}}({\bm{x}},t+\delta t)
 =& \frac{1}{3} f_{\bm{\kappa}}({\bm{x}},t) + \frac{2}{3} f_{\bm{\kappa}}^{(2)}({\bm{x}},t) + 
 \frac{2}{3} \delta t\, L[f_{\bm{\kappa}}^{(2)}({\bm{x}},t)]. \label{eq:rk3}
\end{align}

\subsection{Advection scheme}\label{sec:scheme:WENO5}

The advection term which appears in Eq.~(\ref{ecast}) above, namely
\begin{equation}
 \frac{1}{m} \bm{p_{\bm{\kappa}}} \cdot \nabla f_{\bm{\kappa}} = 
 \frac{1}{m} p_{k_1} \partial_x f_{\bm{\kappa}} + 
 \frac{1}{m} p_{k_2} \partial_y f_{\bm{\kappa}}, \label{eq:advection}
\end{equation}
is computed using the WENO-5 scheme \cite{gan11,jiang96} along each Cartesian coordinate.
Assuming that the flow domain is discretized using 
$1 \le i \le \mathfrak{N}_1$ and $1 \le j \le \mathfrak{N}_2$ nodes on the 
$x$ and $y$ axes, respectively, Eq.~\eqref{eq:advection} becomes:
\begin{equation}
 \left(\frac{1}{m} \bm{p} \cdot \nabla f\right)_{\bm{\kappa}; i, j} = 
 \frac{\mathcal{F}_{\bm{\kappa}; i + 1/2, j} - \mathcal{F}_{\bm{\kappa}; i - 1/2, j}}
 {x_{i+1/2,j} - x_{i-1/2,j}} + 
 \frac{\mathcal{F}_{\bm{\kappa}; i, j + 1/2} - \mathcal{F}_{\bm{\kappa}; i, j - 1/2,}}
 {y_{i,j+1/2} - y_{i,j-1/2}} 
\end{equation}
where 
% the flux $\mathcal{F}_{s+1/2}$ corresponding to the interface  between 
% the cells centred on $x_s$ and $x_{s+1}$ is computed 
% in an upwind-biased approach using the WENO-5 algorithm  
% \cite{gan11}. When the advection velocity is $p_x / m > 0$, whe have
$\mathcal{F}_{\bm{\kappa};i+1/2,j}$ represents the flux of 
$f$ advected with velocity $p_{k_1} / m$ through the vertical interface between 
the cells centered on $\bm{x}_{i,j}$ and $\bm{x}_{i+1,j}$. Similarly, 
$\mathcal{F}_{\bm{\kappa};i, j+1/2}$ represents the flux of $f$ advected 
with velocity $p_{k_2} / m$ through the horizontal interface between 
the cells centered on $\bm{x}_{i,j}$ and $\bm{x}_{i,j+1}$.
The construction of these fluxes is summarized below only for the 
horizontal direction and under the assumption of a positive advection 
velocity $p_{k_1} / m > 0$. In this case, the flux 
$\mathcal{F}_{\bm{\kappa}; i+1/2, j}$ can be computed using the following expression \cite{gan11}:
\begin{equation}
\mathcal{F}_{i+1/2} = \overline{\omega}_1\mathcal{F}^1_{i+1/2} +
\overline{\omega}_2\mathcal{F}^2_{i+1/2} + \overline{\omega}_3\mathcal{F}^3_{i+1/2},
\label{eq:weno5_flux}
\end{equation}
where for brevity, the velocity and vertical indices $\bm{\kappa} = \{k_1, k_2\}$ 
and $j$ were omitted. 
The interpolating functions $\mathcal{F}^q_{i + 1/2}$ ($q = 1,2,3$) 
are given by:
\begin{align}
\mathcal{F}^1_{i+1/2} =& \frac{p_{k_1}}{m} \left(\frac{1}{3}f_{i-2} - \frac{7}{6} f_{i-1} + \frac{11}{6} f_i\right), \nonumber \\
\mathcal{F}^2_{i+1/2} =& \frac{p_{k_1}}{m} \left(-\frac{1}{6}f_{i-1} + \frac{5}{6} f_{i} + \frac{1}{3} f_{i+1}\right), \nonumber \\
\mathcal{F}^3_{i+1/2} =& \frac{p_{k_1}}{m} \left(\frac{1}{3}f_{i} + \frac{5}{6} f_{i+1} - \frac{1}{6} f_{i+2}\right).
\end{align}
% while the weighting factors $\overline{\omega}_q$,
% $q \in \{1,\,2,\,3\}$, are defined as:
The weighting factors $\overline{\omega}_q$ appearing in 
Eq.~\eqref{eq:weno5_flux} are given by:
\begin{equation}
\overline{\omega}_q = \frac{\widetilde{\omega}_q}{\widetilde{\omega}_1+\widetilde{\omega}_2+\widetilde{\omega}_3}, \qquad 
\widetilde{\omega}_q = \frac{\delta_q}{\varsigma^2_q}.
\label{eq:weno5_omega}
\end{equation}
% The ideal weights $\delta_q$ are:
where the ideal weights $\delta_q$ have the following values:
\begin{equation}
 \delta_1 = 1/10, \qquad \delta_2 = 6/10,\qquad \delta_3 = 3/10, 
 \label{eq:weno5_delta}
\end{equation}
% while the indicators of smoothness $\varsigma_q$ are given by:
The indicators of smoothness $\varsigma_q$ can be computed as follows:
\begin{align}
\varsigma_1 =& \frac{13}{12} \left(f_{i-2} -2f_{i-1} + f_i \right)^2 
+ \frac{1}{4} \left( f_{i-2} - 4f_{i-1} + 3f_i \right)^2, 
\nonumber \\
\varsigma_2 =& \frac{13}{12} \left(f_{i-1} -2f_{i} + f_{i+1} \right)^2 
+ \frac{1}{4} \left( f_{i-1} - f_{i+1} \right)^2,
\nonumber \\
\varsigma_3 =& \frac{13}{12} \left(f_{i} -2f_{i+1} + f_{i+2} \right)^2 
+ \frac{1}{4} \left( 3f_{i} -4 f_{i+1} + f_{i+2} \right)^2.
\label{eq:weno5_sigma}
\end{align}

\begin{table}
\begin{center}
\begin{tabular}{r|rrr}
 & $\overline{\omega}_1$ & $\overline{\omega}_2$ & $\overline{\omega}_3$ \\\hline  
$\varsigma_1 = \varsigma_2 = \varsigma_3 = 0$ & $0.1$ & $0.6$ & $0.3$ \\\hline
$\varsigma_2 = \varsigma_3 = 0$ & $0$ & $2/3$ & $1/3$ \\
$\varsigma_3 = \varsigma_1 = 0$ & $1/4$ & $0$ & $3/4$ \\
$\varsigma_1 = \varsigma_2 = 0$ & $1/7$ & $6/7$ & $0$ \\\hline
$\varsigma_1 = 0$ & $1$ & $0$ & $0$\\
$\varsigma_2 = 0$ & $0$ & $1$ & $0$\\
$\varsigma_3 = 0$ & $0$ & $0$ & $1$
\end{tabular}
\end{center}
\caption{The values of the weighting factors $\overline{\omega}_q$ 
% defined in Eq.~
\eqref{eq:weno5_omega} when one, two or all three of the 
indicators of smoothness $\sigma_q$ ($q = 1,2,3$) 
%$\sigma_i$ ($i = 1,2, 3$) functions 
have vanishing values.\label{tab:weno}
}
\end{table}

The computation of the weighting factors 
$\overline{\omega}_q$ \eqref{eq:weno5_omega} implies the 
division between the ideal weights $\delta_q$ \eqref{eq:weno5_delta}
and the indicators of smoothness $\varsigma_q$ \eqref{eq:weno5_sigma}.
In order to avoid division by $0$ when either one, two or all 
three of the indicators of smoothness vanish, it is customary to 
modify Eq.~\eqref{eq:weno5_sigma} by adding a small quantity
$\varepsilon \simeq 10^{-6}$ to $\varsigma_q$. According to 
Ref.~\cite{henrick05}, $\varepsilon$ is a dimensionful quantity 
and its effect on the WENO-5 scheme depends on the typical magnitude 
of the advected function $f$. It can be seen from 
Tab.~\ref{trots} that the ratio 
between the largest weight $(8/15)^2$ and the smallest weight 
$[(7-2\sqrt{10})/60]^2$ is $\sim 2250$, i.e. the set of 
discrete distributions typically spans three orders of magnitude.
Under these circumstances, we follow Refs.~\cite{blaga18prc,busuioc18pre}
and compute the weighting factors $\overline{\omega}_q$ directly 
using Tab.~\ref{tab:weno} in the limiting cases when 
any of the indicators of smoothness vanishes.

\subsection{High order stencils for the gradient and gradient of the Laplacian}
\label{sec:scheme:stencil}

The WENO-5 scheme described in Subsec.~\ref{sec:scheme:WENO5} is of fifth 
order with respect to the lattice spacing for smooth functions \cite{jiang96}. 
The smallest square covering all 
nodes involved in updating a given lattice site comprises 
$7 \times 7 = 49$ lattice sites. It is therefore natural to 
seek the computation of the gradient and gradient of the Laplacian 
appearing in Eq.~\eqref{eq:F_def} using the $49$-point stencils 
described in Refs.~\cite{leclaireJSC2014,patraNMPDE2006}. 
For the reader's convenience, these stencils are summarized below.

Let $\Delta p = p^i - p^w$ be the difference between the ideal and van der Waals pressures. 
Following the discretization of the fluid domain using 
$\mathfrak{N}_1 \times \mathfrak{N}_2$ equal sized square cells of sides $\delta s$, 
the function $\Delta p$ is replaced by a set of time-dependent functions $\Delta p_{i,j}$ 
($1 \le i \le \mathfrak{N}_1$, $1 \le j \le \mathfrak{N}_2$). 
In order to compute the first term in Eq.~\eqref{eq:F_def} corresponding to the 
cell $(i,j)$, the gradient of $\Delta p$ can be obtained using the following procedure 
\cite{leclaireJSC2014}:
\begin{equation}
\begin{pmatrix}
 [\partial_x \Delta p]_{i,j} \\
 [\partial_y \Delta p]_{i,j} 
\end{pmatrix}
  = \frac{1}{\delta s} \sum_{l,m} 
\begin{pmatrix}
 l \\ m
\end{pmatrix}
 \Delta p_{i + l, j + m} w^{\abs{l}, \abs{m}} + O[(\delta s)^6],
 \label{eq:stencil_grad}
\end{equation}
where $l, m \in \{0, \pm 1, \pm 2, \pm 3\}$. The weights $w^{\abs{l}, \abs{m}}$ 
are symmetric with respect to $l$ and $m$ ($w^{\abs{l}, \abs{m}} = 
w^{\abs{m}, \abs{l}}$), having the following values:
\begin{gather}
 w^{1,0} = \frac{31}{70}, \qquad 
 w^{1,1} = \frac{27}{140}, \qquad 
 w^{2,1} = -\frac{3}{70}, \qquad 
 w^{2,2} = \frac{3}{560}, \qquad 
 w^{3,0} = -\frac{1}{630}, \qquad 
 w^{3,1} = \frac{1}{280}, 
\end{gather}
while $w^{0,0} =w^{2,0} = w^{3,2} = w^{3,3} = 0$.
The resulting scheme is $6$th order accurate and $8$th order isotropic 
for smooth functions \cite{leclaireJSC2014}.

The second term in Eq.~\eqref{eq:F_def} involves the gradient of the Laplacian of the 
density $n$, which is replaced using the notation introduced above by a set of 
time-dependent functions $n_{i,j}$ ($1 \le i \le \mathfrak{N}_1$, 
$1 \le j \le \mathfrak{N}_2$). The computation of the gradient of the Laplacian 
is performed using the following procedure:
\begin{subequations}\label{eq:stencil_glap}
\begin{equation}
\begin{pmatrix}
 [\partial_x (\Delta n)]_{i,j}\\
 [\partial_y (\Delta n)]_{i,j}
\end{pmatrix} = \frac{1}{(\delta s)^3} \sum_{l,m} 
\begin{pmatrix}
 S_{i,j} \\ S_{j,i}
\end{pmatrix}
 n_{i+l, j+m} + O[(\delta s)^4],
\end{equation}
where $l, m \in \{0, \pm 1, \pm 2, \pm 3\}$ and the matrix $S_{i,j}$ can be summarized 
as follows:
\begin{equation}
 S_{i,j} = 
 \begin{pmatrix} 
  c_1 & c_5 & c_9 & 0 & -c_9 & -c_5 & -c_1 \\
  c_2 & c_6 & c_{10} & 0 & -c_{10} & -c_6 & -c_2 \\
  c_3 & c_7 & c_{11} & 0 & -c_{11} & -c_7 & -c_3 \\
  c_4 & c_8 & c_{12} & 0 & -c_{12} & -c_8 & -c_4 \\
  c_3 & c_7 & c_{11} & 0 & -c_{11} & -c_7 & -c_3 \\
  c_2 & c_6 & c_{10} & 0 & -c_{10} & -c_6 & -c_2 \\
  c_1 & c_5 & c_9 & 0 & -c_9 & -c_5 & -c_1 
 \end{pmatrix}.
\end{equation}
\end{subequations}
The convention in the above is that $S_{0,0}$ is the central matrix element  
(i.e., $S_{1,2} = -c_{7}$). The coefficients $c_k$ have the following values:
\begin{equation}
 c_1 = -\frac{17}{260}, \quad 
 c_4 = -\frac{59}{240}, \quad 
 c_6 = \frac{13}{120}, \quad 
 c_7 = -\frac{7}{30}, \quad 
 c_8 = \frac{5}{4}, \quad 
 c_{10} = -\frac{7}{60}, \quad 
 c_{11} = \frac{187}{240}, \quad 
 c_{12} = -\frac{59}{20},
\end{equation}
while $c_2 = c_3 = c_5 = c_9 = 0$. Eq.~\eqref{eq:stencil_glap} is 
$4$th order accurate and $6$th order isotropic for smooth functios
\cite{patraNMPDE2006}.

\subsection{Boundary conditions} \label{sec:scheme:bcond}

In this paper, we consider the phase separation in a van der Waals
fluid placed between two parallel walls which are perpendicular 
to the $x$ axis. The flow is always assumed to be periodic along the 
$y$ axis. As already mentioned in Subsec.~\ref{sec:scheme:stencil},
the flow domain is discretized using a $2D$ square lattice with
${\mathfrak{N}}_{1}\times {\mathfrak{N}}_{2}$ nodes. Let 
${\bm{x}} = \delta s [(i - \frac{1}{2}) \bm{e}_1 + (j - \frac{1}{2}) \bm{e}_2]$,
$1\leq i \leq {\mathfrak{N}}_{1}$, $1\leq j \leq {\mathfrak{N}}_{2}$, be
the position vectors of the nodes in this lattice.
The discussion in this section focuses on the 
implementation of the specular and diffuse reflection boundary conditions for the 
distribution functions $f_{\bm{\kappa}} = f_{k_1, k_2}$, as well 
as for the macroscopic fields involved in the computation of the 
force term \eqref{eq:F_def}.

\subsubsection{Specular boundary conditions} \label{sec:scheme:bcond:spec}

During the validation tests considered in Secs.~\ref{sec:planar} and 
\ref{sec:laplace},
concerning a plane interface and the Laplace pressure test, the final 
configuration is considered to be symmetric with respect to the channel centerline,
such that the simulation domain can be reduced 
by implementing {\it specular reflection} along the symmetry planes. 
In particular, let us consider that the center of the channel is 
located at $(i, j) = (1/2, 1/2)$. In order to perform the advection 
of the distribution function $f_{\bm{\kappa}; i, j} = f_{k_1, k_2; i, j}$ 
using Eq.~\eqref{eq:weno5_flux}, the value of $f_{\bm{\kappa}; i, j}$ must 
be defined below the bottom fluid domain boundary, where $j \in \{-2, -1, 0\}$ 
and $1 \le i \le \mathfrak{N}_1$, as well as to the left of the fluid domain 
where $i \in \{-2, -1, 0\}$ and $1 \le j \le \mathfrak{N}_2$.
The specular reflection concept can be implemented along the 
bottom horizontal boundary as follows:
\begin{equation}
 f_{k_1, k_2; i, 0} = f_{{k}_1, \widetilde{k}_2; i, 1}, \qquad 
 f_{k_1, k_2; i, -1} = f_{{k}_1, \widetilde{k}_2; i, 2}, \qquad 
 f_{k_1, k_2; i, -2} = f_{{k}_1, \widetilde{k}_2; i, 3}.
\end{equation}
On the left vertical boundary, the following procedure can be employed:

\begin{equation}
 f_{k_1, k_2; 0, j} = f_{\widetilde{k}_1, {k}_2; 1, j}, \qquad 
 f_{k_1, k_2; -1, j} = f_{\widetilde{k}_1, {k}_2; 2, j}, \qquad 
 f_{k_1, k_2; -2, j} = f_{\widetilde{k}_1, {k}_2; 3, j},
\end{equation}

The values $\widetilde{k}_1$ and $\widetilde{k}_2$ are defined with respect to $k_1$ and $k_2$, such that the corresponding 
velocity component is reverted, i.e.:
\begin{equation}
 p_{\widetilde{k}_1} = -p_{k_1}, \qquad p_{\widetilde{k}_2} = -p_{k_2}.
\end{equation}

\subsubsection{Diffuse reflection boundary conditions} \label{sec:scheme:bcond:diffuse}

Let us now consider the implementation of the diffuse reflection boundary conditions.
For definiteness, we refer to the right wall, which is located 
at $i = \mathfrak{N}_1 + \frac{1}{2}$.
According to the diffuse reflection concept, the distribution function
of the fluid particles that return from a plane wall is the Maxwell-Boltzmann
equilibrium distribution function corresponding to the wall velocity
${\bm{u}}_{w}$ and temperature $T_{w}$ \cite{ambrus16jcp,ambrusPRE2012,ansumaliPRE2002}. 
This amounts to setting the flux 
$\mathcal{F}_{k_1, k_2; \mathfrak{N}_1 + \frac{1}{2}, j}$ \eqref{eq:weno5_flux}
through the interface between 
the fluid and the wall, located at $i = \mathfrak{N}_1 + \frac{1}{2}$, to the 
following value:
\begin{equation}
 \mathcal{F}_{\bm{\kappa}; \mathfrak{N}_1 + \frac{1}{2}, j} = \frac{p_{k_1}}{m} 
 f^{eq}_{w; \bm{\kappa}}, \qquad (p_{k_1} < 0),
 \label{eq:diffuse_flux}
\end{equation}
where $f^{eq}_{w; \bm{\kappa}}$ is computed using Eq.~\eqref{eq:feq} by setting 
$u_x = 0$, $u_y = u_w$, $T = T_w$ and $n = n_w$, where $u_w$ represents the 
vertical velocity of the wall and the wall particle number 
density $n_w$ will be determined below in Eq.~\eqref{eq:nw}.

The flux in Eq.~\eqref{eq:diffuse_flux} can be achieved in the context of the 
WENO-5 scheme discussed in Subsec.~\ref{sec:scheme:WENO5} by fixing the distribution 
functions in the ghost nodes at $i \in \{\mathfrak{N}_1 + 1, \mathfrak{N}_1 + 2, 
\mathfrak{N}_1 + 3\}$ ($1 \le j \le \mathfrak{N}_2$) as follows \cite{busuioc18pre}:
\begin{equation}
 f_{\bm{\kappa}; \mathfrak{N}_1 + 1, j} = 
 f_{\bm{\kappa}; \mathfrak{N}_1 + 2, j} = 
 f_{\bm{\kappa}; \mathfrak{N}_1 + 3, j} = 
 f^{eq}_{w; \bm{\kappa}}, \qquad (p_{k_1} < 0).
\end{equation}

In order to compute the fluxes of the distributions corresponding to 
particles travelling towards the wall, two ghost nodes are required 
inside the wall. The distributions in these ghost nodes are computed 
using a quadratic extrapolation from the fluid domain, as follows:
\begin{equation}
\begin{pmatrix}
 f_{\bm{\kappa}; \mathfrak{N}_1 + 1, j}\\
 f_{\bm{\kappa}; \mathfrak{N}_1 + 2, j}
\end{pmatrix} =
\begin{pmatrix}
 3 \\ 6
\end{pmatrix} f_{\bm{\kappa}; \mathfrak{N}_1, j} - 
\begin{pmatrix}
 3 \\ 8
\end{pmatrix} f_{\bm{\kappa}; \mathfrak{N}_1 - 1, j} + 
\begin{pmatrix}
 1 \\ 3
\end{pmatrix} f_{\bm{\kappa}; \mathfrak{N}_1 - 2, j}, \qquad (p_{k_1} > 0).
\end{equation}

The value of $n_{w}$ in the expression of $f^{eq}_{\bm{\kappa}; w}$ 
% ${\mathfrak{F}}_{\bm{\kappa}, x}^{in}$ above 
is found for each value of $j$ by requiring the total
flux of particles to vanish at the wall:
% \begin{equation}
% {\mathfrak{F}}_{w}^{out} + {\mathfrak{F}}_{w}^{in} = 0,
% \end{equation}
\begin{equation}
 \sum_{k_1, k_2} \mathcal{F}_{\bm{\kappa}; \mathfrak{N}_1 + 1/2, j} = 0,
\end{equation}
such that $n_w$ can be computed using:
\begin{equation}
 n_w = -\frac{\displaystyle \sum_{p_{k_1} > 0, k_2} 
 \mathcal{F}_{\bm{\kappa}; \mathfrak{N}_1 + 1/2, j}}
 {\displaystyle \sum_{p_{k_1} < 0, k_2} g_{k_1}(0, T_w) g_{k_2}(u_w, T_w) p_{k_1}/m},
 \label{eq:nw}
\end{equation}
where $g_{k_\alpha}(u_\alpha, T)$ is defined in Eq.~\eqref{g_alpha}.

\subsubsection{Boundary conditions for the macroscopic fields} 

As discussed in Subsec.~\ref{sec:scheme:stencil}, the computation of the force 
term \eqref{eq:F_def} is performed using $7 \times 7$ stencils. 
In order to employ these stencils near the fluid domain boundary, 
the macroscopic fields ($\Delta p = p^i - p^w$ and $n$) are reflected with respect to the 
coordinate axes, for both the specular and the diffuse reflection boundary conditions. This reflection has to be performed 
at the corners of the ghost domain, i.e. for the bottom left corner, 
where a double reflection occurs as follows:
\begin{equation}
 \Delta p_{1-i, 1-j} = \Delta p_{1-i, j} = 
 \Delta p_{i, 1-j} = \Delta p_{i, j}, \qquad 
 n_{1-i, 1-j} = n_{1-i, j} = n_{i, 1-j} = n_{i, j}.
\end{equation}
where $1 \le i, j \le 3$.

% where
% \begin{equation}
% {\mathfrak{F}}_{w}^{in} (\bm{x}_{1,j},t) =  
% \sum_{\substack{k_{1} = 1 \\ p_{k_{1}} > 0}}^{Q}
% \sum_{k_{2} =1}^{Q}
% {\mathfrak{F}}_{\bm{\kappa}, x}^{in} ({\bm{x}_1},t) =
% \,n_{w}
% \sum_{\substack{k_{1} = 1 \\ p_{k_{1}} > 0}}^{Q}
% \sum_{k_{2} =1}^{Q}
% \frac{\,p_{k_{1}}\,}{m}
% g_{k_{1}}({u_{w}},T_{w})g_{k_{2}}({u_{w}},T_{w}).
% \end{equation}
% Finally, in order to compute the fluxes $\mathfrak{F}^{out}_{\bm{\kappa},x}(\bm{x}_{1,j}, t)$ and 
% $\mathfrak{F}^{in}_{\bm{\kappa}, x}(\bm{x}_{2,j}, t)$ for $p_{k_x} > 0$, the value of 
% $f_{\bm{\kappa}}(\bm{x}_{0,j}, t)$ must be specified. Since we assume that the particles emerging from the wall 
% towards the fluid domain is Maxwellian, we set 
% \begin{equation}
%  f_{\bm{\kappa}}(\bm{x}_{0,j}, t) = n_{w} g_{k_{1}}({u_{w}},T_{w})g_{k_{2}}({u_{w}},T_{w}).
% \end{equation}

% {\color{blue}{
% On both the top ($j={\mathfrak{N}}_{2})$ and the bottom ($j=1$) boundaries
% of the flow domain, 
% in Figure \ref{pspp}, 
% periodic conditions apply.
% and 
% the corresponding {\emph{in}} and {\emph{out}} fluxes in Eq.(\ref{fevol})
% are handled in the same manner as in the bulk nodes of the lattice.
% }}

\section{Planar interface}\label{sec:planar}

% At time $t=0$, we always start with a fluid at rest in local thermal equilibrium, whose temperature equals the critical 
% temperature $T_{c}=1.0$. The
% initial value of the particle number density in the lattice nodes fluctuates
% around the critical value $n_{c}=1$  with random deviations of
% at most $0.01\times n_{c}$. 
% In this arrangement, the energy
% is extracted through the walls and the liquid and vapor phases start to
% separate. 

In this section, the capabilities of our thermal models are discussed in the 
simple case of a planar interface. Complete homogeneity is assumed along 
the vertical ($y$) direction, such that the flow domain becomes essentially 
one-dimensional. The number of grid points in this case 
is $\mathfrak{N}_1 \times 1$ and the spatial derivatives are 
computed only along the horizontal direction. In particular, the 
gradient of $\Delta p = p^i - p^w$ \eqref{eq:stencil_grad} reduces to:
\begin{equation}
 (\partial_x \Delta p)_i = \frac{1}{\delta s}\left( 
 \frac{1}{60} \Delta p_{i+3} - \frac{3}{20} \Delta p_{i+2} + 
 \frac{3}{4} \Delta p_{i+1} - \frac{3}{4} \Delta p_{i-1} +
 \frac{3}{20} \Delta p_{i-2} - \frac{1}{60} \Delta p_{i-3}\right).
\end{equation}
Similarly, the stencil \eqref{eq:stencil_glap} for the computation of 
$\nabla (\Delta n)$ reduces to:
\begin{equation}
 [\nabla (\Delta n)]_i = \frac{1}{(\delta s)^3} \left(
 -\frac{1}{8} n_{i+3} + n_{i+2} - \frac{13}{8} n_{i+1} +
 \frac{13}{8} n_{i-1} - n_{i-2} + \frac{1}{8} n_{i-3}\right).
\end{equation}

The analysis presented in this section concerns only the stationary state, in which we assume 
that the gas phase occupies the central half of the channel, while the liquid phase is confined to 
the vicinity of the walls. The stationary state is assumed to be symmetric with respect to the 
center of the channel. Assuming that the time derivatives 
in Eqs.~\eqref{eq:macro} vanish, Eq.~\eqref{eq:macro_n} shows that 
$u_x = 0$, while Eq.~\eqref{eq:macro_e} shows that $\partial_x q_x = 0$, since 
the viscous part of the stress tensor $\Pi_{\alpha\beta}$ \eqref{eq:CE} vanishes.
Since the heat flux must vanish at the center of the domain due to symmetry, 
$\partial_x q_x = 0$ implies $q_x = 0$ throughout the fluid domain. Finally, 
Fourier's law \eqref{eq:CE} shows that $T = T_w$ everywhere inside the fluid 
domain. The interface shape can be found by solving the stationary 
limit of Eq.~\eqref{eq:macro_u_waals}:
\begin{equation}
 \partial_x p^w - \sigma \partial_x (\Delta n) = 
 \partial_x \left(p^w - \sigma \Delta n\right) = 0.
\end{equation}
An approximate solution due to Wagner and Pooley \cite{wagner07} for the 
interface profile in the right half of the channel is:
\begin{equation}
 n(x) = n_g + \frac{n_l - n_g}{2}\left[1 + \tanh \frac{(x - x_0)}{\xi}\right],
 \label{eq:tanh}
\end{equation}
where $n_g$ and $n_l$ are the gas and liquid densities, $\xi$ is the interface 
width and $x_0$ is the interface position. This solution loses validity when $T$ is
 smaller than $1$.

In order to save computational time, only the right half of the 
channel is simulated, while specular reflection boundary conditions are imposed at the left boundary,
as discussed in Sec.~\ref{sec:scheme:bcond:spec}. The wall is located at 
$i_w = \mathfrak{N}_x + \frac{1}{2}$, where $x_{i_w} = 0.5$, such that the number of nodes is 
$\mathfrak{N}_x = \mathfrak{N} / 2$ and the lattice spacing is 
$\delta s = 1/\mathfrak{N} = 1 / 2\mathfrak{N}_x$. 
The system is initialized with a fluid at constant temperature $T = T_w < 1$ 
with the density profile given in Eq.~\eqref{eq:tanh}, where 
$n_g$ and $n_l$ are the liquid and gas densities at $T = T_w$ predicted through the 
well-known Maxwell construction (also known as the equal area construction rule),
a procedure described in Refs.~\cite{prigogine,sekerka,kruger17}. 
The interface width $\xi$ in Eq.~\eqref{eq:tanh} 
is approximated using the following expression:
\begin{equation}
 \xi_w = \sqrt{\frac{8 \kappa}{9(1 - T_w)}}.
 \label{eq:xiwagner}
\end{equation}
The interface position is set to $x_0 = 0.25$.

Since Eq.~\eqref{eq:tanh} is not an exact solution, after the initialization the fluid undergoes an interface adjustment which causes the temperature to rise 
inside the fluid domain. Due to the extraction of the heat through the 
lateral walls, the fluid temperature near the wall remains close to $T_w$. Around the
liquid-gas interface, where the fluid density is not constant, heat generation occurs due to the 
non-vanishing spurious velocity, which is known to plague multiphase 
simulations \cite{gonnellaPRE2007,sofoneaPRE2004,cristeaijmpc2003,sofoneajcp2003,sofoneaijmpc2005,sofoneaPRE2018,hou1997,teng2000}. 
The stability and accuracy of our simulations is 
directly improved when the magnitude of these effects is reduced. Thus, 
Subsec.~\ref{sec:planar:conv} is devoted to the analysis with respect to the 
lattice spacing and time step of the maximum 
temperature difference $\Delta T = T - T_w$, as well as of the
maximum value of the spurious velocity observed in the stationary state.
This test is performed at $T_w = 0.8$, which is considered the working temperature 
in this paper. After choosing a suitable grid, an analysis of the 
robustness of our simulations is performed in Subsec.~\ref{sec:planar:profiles} 
by considering various values of $T_w$. This analysis is important in order
to highlight the validity domain of our simulations. 
A further validation test is performed in Subsec.~\ref{sec:planar:xi}, 
where the width of the planar interface is discussed. 
Finally, the phase diagram is discussed in Subsec.~\ref{sec:planar:maxwell}.

\subsection{Grid convergence}\label{sec:planar:conv}

\begin{figure}
\begin{center}
\begin{tabular}{cc}
\includegraphics[width=220pt]{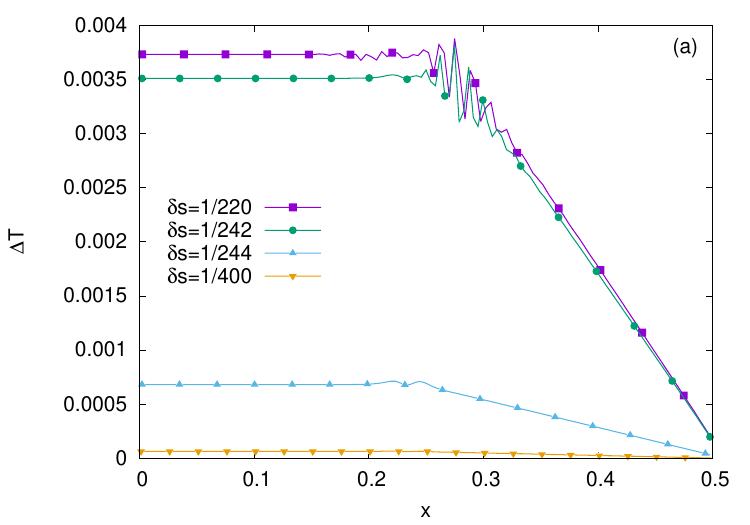} &
\includegraphics[width=220pt]{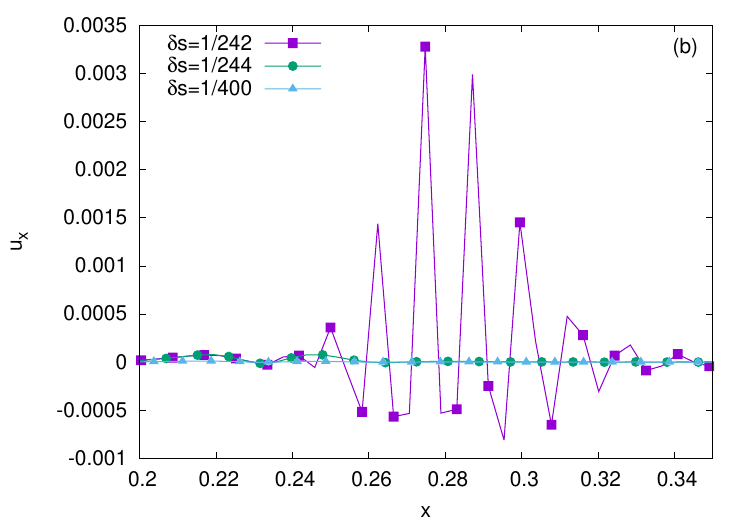} \\
\includegraphics[width=220pt]{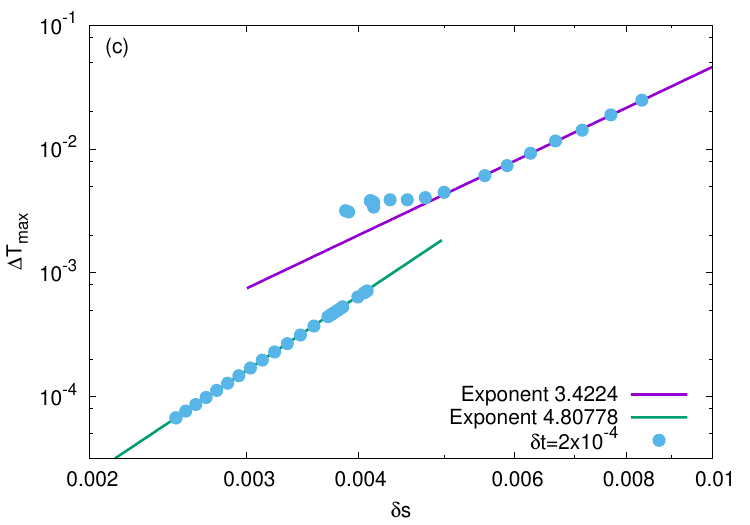} &
\includegraphics[width=220pt]{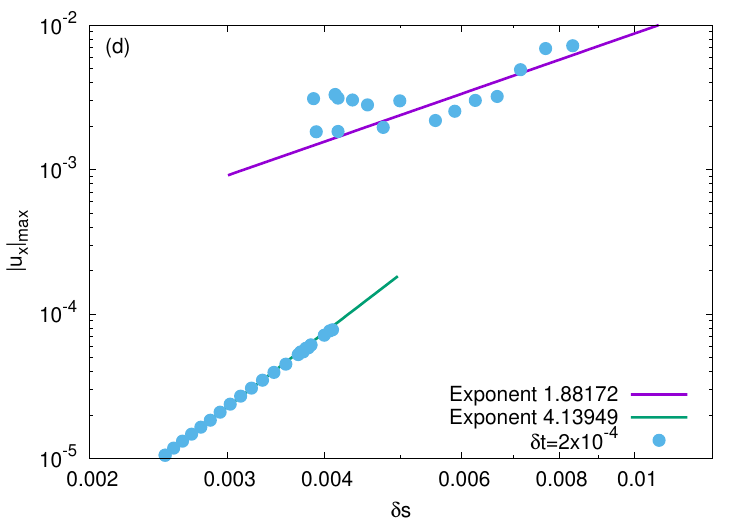}
\end{tabular}
\end{center}
\caption{
Profiles of (a) the temperature difference $\Delta T=T-T_w$ and (b) the 
velocity $u_x$ for various values of the lattice spacing $\delta s = 1/2\mathfrak{N}_x$.
The dependence of the maximum values (c) of the temperature difference $(\Delta T)_{\rm max}$ 
and (d) of the absolute value of the fluid velocity $|u_x|_{\rm max}$ on the lattice spacing 
$\delta s$. The simulation parameters are $\delta t=2\times 10^{-4}$, $\sigma = 10^{-4}$ and 
$T_w=0.8$.
\label{fig:convds}}
\end{figure}

To find the convergence order of the numerical scheme employed in this paper, we performed 
a series of simulations with constant time step $\delta t = 2\times 10^{-4}$ 
at $T_w = 0.8$ and $\sigma = 10^{-4}$, for various values of the lattice spacing 
$\delta s$. Figure~\ref{fig:convds} shows the general decrease of 
the temperature difference $\Delta T = T - T_w$ and of the spurious velocity $u_x$ 
when the lattice spacing $\delta s$ is decreased. The half-channel profiles of $\Delta T$ and $u_x$ are 
shown in Figures~\ref{fig:convds}(a) and (b) for various values of $\delta s$. It 
can be seen that both $\Delta T$ and $u_x$ exhibit strong oscillations at 
large lattice spacings (i.e., when $\delta s \gtrsim 1/242$), which are suddenly 
damped when $\delta s$ decreases under a certain threshold value 
(i.e., when $\delta s \lesssim 1/244$).
A more quantitative analysis of the $\delta s$ dependence of these spurious 
effects can be made at the level of the maximum temperature differece 
$(\Delta T)_{\rm max} = T_{\rm max} - T_w$ and 
maximum absolute value of the velocity $|u_x|_{\rm max}$. Contrary to expectations, 
Figs.~\ref{fig:convds}(c) and (d) reveal two exponents. The first 
corresponds to large values of $\delta s$, when the corresponding profiles 
are plagued by high amplitude oscillations, and has the values 
$\sim 3.42$ for $(\Delta T)_{\rm max}$ and $\sim 1.88$ 
for $|u_x|_{\rm max}$. The second 
exponent refers to the case when the oscillations magnitude is small, having the 
values $\sim 4.80$ for $(\Delta T)_{\rm max}$ and $\sim 4.14$ for $u_{x; {\rm max}}$.

We thus conclude that, in order to perform accurate simulations at $T_w = 0.8$ 
and $\sigma = 10^{-4}$, the lattice spacing should be decreased below $1/244$.
Unless otherwise stated, we will employ $\delta s = 1/320$ for the 
rest of the simulations presented in this paper.

Before ending this subsection, it is worth mentioning that the convergence with 
respect to the time step is much less instructive. This is because, for a fixed value 
of the lattice spacing, the time step is constrained via the CFL condition:
\begin{equation}
 {\rm CFL} = \frac{p_{x; {\rm max}} \delta t}{m \delta s} < 1.
\end{equation}
In particular, for the fifth quadrature order model employed in this paper, 
$p_{x; \rm max} \simeq 2.86$, such that for $\delta s = 1/320$, 
the time step is constrained via $\delta t < 10^{-3}$. Already at this value,
the error due to the time integration is smaller than the error 
introduced by the spatial discretization, such that further 
decreasing the time step does not seem to significantly improve the 
numerical results and the convergence test in this particular case 
is inconclusive. For the rest of this paper, we employ 
$\delta t = 2\times 10^{-4}$.

\subsection{Stationary profiles at various temperatures}\label{sec:planar:profiles}

\begin{figure}[t]
\begin{center}

\begin{tabular}{ccc}
 \includegraphics[width=140pt]{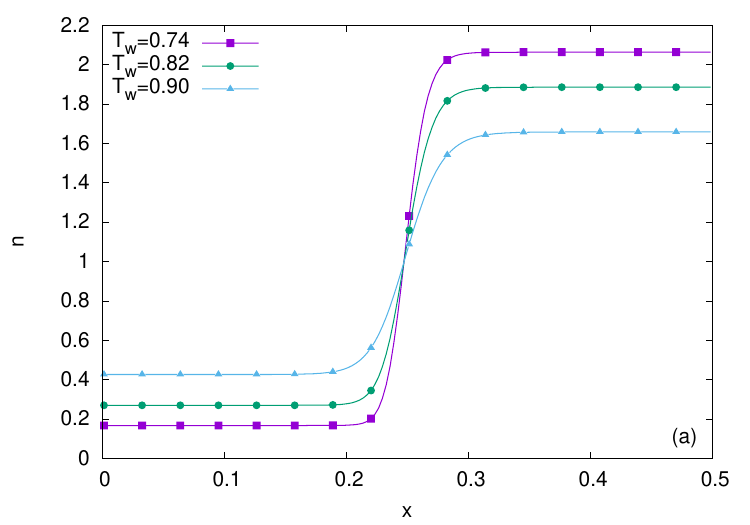}&
 \includegraphics[width=140pt]{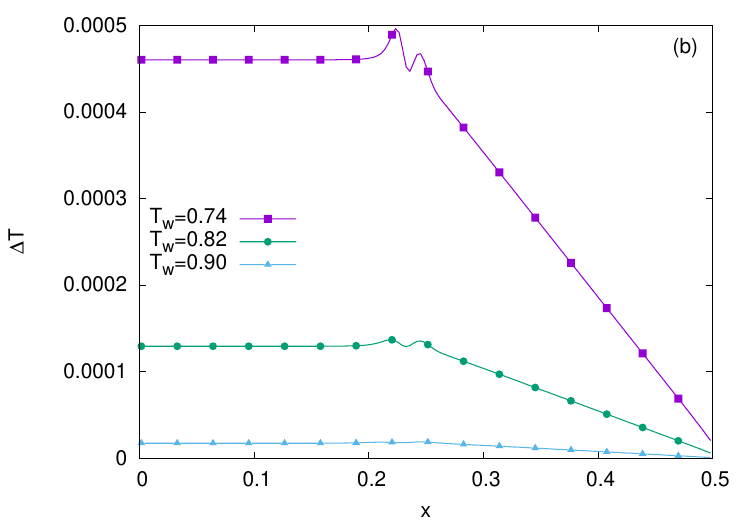}&
 \includegraphics[width=140pt]{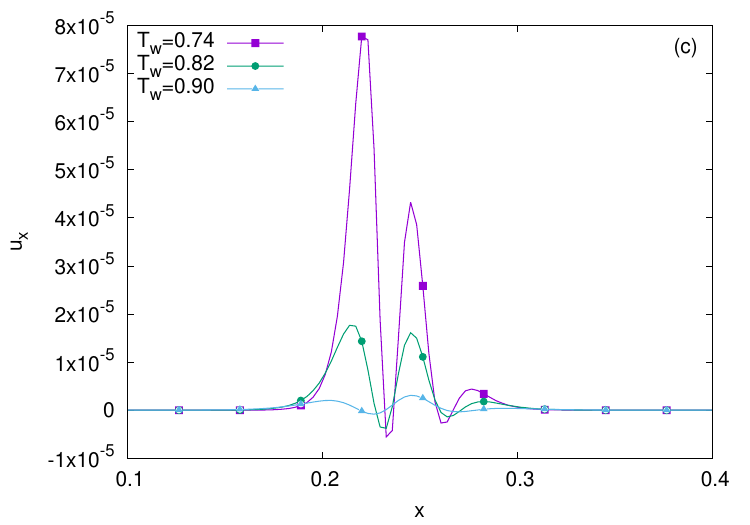}
\end{tabular}
\end{center}

\caption{
Steady state profiles of (a) the density $n(x)$, (b) temperature difference $\Delta T(x) = T(x)-T_w$ 
and (c) the horizontal velocity component $u_x(x)$ for three
values of the wall temperature $T_w=0.74,0.82,0.90$. The time step and lattice spacing 
were set to $\delta t = 2\times 10^{-4}$ and $\delta s = 1/320$, while 
$\sigma = 10^{-4}$ and $\tau = 5\times 10^{-3}$.
\label{fig:profiles}}
\end{figure}

\begin{figure}
\begin{center}
\begin{tabular}{cc}
\includegraphics[width=220pt]{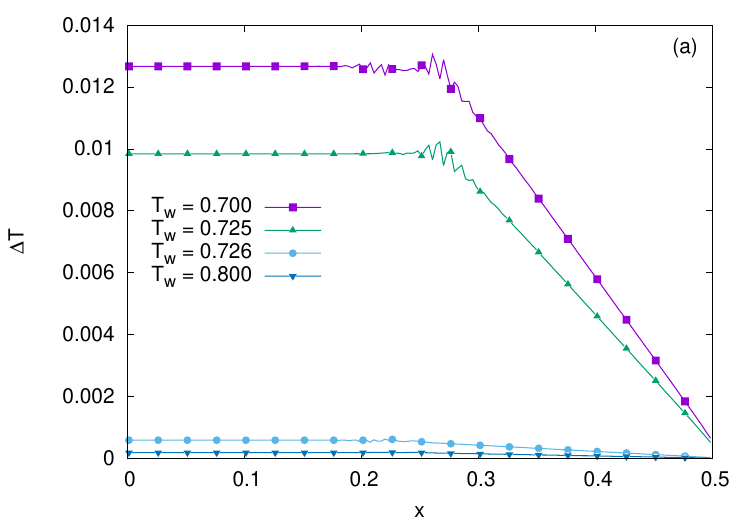} &
\includegraphics[width=220pt]{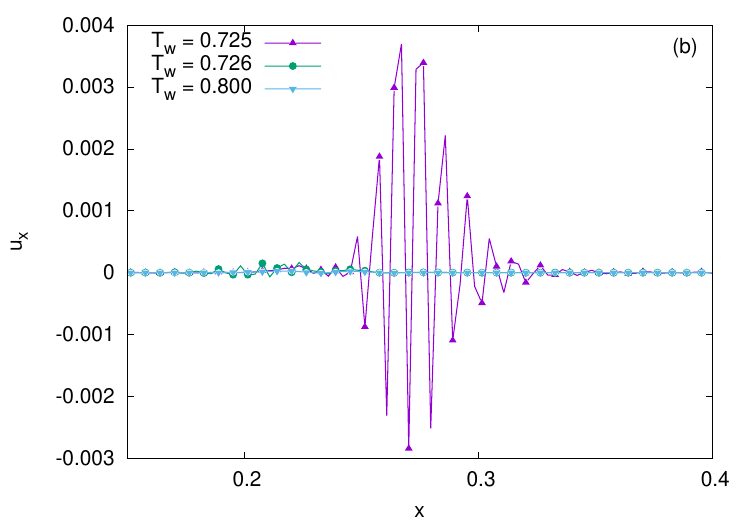} \\
\includegraphics[width=220pt]{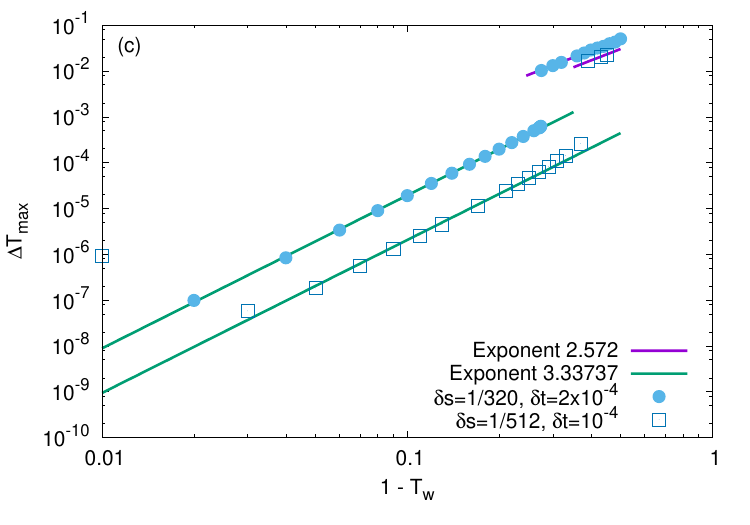} &
\includegraphics[width=220pt]{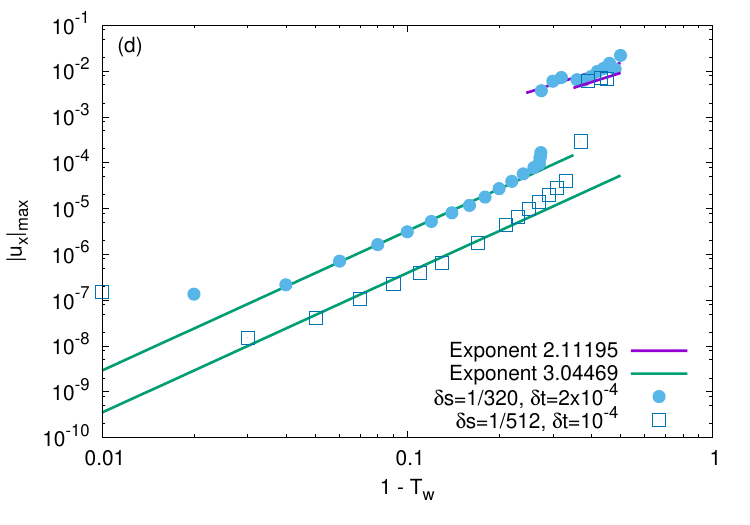} 
\end{tabular}
\end{center}
\caption{
Profiles of (a) the temperature difference $\Delta T=T-T_w$ and (b) the 
velocity $u_x$ for various values of the wall temperature $T_w$. 
The dependence of the maximum values 
(c) $(\Delta T)_{\rm max}$ and (d) $u_{x;{\rm max}}$ with respect to $1 - T_w$.
The simulation parameters are $\tau = 5\times 10^{-3}$ and $\sigma = 10^{-4}$.
The time step and lattice spacing for (a) and (b) 
are $\delta t=2\times 10^{-4}$ and $\delta s = 1/320$.
\label{fig:profiles_var}}
\end{figure}

Figure~\ref{fig:profiles} shows the density, temperature and velocity profiles at various 
values of the wall temperature $T_w$ when $\delta s = 1/320$, 
$\delta t = 2\times 10^{-4}$, $\sigma = 10^{-4}$ and $\tau=5\times 10^{-3}$. 
The magnitude of the spurious velocities is less than $10^{-4}$, while the 
maximum temperature difference is smaller than $5 \times 10^{-4}$ 
even at $T_w = 0.74$. It can also be seen that for lower values of $T_w$, the density gradient along the interface, and hence the amplitude of the spurious 
velocity and the temperature difference increase, as already observed in 
Refs.~\cite{gonnellaPRE2007,cristeaijmpc2003}.

To better assess the range within which our models can be reliably used for 
simulations, a series of computer runs were performed with $\sigma = 10^{-4}$
by varying the wall temperature 
between $0.5$ and $0.99$. Figures~\ref{fig:profiles_var}(a) and (b) show the 
profiles of $\Delta T$ and $u_x$ for various values of the temperature $T_w$, 
obtained for $\delta s = 1/320$ and $\delta t = 2 \times 10^{-4}$.
A sudden decrease in the magnitudes of $\Delta T$ and $u_x$ can be observed when 
the wall temperature is increased from $T_w = 0.725$ to $T_w = 0.726$. 
This sudden change of magnitude is visible also in Figs.~\ref{fig:profiles_var}(c) and 
(d), where the maximum values $(\Delta T)_{\rm max}$ and $u_{x; {\rm max}}$ are represented 
with respect to $1 - T_w$. These figures show two sets of simulation results, 
the first corresponding to $\delta s = 1/320$ and $\delta t =2\times 10^{-4}$, 
while the second corresponds to $\delta s = 1/512$ and $\delta t = 10^{-4}$.
Two apparently disjoint regimes can be observed, 
corresponding to the cases when the fluctuations of the 
amplitudes are large (small $T_w$) 
or small (large $T_w$).
As expected, at fixed $T_w$, the values of $(\Delta T)_{\rm max}$ and 
$u_{x; {\rm max}}$ decrease when the lattice spacing is decreased. 
It can be seen that the point $T_w = 0.8$ lies inside the region of smaller 
errors, thus we conclude that the grid spacing $\delta s= 1/320$ and 
the time step $\delta t = 2\times 10^{-4}$ are suitable for the simulation 
of one-dimensional thermal phase separation between parallel plates having the temperature 
$T_w = 0.8$.

\subsection{Interface width test}\label{sec:planar:xi}

\begin{figure}
\begin{center}
\begin{tabular}{cc}
\includegraphics[width=220pt]{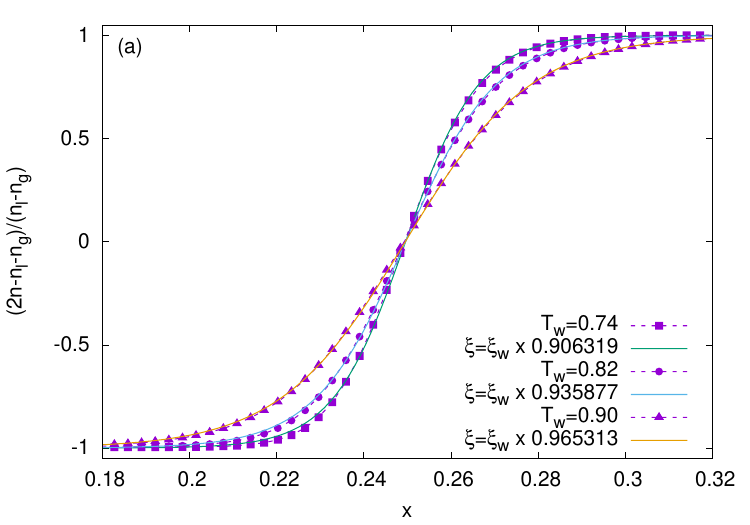} &
\includegraphics[width=220pt]{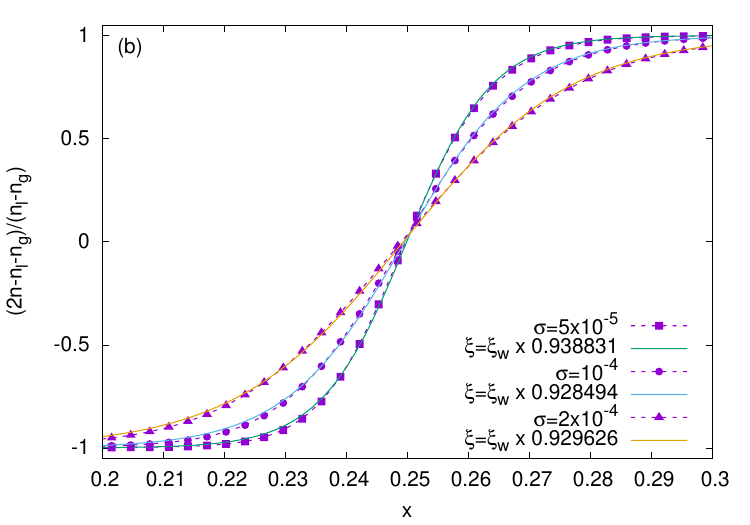} \\
\includegraphics[width=220pt]{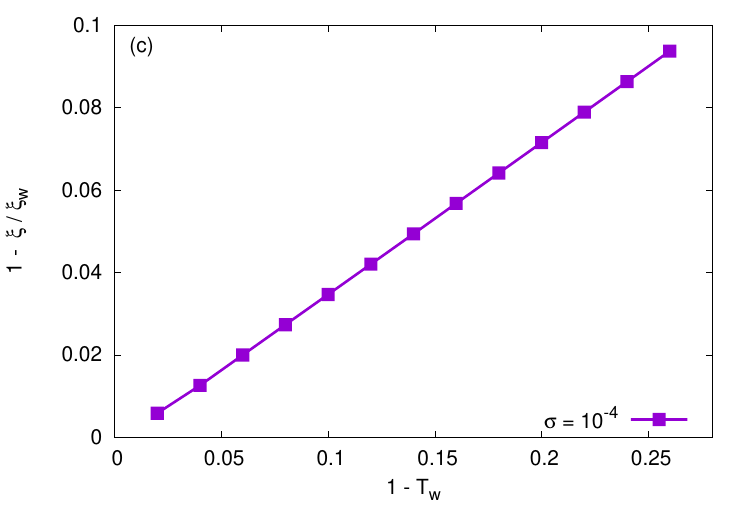} &
\includegraphics[width=220pt]{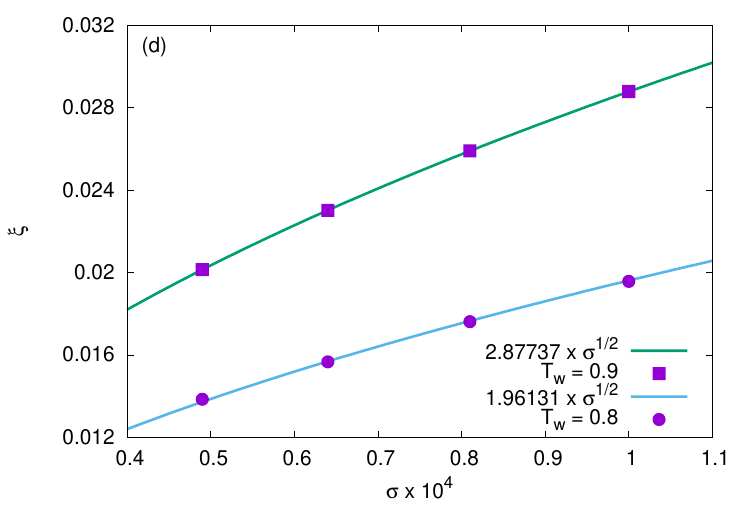} 
\end{tabular}
\end{center}
\caption{
Density profiles for (a) various wall temperatures $T_w = 0.74, 0.82, 0.9$ at 
$\sigma = 10^{-4}$ and (b) various values of $\sigma = 5 \times 10^{-5}, 10^{-4}, 
2 \times 10^{-4}$ at $T_w = 0.8$. The continuous lines are drawn using the 
best fit of Eq.~\eqref{eq:tanh} to the numerical data (dotted lines and 
points) as described in Subsec.~\ref{sec:planar:xi}. The best fit 
value of the interface width $\xi$ is compared with the approximate formula 
$\xi_w$ \eqref{eq:xiwagner}. (c) Relative deviation of the best fit 
value of $\xi$ compared to $\xi_w$ as a function of the temperature 
difference $1 - T_w$ measured from the critical point. (d) 
Best fit value of $\xi$ (points) for various values of $\sigma$ at $T_w = 0.8$ 
and $0.9$, fitted by a square root dependence on $\sigma$ (solid lines).
The fit coefficients are displayed in the plot legend. 
All simulations were performed using $\delta s = 1/320$ and 
$\delta t = 2\times 10^{-4}$.
\label{fig:xi}
}
\end{figure}

In order to study the properties of the interface between the gas and liquid 
phases, the density profile is investigated in the stationary state, for 
various values of the wall temperature $T_w$ and of the surface tension parameter 
$\sigma$. In particular, we aim to characterize the interface shape using the 
approximate formula \eqref{eq:tanh}. The gas density $n_g$, liquid 
density $n_l$, interface location $x_0$ and interface width $\xi$ are 
obtained by performing a four-parameter nonlinear fit of Eq.~\eqref{eq:tanh}.
The first two plots in Fig.~\ref{fig:xi} represent the nondimensionalized 
density $\nu = (2n - n_l - n_g) / (n_l - n_g)$, for: (a) various values 
of the wall temperature $T_w$ and $\sigma = 10^{-4}$; (b) various values of $\sigma$ at $T_w = 0.8$. The values of 
$n_g$ and $n_l$ are determined separately for each data set 
as described above. The numerical results are compared with the 
approximate formula \eqref{eq:tanh}. The legend 
gives the ratio between the interface width $\xi$ obtained 
using the nonlinear fit and the value given in Eq.~\eqref{eq:xiwagner}. 
It can be seen that the ratio approaches $1$ as $T_w$ approaches 
the critical temperature. 

In order to assess the validity of Eqs.~\eqref{eq:tanh} and \eqref{eq:xiwagner} 
away from the critical point, Fig.~\ref{fig:xi}(c) shows the relative 
difference $1 - \xi / \xi_w$ with 
respect to the distance from the critical point, measured by $1 - T_w$.
It can be seen that the formula loses validity as the departure from the 
critical point is increased. In particular, for $T_w = 0.8$ the relative departure 
from the predicted value is $7\%$.

Finally, Fig.~\ref{fig:xi}(d) tests the linear dependence of $\xi$ on 
$\sqrt{\sigma}$ for $T_w = 0.8$ and $0.9$. This dependence is 
tested in two steps, as follows. First, the value of 
$\xi$ is obtained via the four-parameter numerical fit described above.
Next, the values of $\xi$ corresponding to various values of $\sigma$ 
for the same value of $T_w$ are used to perform a fit of the law 
$\xi = \alpha \sqrt{\sigma}$. Excellent agreement is found, while 
the values of $\alpha$ found at $T_w = 0.9$ and $0.8$ are 
within $3.5\%$ and $7\%$ departure from the value predicted 
through Eq.~\eqref{eq:xiwagner}.

\subsection{Phase diagram}\label{sec:planar:maxwell}

\begin{figure}
\begin{center}
 \includegraphics[width=440pt]{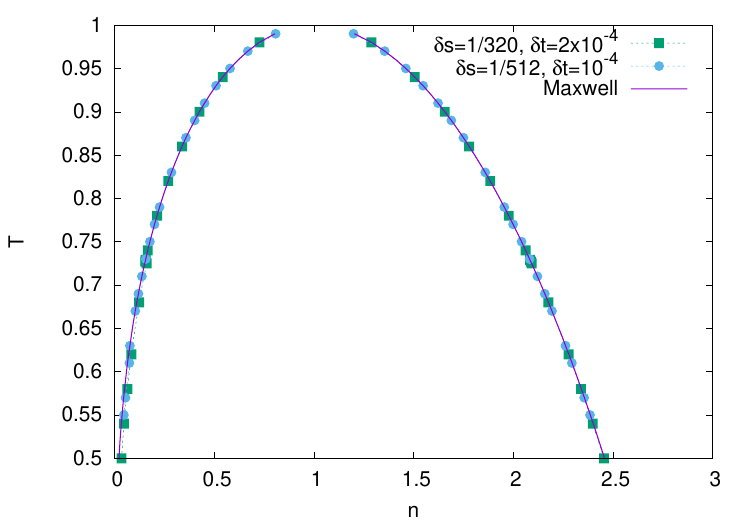}
\end{center}
\caption{Phase diagrama recovered using $\sigma = 10^{-4}$ 
and $(\delta s, \delta t) \in \{(1/320, 2\times 10^{-4}),
(1/512, 10^{-4})\}$ as compared to the Maxwell construction 
\cite{prigogine,sekerka,kruger17}.
\label{fig:phasediag}}
\end{figure}

Figure~\ref{fig:phasediag} shows the liquid-vapour phase diagram, as recovered 
with our model using $\sigma = 10^{-4}$ and two sets of values for the lattice 
spacing and time step, namely $(\delta s, \delta t) \in \{(1/320, 2\times 10^{-4}),
(1/512, 10^{-4})\}$. The values of the density are collected from the first lattice node 
near the wall for the liquid phase ($n_l$) and from the center of the channel for the 
vapour phase ($n_g$). Very good agreement is observed
between the density values obtained with our model and those obtained using 
Eq.~\eqref{vdw} via the Maxwell construction. The details of the Maxwell construction
are given in Refs.~\cite{prigogine,sekerka,kruger17}.

\section{Laplace pressure test}\label{sec:laplace}

\begin{figure}[t]
\centering
\begin{tikzpicture}
\fill [black!50!white] (0,0) rectangle (40mm,40mm);
\draw [line width=2,dashed] (0,0) rectangle (40mm,40mm);
\draw [line width=2] (40mm,0mm) -- (40mm,40mm);

\foreach \y in {1,2,...,10} {
    \draw [line width=1.2] (40mm,\y*4mm-4mm) -- (43mm,\y*4mm);
}
\foreach \y in {1,2,...,10} {
    \draw [line width=1.2] (43mm,\y*4mm-4mm) -- (40mm,\y*4mm);
}

\draw [fill=black!10!white] (20mm,0) arc (0:90:20mm)--(0,0)--(20mm,0);

\node at (20mm,-5mm) {Specular};
\node [rotate=90] at (-5mm,20mm) {Specular};
\node at (20mm,42.5mm) {Specular};

% \fill[black!20!white] (0mm,0mm) rectangle (80mm,40mm);

\node at (55mm,20mm) {Diffuse};
\node at (55mm,15mm) {$(T_w=0.8)$};
\node at (32mm,35mm) {$T=T_w$};
\node at (32mm,30mm) {$n=n_l$};
\node at (8mm,8mm) {$T=T_w$};
\node at (8mm,3mm) {$n=n_g$};

\draw [<->,dashed] (20mm,0)--(20mm,20mm);
\draw [<->,dashed] (0,20mm)--(20mm,20mm);
\node at (10mm,22mm) {$R=R_0$};

\end{tikzpicture}
\caption{Geometry for the Laplace pressure test. The system is initialized 
as described in Sec.~\ref{sec:laplace}.}
\label{fig:laplace_setup}
\end{figure}
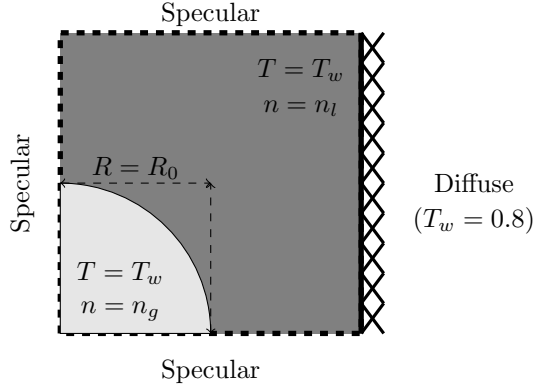

\begin{figure}
\begin{center}
\begin{tabular}{cc}
\includegraphics[width=220pt]{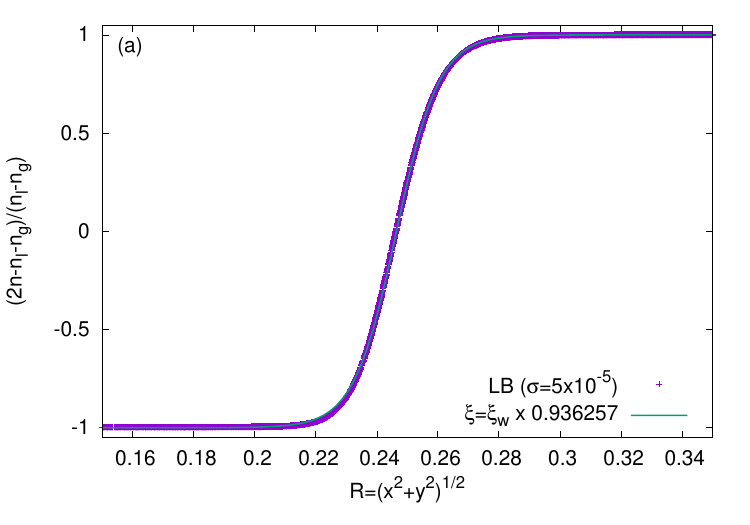} &
\includegraphics[width=220pt]{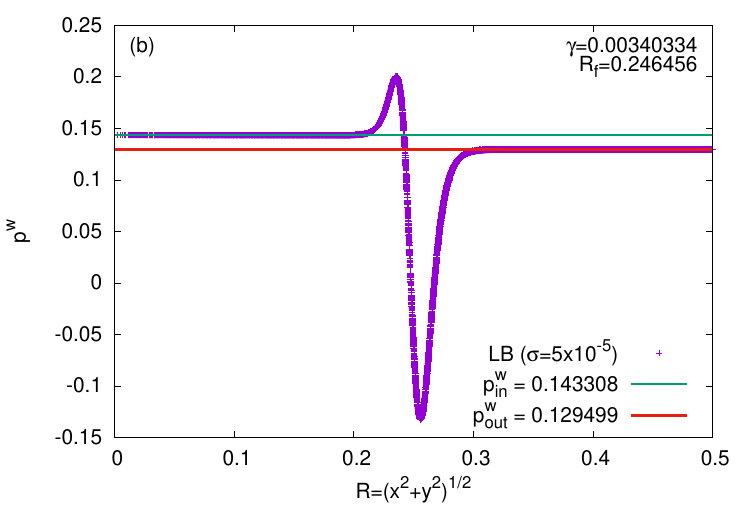}
\end{tabular}
\begin{tabular}{c}
\includegraphics[width=220pt]{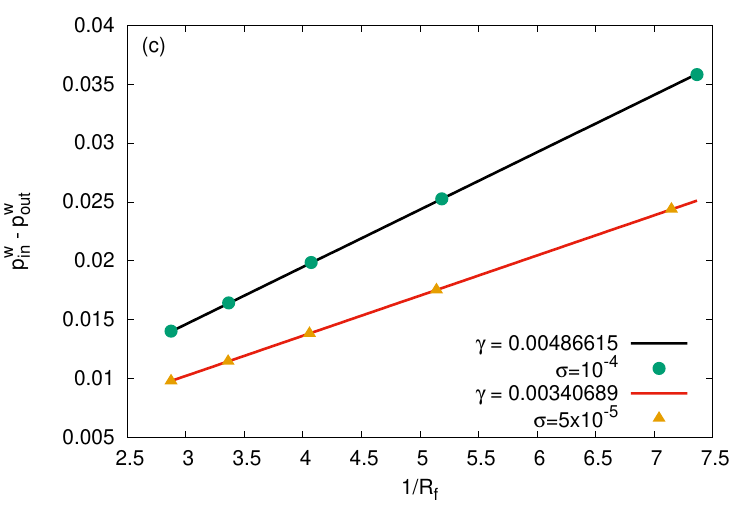}
\end{tabular}
\end{center}
\caption{
(a) Nondimensionalized density $\nu = (2n - n_l - n_g) / (n_l - n_g)$ 
with respect to radial distance $R = \sqrt{x^2 + y^2}$
(points) and the best fit of Eq.~\eqref{eq:tanh} (lines). 
The best fit curve is overlapped with the numerical results.
(b) Non-ideal (Van der Waals) pressure with respect to 
$R$ to the bubble centre (points) and 
numerical fits for the pressures inside ($p^w_{\rm in}$)
and outside ($p^w_{\rm out}$) the bubble.
(c) Pressure difference with respect to the inverse bubble radius 
in the stationary state ($R_f^{-1}$) with the legend indicating the 
fitted surface tension value $\gamma \equiv \gamma_{\rm Lap}$. 
All simulations were performed using $\delta s = 1/320$ and 
$\delta t = 2\times 10^{-4}$.
\label{fig:laplace}
}
\end{figure}

\begin{table}
\begin{center}
% \begin{tabular}{|r|rrr|}
%  \hline
%  $\sigma$ & $\gamma_{\rm plane}$ & $\gamma_{\rm avg}$ & $\gamma_{\rm lin}$ \\\hline
%  $5 \times 10^{-5}$ & $3.386731 \times 10^{-3}$ & $3.41794 \times 10^{-3}$ & $3.41876 \times 10^{-3}$ \\
%  $10^{-4}$ & $4.883771 \times 10^{-3}$ & $4.518318 \times 10^{-3}$ & $4.53590 \times 10^{-3}$\\\hline
% \end{tabular}
\begin{tabular}{|r|rr|}
 \hline
 $\sigma$ & $\gamma_{\rm pl}$ & $\gamma_{\rm Lap}$ \\\hline
 $5 \times 10^{-5}$ & $3.386731 \times 10^{-3}$ & $3.40689 \times 10^{-3}$ \\
 $10^{-4}$ & $4.883771 \times 10^{-3}$ & $4.86615 \times 10^{-3}$ \\\hline
\end{tabular}
\end{center}
\caption{Summary of the values for the surface tension $\gamma_{\rm pl}$
and $\gamma_{\rm Lap}$ obtained by evaluating numerically Eq.~\eqref{eq:gamma_int} 
for the planar interface profiles shown in Fig.~\ref{fig:xi}(b) and 
using the Laplace pressure test, as shown in Fig.~\ref{fig:laplace}(c).
\label{tab:laplace}}
\end{table}

In the case of circular droplets or bubbles, the pressure inside the 
interface is larger than the pressure outside of it. In this section, we 
consider the system described in Fig.~\ref{fig:laplace_setup}, where a gas bubble 
of radius $R_0$ located at the center of the channel is immersed inside 
the liquid phase. At initial time, the fluid is assumed to be everywhere in 
thermal equlibrium ($f = f^{eq}$) at the wall temperature ($T = T_w$). The 
density is initialized according to Eq.~\eqref{eq:tanh}, where $n_g$ and $n_l$ 
are the gas and liquid densities obtained via the Maxwell construction \cite{prigogine,sekerka,kruger17} 
at $T = T_w$. The interface width $\xi$ is computed using the approximate formula 
\eqref{eq:xiwagner}, while the coordinate $x$ is replaced
by the distance $R = \sqrt{x^2 + y^2}$ measured from the 
center of the channel. The initial interface location is at $x_0 = R_0\in\{0.15,0.2,0.25,0.3,0.35\}$.

Assuming that $R_0$ is sufficiently large for the 
bubble to be stable, thermodynamic equlibrium is reached in the stationary 
state, such that $T = T_w$ throughout the domain. The pressure difference 
between the inside and outside of the interface satisfies:
\begin{equation}
 p^w_{\rm in} - p^w_{\rm out} = \frac{\gamma}{R_f},
 \label{eq:Laplace}
\end{equation}
where $\gamma$ is the surface tension and $R_f$ is the bubble radius 
after the interface is stabilized. 

Since we are interested only in the stationary state which we assume to 
be symmetric with respect to the horizontal and vertical lines that intersect 
at the channel center, the computational demand can be reduced by implementing 
the specular reflection boundary conditions described in 
Sec.~\ref{sec:scheme:bcond:spec} on the bottom, left and top domain boundaries,
as indicated in Fig.~\ref{fig:laplace_setup}. The channel center is located 
at $x = y = 0$, while the wall and the top boundary are located
at $x = 0.5$ and $y = 0.5$, respectively. The time step is set 
to $\delta t = 2\times 10^{-4}$ and the lattice spacing 
is $\delta s = 1/320$, such that the simulations are performed on 
a square domain of size $160 \times 160$. For this test case, 
all simulations were performed with $T_w = 0.8$ and $\tau=5\times10^{-3}$.

The density and non-ideal (Van der Waals) pressure are shown in Figs.~\ref{fig:laplace}
(a) and (b), respectively, for $\sigma = 5\times10^{-5}$. Good agreement can be observed
between the fitted value for $\xi$ ($\sim 0.935 \xi_w$) 
and that obtained in Fig.~\ref{fig:xi}(b) for the planar interface 
($\sim 0.937 \xi_w$). The surface tension is computed by multiplying the fitted 
value $R_f$ for the location of the interface by the difference 
$p^w_{\rm in} - p^w_{\rm out}$ between the van der Waals pressure 
\eqref{vdw} inside ($p^w_{\rm in}$) and outside ($p^w_{\rm out}$) of 
the bubble. Figure~\ref{fig:laplace}(c) represents this difference with respect 
to $R_f^{-1}$, for two values of $\sigma$. 
A linear fit gives the value of the surface tension $\gamma_{\rm Lap}$. 

Alternatively, the surface tension can be 
computed in the context of a planar interface 
using the following formula:
\begin{equation}
 \gamma_{\rm pl} = \sigma \int \left(\frac{dn}{dx}\right)^2 dx.
 \label{eq:gamma_int}
\end{equation}
The integration domain is understood to cross only one interface. 
For this purpose, the profiles presented in Fig.~\ref{fig:xi}(b) can 
be used to obtain the value $\gamma_{\rm pl}$ for the planar interface
for the values of $\sigma$ considered in Fig.~\ref{fig:laplace}(c). 
The numerical results obtained using the Laplace pressure test 
and the planar interface are summarized in Tab.~\ref{tab:laplace}.
The relative error $\gamma_{\rm Lap}/\gamma_{\rm pl}-1$ is below
$1\%$ for both $\sigma=5\times 10^{-5}$ and $\sigma=10^{-4}$.

\section{Transport coefficients, sound speed and Galilean invariance}\label{sec:galilei}

We further investigate the features of our models in several contexts. 
First, we demonstrate the ability of our models to correctly recover the transport 
coefficients of the fluid that we are simulating in the context of the damping 
of transversal (shear) and longitudinal (sound) waves. Galilean invariance of LB models 
has been discussed in many studies throughout the past two decades 
\cite{lallemandPRE2003,luo2000,shanEPL2008,qian98,DELLAR13,geierjcp2017,geierc&p2018,shan2018}.
In this section, we evaluate the 
sensitivity of our models to Galilei transformations, by considering the wave damping 
problems at non-vanishing background fluid velocities. We conclude this section 
by presenting a study of the evolution of a gas bubble enclosed between parallel walls 
kept at constant temperature for various values of the background velocity.
In this section, we use the tilde ($\widetilde{\hspace{0.5em}}$) to denote 
time-dependent amplitudes. This notation should not be confused with the one 
introduced in Subsec.~\ref{sec:model:nondim} for dimensional quantities.

\subsection{Shear waves}\label{sec:galilei:shear}

\begin{figure}
\begin{center}
\begin{tabular}{cc}
\includegraphics[width=220pt]{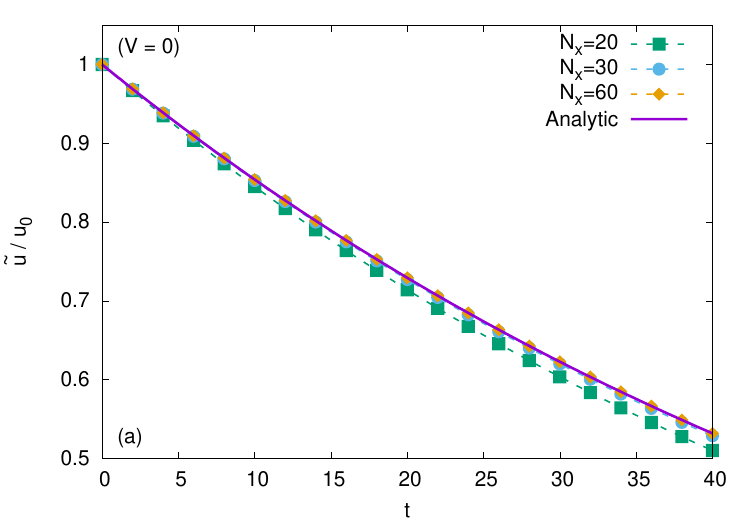} &
\includegraphics[width=220pt]{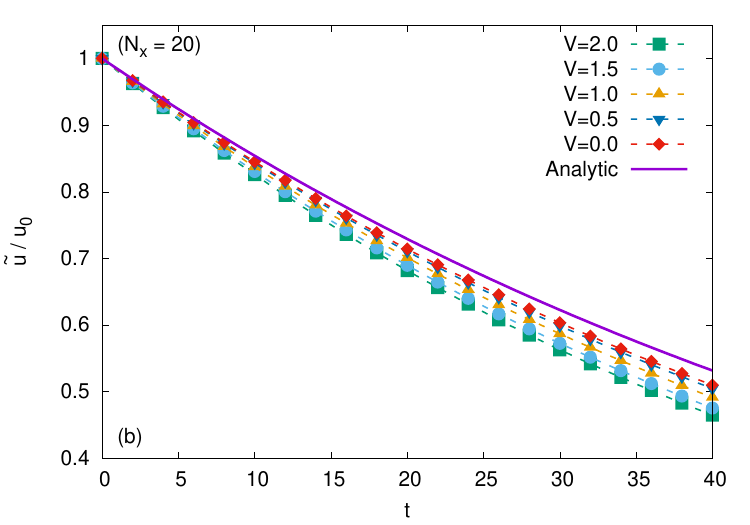} \\
\includegraphics[width=220pt]{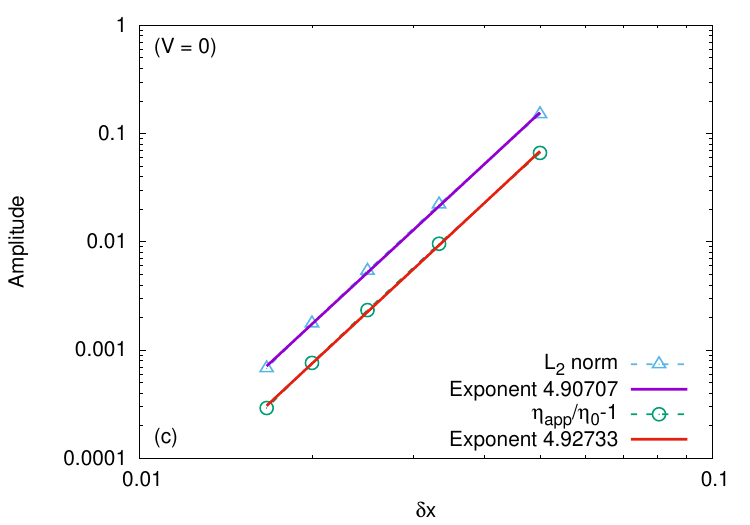} &
\includegraphics[width=220pt]{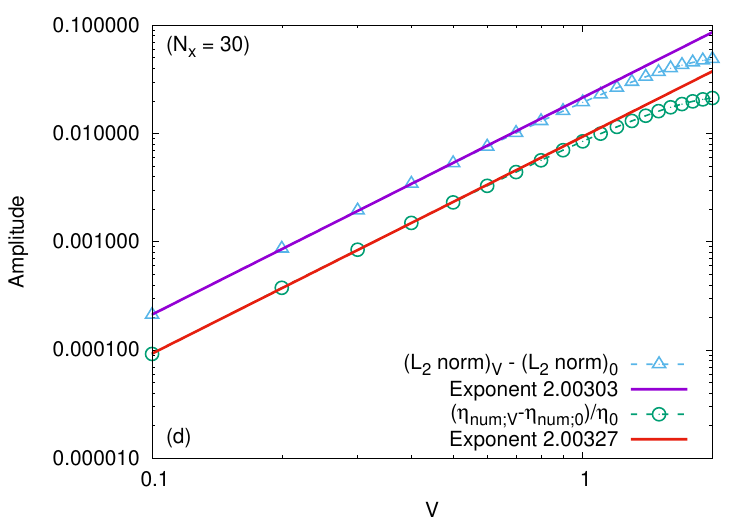} 
\end{tabular}
\end{center}
\caption{
(a,b) Comparison between the numerical results for the evolution of the amplitude 
$\widetilde{u}$ in the shear wave problem and the analytic result \eqref{eq:shear_sol}
for (a) stationary case $V = 0$ and various number of nodes $\mathfrak{N}_x = 20$, $30$ and $60$; 
(b) Fixed number of nodes $\mathfrak{N}_x = 20$ and various longitudinal velocities 
$V = 0$, $0.5$, $1$, $1.5$ and $2$.
(c) Exponents for the $\delta x = 1/\mathfrak{N}_x$ dependence of 
the $L_2$ norm \eqref{eq:shear_L2} and of 
$\eta_{\rm app}/\eta_0 -1$ computed in the stationary case $V = 0$.
(d) Exponent for the $V$ dependence of 
$(L_2)_{V} - (L_2)_{0}$ and $(\eta_{\rm app;V} - \eta_{\rm app;0}) /\eta_0$,
computed for $\mathfrak{N}_x = 30$ at various values of the longitudinal velocity 
$V$ (the subscript $0$ denotes the laboratory frame, where $V = 0$).
\label{fig:shear}
}
\end{figure}

A popular benchmark of lattice Boltzmann models is the damping of 
shear waves \cite{sofoneajcp2003,sofoneaPRE2018,shanEPL2008,geierc&p2018}. 
The system is considered to be homogeneous along the 
$y$ direction. At initial time, the system is considered to be in 
local thermal equilibrium at constant density and temperature, while 
the velocity along the $y$ axis is initialised according to:
\begin{equation}
 u_y(t = 0) = u_0 \sin k x,
\end{equation}
where $k = 2\pi / L$ is the wavenumber of the shear wave, $L$ is 
its wavelength and we use $u_0 = 10^{-3}$. Without loss of generality, 
we set $L = 1$ and choose a number of $\mathfrak{N}_x$ cells to discretise the system 
along the $x$ direction. The coordinate of the center of cell $i$ ($1 \le i \le \mathfrak{N}_x$) 
is 
\begin{equation}
 x_i = \frac{i-0.5}{\mathfrak{N}_x} - 0.5,
\end{equation}
such that at initial time, we set $u_{y;i}(t = 0) = u_0 \sin k x_i$.

Considering that $u_0$ is a small quantity, the continuity, Navier-Stokes and 
temperature equations \eqref{eq:macro} reduce to:
\begin{equation}
 \partial_t u_y - \frac{\eta_0}{\rho_0} \partial^2_x u_y = 0,
 \label{eq:shear_dtuy}
\end{equation}
while $n(t) = n_0 = {\rm const}$ and $T(t) = T_0 = {\rm const}$. 
Assuming that for $t > 0$, $u_y = \widetilde{u}(t) \sin kx$, 
Eq.~\eqref{eq:shear_dtuy} yields
\begin{equation}
 \widetilde{u}_{\rm lin}(t) = u_0 e^{-\nu_{\rm an} t}, \qquad 
 \nu_{\rm an} = \frac{k^2 \eta_0}{\rho_0}, \qquad 
 \eta_0 = \tau n_0 T_0,
 \label{eq:shear_sol}
\end{equation}
where the subscript ${\rm lin}$ refers to the analytic solution derived 
in the linearised limit of the macroscopic equations \eqref{eq:macro}.
According to the Galilean invariance of the 
theory, the solution \eqref{eq:shear_sol} should be valid also when seen by an observer
travelling towards negative values of $x$ with a constant velocity $V$, according 
to $X(t) = -V t$, expressed in the laboratory frame.
In the inertial frame of this observer, the transverse velocity profile
becomes:
\begin{equation}
 u_y = \widetilde{u}(t) \sin [k (x - V t)],
 \label{eq:shear_sol_V}
\end{equation}
where $\widetilde{u}(t)$ is given in Eq.~\eqref{eq:shear_sol} and 
$u_x = V$ everywhere in the fluid. In order to recover the amplitude $\widetilde{u}(t)$ 
during our simulations, we use 
\begin{equation}
 \widetilde{u}(t) = \frac{2}{L} \int_{-L/2}^{L/2} dx\, u_y \sin[k(x - V t)]. 
 \label{eq:shear_ut}
\end{equation}

\begin{table}
\begin{center}
\begin{tabular}{l|rrr|rrr}
 $V$ & \multicolumn{3}{c}{$\eta$ order} & \multicolumn{3}{c}{$L_2$ order} \\
 & Ideal gas & Vapour & Liquid
 & Ideal gas & Vapour & Liquid
 \\\hline
 $0.0$ & $4.927334$ & $4.927341$ & $4.927331$ & $4.907067$ & $4.907078$ & $4.907064$ \\
 $0.5$ & $4.914673$ & $4.914681$ & $4.914674$ & $4.892658$ & $4.892671$ & $4.892660$ \\
 $1.0$ & $4.893496$ & $4.893502$ & $4.893497$ & $4.864938$ & $4.864947$ & $4.864940$ \\
 $2.0$ & $4.876409$ & $4.876413$ & $4.876410$ & $4.831781$ & $4.831786$ & $4.831781$
\end{tabular}
\end{center}
\caption{Exponents of $\delta x = 1/N_x$ for the relative error 
$[(\eta_{\rm app} / \eta_0) - 1]$ and for the $L_2$ norm \eqref{eq:shear_L2} 
for the ideal gas and vapour and liquid phases of the van der Waals fluid,
for various longitudinal velocities, in the context of the 
damping of shear waves.\label{tab:shear_ux}}
\end{table}

\begin{table}
\begin{center}
\begin{tabular}{l|rr|rr|rr}
 $N_x$ & \multicolumn{2}{c}{Ideal gas} & \multicolumn{2}{c}{Vapour} & \multicolumn{2}{c}{Liquid} \\
 & $\eta$ order & $L_2$ order
 & $\eta$ order & $L_2$ order
 & $\eta$ order & $L_2$ order\\\hline
 $20$ & $1.848778$ & $1.848780$ & $1.848780$ & $1.845612$ & $1.845614$ & $1.845614$\\
 $30$ & $2.003266$ & $2.003267$ & $2.003280$ & $2.003027$ & $2.003027$ & $2.003041$ \\
 $40$ & $1.978353$ & $1.978370$ & $1.978403$ & $1.978272$ & $1.978290$ & $1.978323$ \\
 $50$ & $1.987273$ & $1.987361$ & $1.987420$ & $1.987260$ & $1.987352$ & $1.987413$ \\
 $60$ & $1.982981$ & $1.983316$ & $1.983342$ & $1.983050$ & $1.983397$ & $1.983421$ 
\end{tabular}
\end{center}
\caption{Exponents of $V$ for the differences between the values measured 
at longitudinal velocity $V$ and at rest $V = 0$ of the relative 
apparent shear viscosity 
$[(\eta_{\rm app; V} - \eta_{\rm app; V=0}) / \eta_0]$ ($\eta$ order) 
and $L_2$ norm $(L_{2;V} - L_{2;V=0})$ ($L_2$ order), for 
the ideal gas and the vapour and liquid phases of the van der Waals fluid,
in the context of the damping of shear waves.
\label{tab:shear_nx}}
\end{table}

Throughout this section, we set the background temperature to $T_0 = 0.8$, 
$\tau = 5 \times 10^{-4}$ in order to ensure that our simulations lie in the 
hydrodynamic regime, $\delta t = 2 \times 10^{-4}$ and $u_0 = 10^{-3}$ in order 
to ensure the validity of the linearisation ansatz which leads to 
Eq.~\eqref{eq:shear_sol}. The time variable is discretised using 
$N_t = 2\times 10^5$ values $t_n = n \delta t$ ($1 \le n \le N_t$) in addition 
to the initial time $t_0 = 0$. We further consider a division with respect to 
$\Delta t = 1000 \delta t$, giving rise to $S = 200$ values 
$\hat{t}_s = s \Delta t = 1000s \times \delta t$
($1 \le s \le S$). For each value of $s$, the quantity $\widetilde{u}_s$  
is computed using a discrete analogue of Eq.~\eqref{eq:shear_ut}:
\begin{equation}
 \widetilde{u}_s = \frac{2}{\mathfrak{N}_x} \sum_{i = 1}^{\mathfrak{N}_x} 
 u_{y;s;i} \sin[k (x_i - V \hat{t}_s)],
\end{equation}
where $u_{y;s;i}$ is the value of $u_{y;i}$ at time 
$t = \hat{t}_s = 1000s \times \delta t = 0.2 s$.

In order to perform quantitative analyses, the values $\widetilde{u}_s$ are used 
to perform a numerical fit of 
Eq.~\eqref{eq:shear_sol} which allows the parameter $\nu_{\rm app}$ to be extracted,
using which the apparent viscosity can be computed via
$\eta_{\rm app} = \rho_0 \nu_{\rm app} / k^2$. 
The second type of quantitative analysis concerns the 
$L_2$ norm of the relative difference between $\widetilde{u}_s$ and the expected 
value $\widetilde{u}(t)$ \eqref{eq:shear_sol}, which we compute using the trapezoidal 
rule:
\begin{equation}
 L_2 = \left\{\int_{0}^{t_{\rm f}} \frac{dt}{t_{\rm f}} 
 \left[\frac{\widetilde{u}_{\rm num}(t)}{\widetilde{u}_{\rm lin}(t)} - 1 \right]^2\right\}^{1/2}
 \simeq \left\{\frac{1}{S} \sum_{s = 0}^{S} \mathfrak{f}_s
 \left[\frac{\widetilde{u}_s}{\widetilde{u}_{\rm lin}(\hat{t}_s)} - 1\right]^2\right\}^{1/2},
 \label{eq:shear_L2}
\end{equation}
where $\mathfrak{f}_s = 0.5$ when $s = 0$ and $s = S$ and 
$\mathfrak{f}_s = 1$ for $1 \le s < S$.

We consider three batches of simulations. The first corresponds to the case of the ideal 
gas at unit density ($n_0 = n_{0;{\rm ideal}} = 1$), when the forcing term in Eq.~\eqref{evolution} 
is not taken into account. The second and third batches correspond to the cases of 
the van der Waals vapour ($n_0 = n_{0;{\rm g}} \simeq 0.2397$) and liquid 
($n_0 = n_{0; {\rm l}} \simeq 1.933$) phases, respectively. 
For each simulation batch, we consider discretisations with $\mathfrak{N}_x = 20$, 
$30$, $40$, $50$ and $60$ points. For each value of $\mathfrak{N}_x$, we consider velocities $V$ 
ranging from $0$ (laboratory frame) to $2$, with a step of $0.1$. Since in this problem, 
the gradients of the density and temperature (and hence, of the pressure) are 
expected to vanish, the numerical results for these three media are very similar.

Figure~\ref{fig:shear} shows in the top panels the typical 
time dependence of the amplitude $\widetilde{u}(t)$, while in the 
bottom panels, the convergence tests are presented, as discussed below.
For simplicity, only the results for the first batch of simulations 
are shown (the case of the ideal gas).

In panel (a) of Fig.~\ref{fig:shear}, the longitudinal velocity is $V = 0$ (laboratory frame)
and the domain is discretised using various number of nodes. It can be seen that already 
at $\mathfrak{N}_x=  30$, a reasonable agreement is found compared to the analytic 
formula \eqref{eq:shear_sol}.
In panel (b), various values of the overall longitudinal velocity $V$ are considered, 
while keeping $\mathfrak{N}_x = 20$ in order to enhance the differences between the 
various numerical results. It can be seen that the results deteriorate as $V$ is increased. 

In the bottom panels of Fig.~\ref{fig:shear}, the error
in $[(\eta_{\rm app} / \eta_0) - 1]$ and in the $L_2$ norm 
computed using Eq.~\eqref{eq:shear_L2} are presented. 

In panel (c) of Fig.~\ref{fig:shear}, 
the longitudinal velocity is set to $V = 0$ and the number of nodes $\mathfrak{N}_x$ 
is varied. A numerical fit of $[(\eta_{\rm app} / \eta_0) - 1]$ and $L_2$ 
as functions of $a (\delta x)^\gamma$ gives values of $\gamma$ close to $5$, 
confirming that the WENO-5 scheme employed in this paper is fifth-order accurate, 
as also shown in Ref.~\cite{jiang96}. The analysis discussed above is 
performed for overall longitudinal velocities $V = 0$, $0.5$, $1$ and $2$ 
for the ideal gas and for the vapour and liquid phases of the van der Waals fluid 
and the results are reported in Tab.~\ref{tab:shear_ux}. It can be seen that the 
decrease of the exponents with $V$ is insignificant (less than $2\%$ difference 
between the cases $V = 0$ and $V = 2$). As expected, the exponents 
at fixed values of $V$ are very similar for the three media considered herein.

Finally, panel (d) of Fig.~\ref{fig:shear} measures the effects of 
increasing the longitudinal velocity on the numerical viscosity and 
the $L_2$ norm. The number of nodes is kept fixed at $\mathfrak{N}_x = 30$.
Since at $V = 0$, the error compared to the analytic estimates is finite, 
the influence of $V$ can be isolated by considering the 
numerical results for ${\eta_{\rm app;V}} / \eta_0$ and $L_{2;V}$ 
obtained at finite $V$ relative to their values when $V = 0$. 
A numerical fit of the scaling law $a V^\gamma$ shows that 
the differences $(\eta_{{\rm app};V} - \eta_{{\rm app};0})/\eta_0$ and 
$L_{2;V} - L_{2;0}$ grow with exponent $\gamma \simeq 2$. Since this 
scaling holds only for small values of $V$, the fits are performed for 
$0 < V \le 0.5$.
Further results for this test are shown in Tab.~\ref{tab:shear_nx}, 
where the number of grid points $\mathfrak{N}_x$ is varied from $20$ to $60$. 
It can be seen that the exponent $\gamma$ is very close to $2$ 
for all tested cases. As expected, the difference between the results 
obtained for the ideal gas and the vapour and liquid phases of the 
van der Waals fluid is negligible.

\subsection{Longitudinal waves}\label{sec:galilei:sound}

\begin{figure}[t]
\begin{center}

\begin{tabular}{ccc}
 \includegraphics[width=140pt]{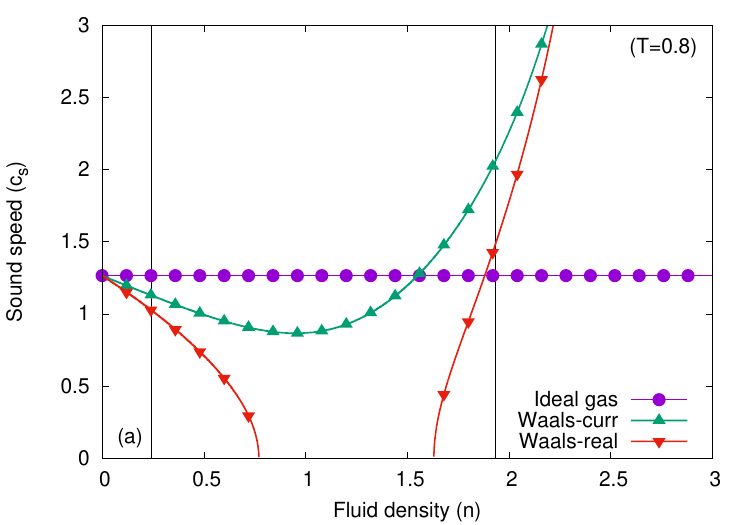}&
 \includegraphics[width=140pt]{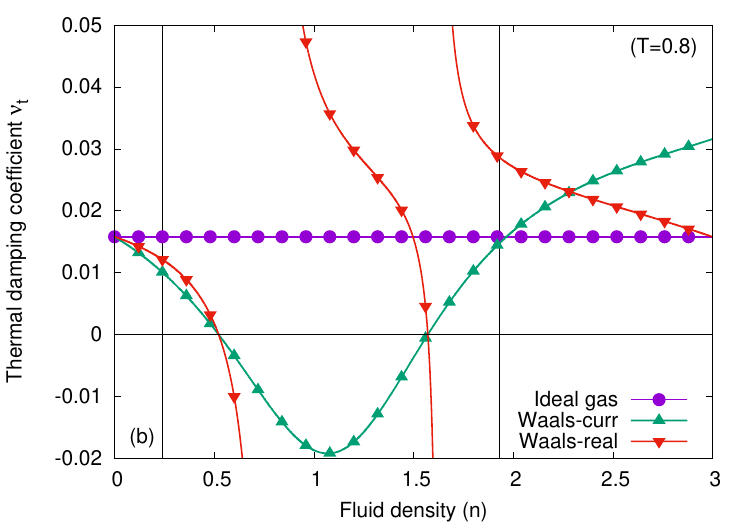}&
 \includegraphics[width=140pt]{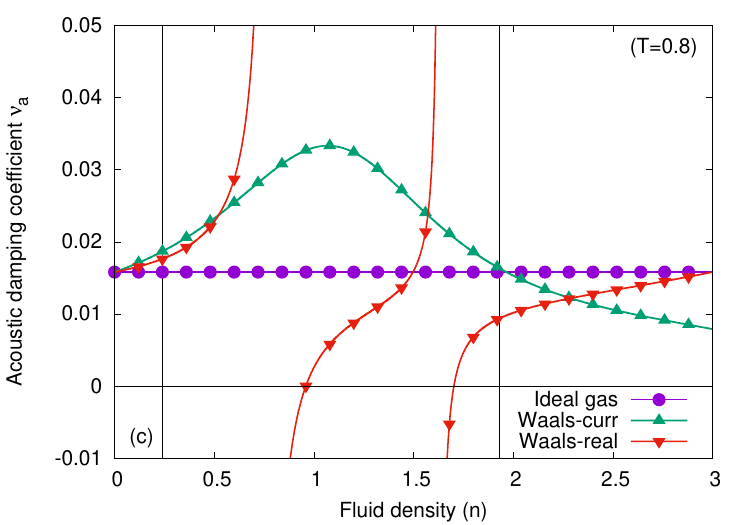}
\end{tabular}
\end{center}

\caption{
Comparison with respect to the fluid density $n$ at fixed temperature $T = 0.8$ of 
(a) the speed of sound and (b,c) the thermal and acoustic damping coefficients 
$\nu_t$ and $\nu_a$ for the cases of the ideal gas ($P_{\eta} = P_{\kappa} = p^i$),
van der Waals gas ($P_{\eta} = P_{\kappa} = p^w$) and the current model 
($P_{\eta} = p^w$ and $P_{\kappa} = p^i$). The vertical lines correspond 
to the densities in the vapour and liquid phases of the van der Waals fluid. 
The intersections between the horizontal line and $\nu_t$ 
represent the boundaries of the spinodal decomposition region.
\label{fig:long_coef}}
\end{figure}

\begin{figure}
\begin{center}
\begin{tabular}{cc}
 \includegraphics[width=220pt]{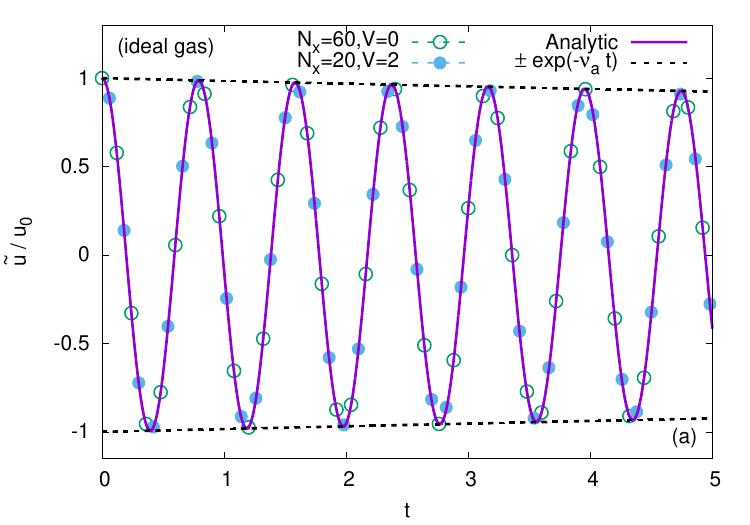} &
 \includegraphics[width=220pt]{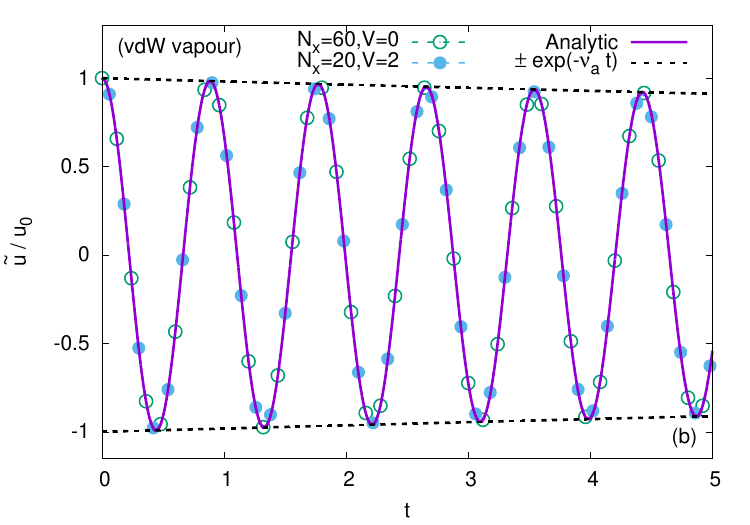} 
\end{tabular}
\begin{tabular}{c}
 \includegraphics[width=220pt]{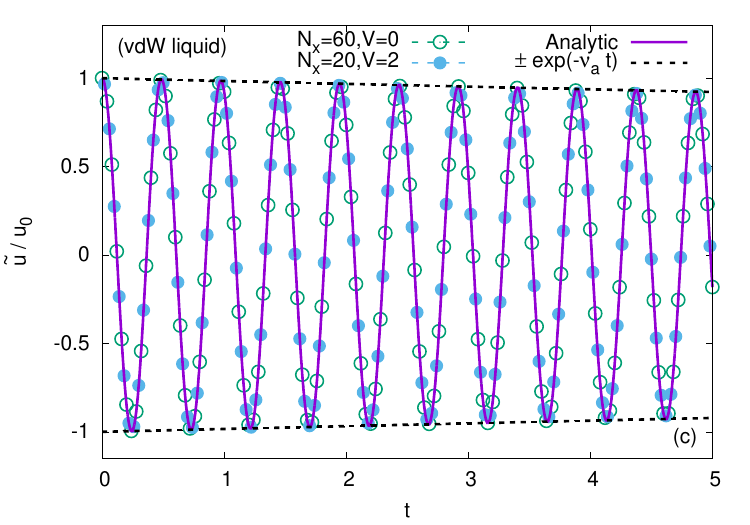} 
\end{tabular}
\end{center}
\caption{
Comparison between the time evolution of the normalised velocity amplitude 
$\widetilde{\delta u}(t) / \delta u_0$ of a longitudinal wave obtained numerically 
with $\mathfrak{N}_x = 20$ in the laboratory frame $V = 0$ (dotted lines 
and filled circles) and $\mathfrak{N}_x = 60$ at $V = 2$ (dotted lines and hollow circles) 
and the analytic prediction \eqref{eq:long_usol} (solid lines), as well as 
the overall acoustic damping factor $\pm e^{-\nu_a t}$ with $\nu_a$ 
computed from Eq.~\eqref{eq:long_coeff}, for the ideal gas (a) and 
vapour (b) and liquid (c) phases of the van der Waals fluid.
\label{fig:long_evol}}
\end{figure}

\begin{figure}
\begin{center}
\begin{tabular}{cc}
 \includegraphics[width=220pt]{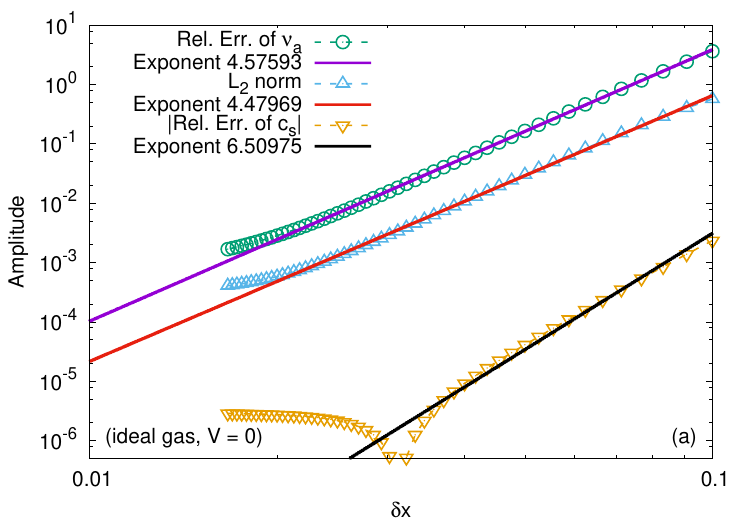} &
 \includegraphics[width=220pt]{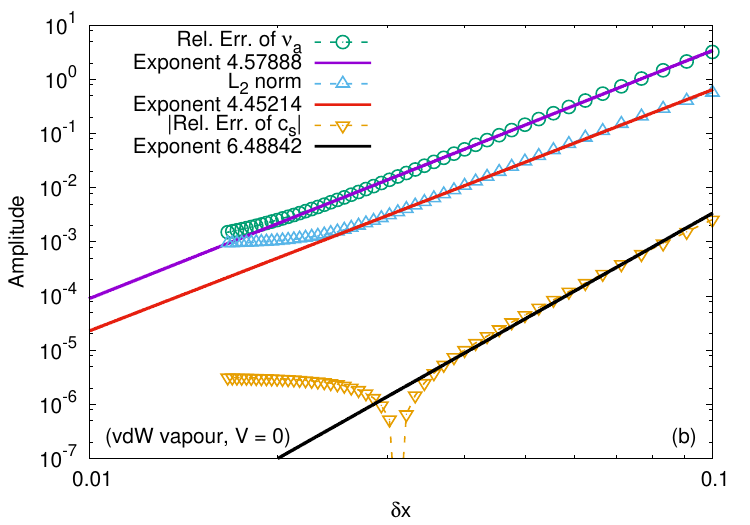}
\end{tabular}
\begin{tabular}{c}
 \includegraphics[width=220pt]{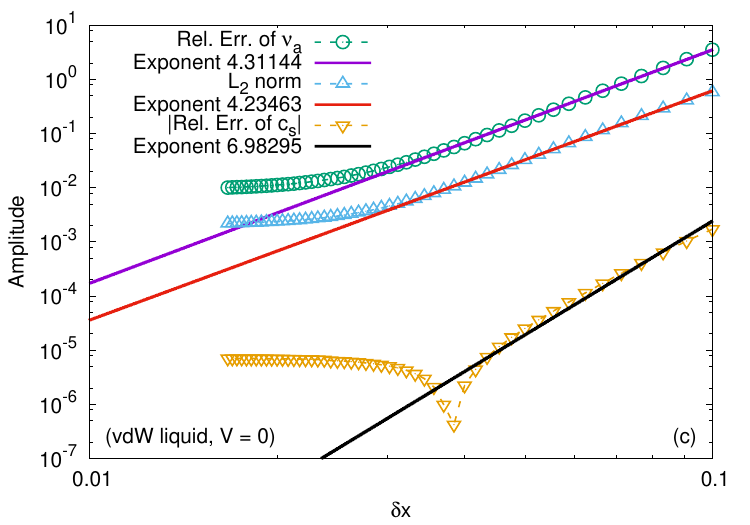}
\end{tabular}
\end{center}
\caption{
Exponents in the context 
of the damping of a longitudinal wave
for the $\delta x = 1/\mathfrak{N}_x$ dependence of 
relative errors of $\nu_a$ and $c_s$, computed as 
$(\nu_{a;{\rm app}}/\nu_{a;{\rm lin}})-1$
and $|(c_{s;{\rm app}} / c_{s;{\rm lin}}) - 1|$,
where $\nu_{a;{\rm app}}$ and $c_{s;{\rm app}}$ 
are the values of $\nu_a$ and $c_s$ obtained by 
fitting Eq.~\eqref{eq:long_usol} to the numerical 
data, while $\nu_{a;{\rm lin}}$ and $c_{s;{\rm lin}}$ 
are given in the linear approximation through Eq.~\eqref{eq:long_coeff}.
The $L_2$ norm is computed according to Eq.~\eqref{eq:long_L2}.
The results corresponding to $\mathfrak{N}_x \le 30$ are fitted by a power law and 
the exponents are shown in the caption, separately for 
the ideal gas (a) and for the vapour (b) and liquid (c) phases 
of the van der Waals fluid.
\label{fig:long_errnx}}
\end{figure}

\begin{figure}
\begin{center}
\begin{tabular}{cc}
 \includegraphics[width=220pt]{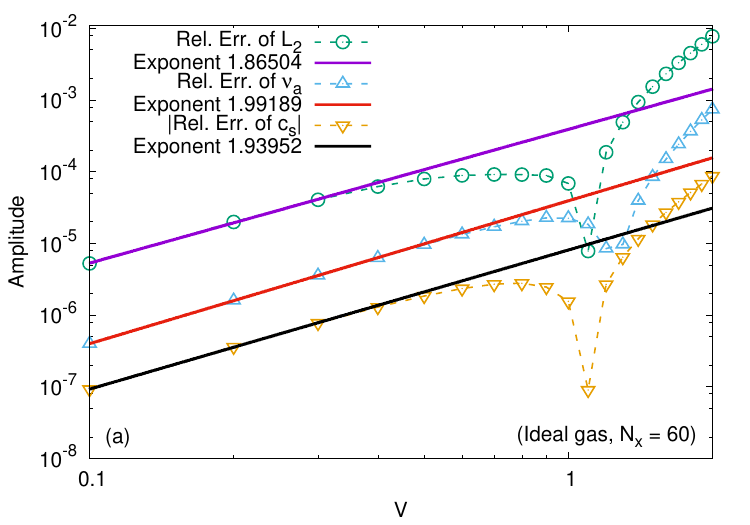} &
 \includegraphics[width=220pt]{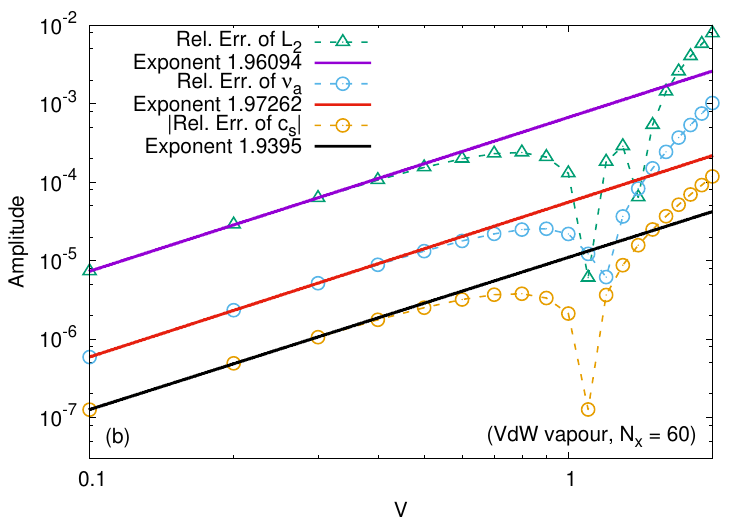}
\end{tabular}
\begin{tabular}{c}
 \includegraphics[width=220pt]{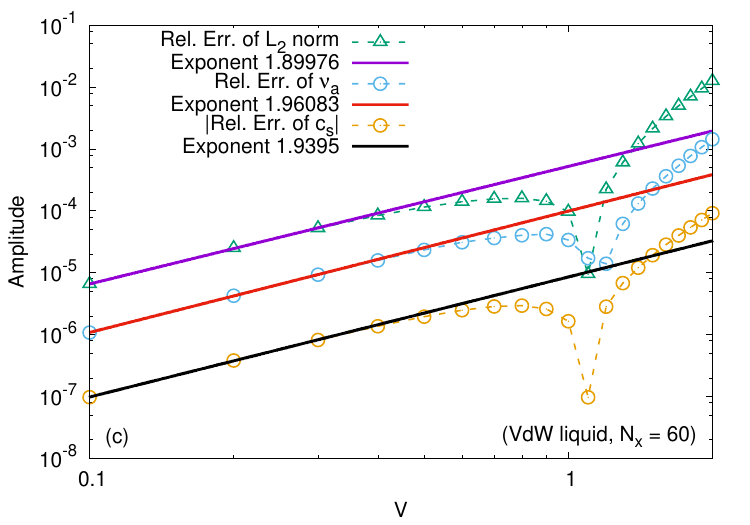}
\end{tabular}
\end{center}
\caption{
Exponents in the context 
of the damping of a longitudinal wave
for the $V$ dependence of the following relative errors:
$(L_2)_{V} - (L_2)_{0}$; $(\nu_{a;{\rm app};V}-\nu_{d;{\rm app};0})/{\nu_{a;{\rm lin}}}$;
and $|c_{s;{\rm app};V}-c_{s;{\rm app};0}|/c_{s;{\rm lin}}$, where the 
quantities bearing the subscript $V$ are obtained by numerically fitting 
Eq.~\eqref{eq:long_usol} to the data obtained when the background longitudinal 
velocity has value $V$, while $0$ corresponds to the laboratory frame ($V = 0$). 
The quantities bearing the subscript ${\rm lin}$ refer to the solution of the 
linearised equations derived in Eq.~\eqref{eq:long_coeff}.
The results are obtained using $\mathfrak{N}_x = 60$ nodes.
\label{fig:long_errux}}
\end{figure}

\begin{table}
\begin{center}
\begin{tabular}{l|rrr|rrr|rrr}
 $V$ & \multicolumn{3}{c}{$\nu_a$ order} & \multicolumn{3}{c}{$L_2$ order} & \multicolumn{3}{c}{$c_s$ order}  \\
 & Ideal gas & Vapour & Liquid
 & Ideal gas & Vapour & Liquid
 & Ideal gas & Vapour & Liquid
 \\\hline
 $0.0$ & $4.576$ & $4.579$ & $4.311$ & $4.480$ & $4.452$ & $4.235$ & 6.510 & 6.488 & 6.983\\ 
 $0.5$ & $4.565$ & $4.571$ & $4.294$ & $4.469$ & $4.439$ & $4.221$ & 5.960 & 5.860 & 7.423\\
 $1.0$ & $4.533$ & $4.540$ & $4.258$ & $4.433$ & $4.407$ & $4.178$ & 5.906 & 5.803 & 7.018\\
 $2.0$ & $4.554$ & $4.533$ & $4.481$ & $4.095$ & $4.004$ & $3.781$ & 1.515 & 0.942 & 1.479
\end{tabular}
\end{center}
\caption{Exponents in the context 
of the damping of a longitudinal wave of $\delta x = 1/\mathfrak{N}_x$ for 
$(\nu_{a;{\rm app}} / \nu_{a;{\rm lin}}) - 1$,
$(c_{s;{\rm app}} / c_{s;{\rm lin}}) - 1$ and the
$L_2$ norm, for various values of the background velocity $V$.
\label{tab:long_errux}}
\end{table}

\begin{table}
\begin{center}
\begin{tabular}{l|rrr|rrr|rrr}
 $\mathfrak{N}_x$ & \multicolumn{3}{c}{$\nu_a$ order} & \multicolumn{3}{c}{$L_2$ order} & \multicolumn{3}{c}{$c_s$ order}  \\
 & Ideal gas & Vapour & Liquid
 & Ideal gas & Vapour & Liquid
 & Ideal gas & Vapour & Liquid
 \\\hline
 $20$ & $1.951$ & $1.955$ & $1.944$ & $1.962$ & $1.970$ & $1.928$ & 1.807 & 1.796 & 2.164\\
 $30$ & $1.902$ & $1.902$ & $1.903$ & $1.874$ & $0.133$ & $1.884$ & 1.940 & 1.939 & 1.937\\
 $40$ & $1.889$ & $1.884$ & $1.867$ & $1.822$ & $1.982$ & $1.891$ & 1.940 & 1.939 & 1.770\\
 $50$ & $1.767$ & $1.326$ & $2.076$ & $1.840$ & $1.965$ & $1.897$ & 1.940 & 1.939 & 1.940\\
 $60$ & $1.992$ & $1.973$ & $1.961$ & $1.865$ & $1.961$ & $1.900$ & 1.940 & 1.939 & 1.940
\end{tabular}
\end{center}
\caption{Exponents in the context 
of the damping of a longitudinal wave of $V$ for the differences 
$(\nu_{a;{\rm app};V}-\nu_{a;{\rm app};0})/{\nu_{a;{\rm lin}}}$;
$(L_2)_{V} - (L_2)_{0}$; and 
$(c_{s;{\rm app};V}-c_{s;{\rm app};0})/c_{s;{\rm lin}}$, where the 
quantities bearing the subscript $V$ are obtained by numerically fitting 
Eq.~\eqref{eq:long_usol_2} to the data obtained when the background longitudinal 
velocity has value $V$ (the subscript $0$ denotes the laboratory frame, 
in which $V = 0$). The quantities bearing the subscript ${\rm lin}$ refer 
to the solution of the linearised equations derived in Eq.~\eqref{eq:long_coeff}.
\label{tab:long_errnx}}
\end{table}

We now turn to another important problem in fluid dynamics which 
concerns the study of longitudinal waves 
\cite{shan2018,cheng81,cercignani88,faber95,sharipov08,
wang12,sharipov16,ambrus18prca}. 
The propagation of longitudinal 
waves induces fluctuations in the macroscopic properties of the fluid, 
the amplitudes of which decay due to viscous and thermal dissipation.
Considering that the wave propagates through a background state 
characterised by $n_0$, $T_0$ and $u_{x;0} = V$, the macroscopic quantities 
can be written as:
\begin{equation}
 n(x, t) = n_0 + \delta n(x, t), \qquad 
 T(x, t) = T_0 + \delta T(x, t), \qquad 
 u(x, t) = V + \delta u(x, t).
\end{equation}
The background velocity $V$ is introduced above to enable the solution 
of the longitudinal wave problem to be considered in any Galilean frame 
in motion along the wave's direction of propagation. For simplicity, 
we set $V = 0$ in the following, since $V$ can be restored at the end 
of the calculation using the principle of Galilean invariance.

Taking the limit when the wave amplitudes $\delta n$, $\delta T$ and $\delta u$ are 
small, the macroscopic equations \eqref{eq:macro} can be linearised as follows:
\begin{align}
 \partial_t \delta n + n_0 \partial_x \delta u =& 0, \nonumber\\
 \rho_0 \partial_t \delta u + \partial_x \delta P_{\eta} - 
 \eta_0 \partial^2_x \delta u - n_0 \sigma \partial_x^3 \delta n =& 0, \nonumber\\
 n_0 \partial_t \delta T - \kappa_{T;0} \partial_x^2 \delta T + 
 P_{\kappa;0} \partial_x \delta u =& 0.
 \label{eq:sound_macro}
\end{align}
We remind the reader that 
the heat capacity at constant volume is $C_V = 1$ in our 
two-dimensional framework, corresponding to an adiabatic index $\gamma = 2$.
In the above, $\rho_0 = m n_0$, $\sigma$ is the surface tension parameter, 
while $\eta_0 = \tau n_0 T_0$ and $\kappa_{T;0} = 2\tau n_0 T_0 / m$ are the shear viscosity 
and thermal conductivity for the BGK model discussed in Sec.~\ref{sec:model:macro}.
The pressures $P_\eta$ and $P_\kappa$ appearing in the Navier-Stokes and heat equations 
are left unspecified in order to allow the same framework to be applied for 
the ideal and van der Waals fluids. Their values at $n = n_0$ and $T = T_0$
are denoted by $P_{\eta;0}$ and $P_{\kappa;0}$, while the perturbation
$\delta P_{\eta} = P_\eta - P_{\eta;0}$ can be written as:
\begin{equation}
 \delta P_{\eta} = \left(\frac{\partial P_{\eta}}{\partial n}\right)_0 \delta n + 
 \left(\frac{\partial P_{\eta}}{\partial T}\right)_0 \delta T.
\end{equation}

Since all the relations in Eq.~\eqref{eq:sound_macro} are linear and homogeneous 
with respect to the perturbation amplitudes, a harmonic decomposition can be made.
Let $k = 2\pi / L$ be the wave number of a longitudinal wave with wavelength $L$.
In the laboratory frame ($V = 0$), the following ansatz can be made:
\begin{equation}
 \delta u = \widetilde{\delta u}(t) \sin kx, \qquad
 \delta n = \widetilde{\delta n}(t) \cos kx, \qquad 
 \delta T = \widetilde{\delta T}(t) \cos kx,
 \label{eq:sound_harmonic}
\end{equation}
where the amplitudes $\widetilde{\delta u}$, $\widetilde{\delta n}$ and 
$\widetilde{\delta T}$ depend only on time. The analysis of the wave damping 
can be made at the level of independent modes by writing these amplitudes as 
follows:
\begin{equation}
 \widetilde{\delta u} = \sum_{\nu} e^{-\nu t} \delta u_\nu, \qquad 
 \widetilde{\delta n} = \sum_{\nu} e^{-\nu t} \delta n_\nu, \qquad 
 \widetilde{\delta T} = \sum_{\nu} e^{-\nu t} \delta T_\nu,
 \label{eq:sound_modes}
\end{equation}
where the coefficients $\delta u_\nu$, $\delta n_\nu$ and $\delta T_\nu$ 
are constant. 
% The real part of $\nu$ is responsible for the 
% dissipative damping of the wave, while its imaginary part is related 
% to the wave propagation speed (the speed of sound).

Substituting Eqs.~\eqref{eq:sound_harmonic} and \eqref{eq:sound_modes} 
into Eq.~\eqref{eq:sound_macro} gives:
\begin{align}
 \nu \delta n_\nu =& k n_0 \delta u_\nu, \nonumber\\
 (\rho_0 \nu - \eta_0 k^2)\delta u_\nu =& -k \left\{
 \left[\left(\frac{\partial P_{\eta}}{\partial n}\right)_0 + n_0 \sigma k^2\right] \delta n_\nu + 
 \left(\frac{\partial P_{\eta}}{\partial T}\right)_0 \delta T_\nu\right\},\nonumber\\
 (n_0 \nu - \kappa_{T;0} k^2) \delta T_\nu =& k P_{\kappa;0} \delta u_\nu.
 \label{eq:sound_macro_nu}
\end{align}
% The above relations can be combined to yield:
% \begin{equation}
%  \delta u_\nu \left[\rho_0 \nu - \eta k^2 + \frac{k^2 n_0}{\nu} 
%  \left[\left(\frac{\partial P_{\eta}}{\partial n}\right)_0 + n_0 \sigma k^2\right] + 
%  \frac{k^2 P_{\kappa;0}}{n_0 \nu - \kappa_T k^2} 
%  \left(\frac{\partial P_{\eta}}{\partial T}\right)_0\right] = 0.
% \end{equation}
Setting $\delta u_\nu = 0$ yields 
the trivial solution $\delta n_\nu = \delta T_\nu = 0$. 
Thus, non-trivial dynamics occur only when:
\begin{multline}
 \nu^3 - \frac{\nu^2 k^2}{\rho_0} (\eta_0 + m \kappa_{T;0}) + 
 \frac{\nu k^2}{m} \left[\left(\frac{\partial P_{\eta}}{\partial n}\right)_0 + 
 n_0 \sigma k^2 + \frac{P_{\kappa;0}}{n_0^2} \left(\frac{\partial P_{\eta}}{\partial T}\right)_0 + 
 \frac{\eta_0 \kappa_{T;0} k^2}{n_0^2}\right]\\ 
 - \frac{k^4 \kappa_{T;0}}{\rho_0}\left[
 \left(\frac{\partial P_{\eta}}{\partial n}\right)_0 
 + n_0 \sigma k^2\right] = 0.
 \label{eq:sound_nu3}
\end{multline}
The above equation is cubic with respect to $\nu$, thus it admits at least one 
real solution, which corresponds to the thermal mode $\nu_t$.
The other two roots  $\nu_\pm$, corresponding to the acoustic modes, 
must be complex in order to allow the wave to propagate. 
Writing $\nu_{\pm} = \nu_a \pm i \nu_s$, we see that $\nu_a$
induces acoustic dissipation, while $\nu_s = k c_{s;0}$ is related to 
the speed of sound $c_{s;0}$ at the background parameters. 
In order to derive expressions 
for the allowed values of $\nu$, we remember that $\eta_0$ and $\kappa_{T;0}$ 
are proportional to $\tau$, which is assumed to be a 
small number in order for the flow to remain within the hydrodynamic regime.
Thus, we seek solutions of the form:
\begin{equation}
 \nu = i \nu_0 + \nu_1 \tau + O(\tau^2),\label{eq:sound_nu_ansatz}
\end{equation}
where the imaginary unit $i$ was inserted in front of the leading order term. 
Inserting Eq.~\eqref{eq:sound_nu_ansatz} 
into Eq.~\eqref{eq:sound_nu3} gives, to orders $\tau^0$ and $\tau^1$,
\begin{equation}
 \nu_0(\nu_0^2 - k^2 c_{s;0}^2) = 0, \qquad 
 \tau \nu_1 (k^2 c_{s;0}^2 - 3\nu_0^2) + \frac{\nu_0^2 k^2}{\rho_0}(\eta_0 + m \kappa_{T;0}) 
 - \frac{\kappa_{T;0} k^4}{\rho_0} \left[\left(\frac{\partial P_{\eta}}{\partial n}\right)_0 
 + n_0 k^2 \sigma\right] = 0,\label{eq:sound_nu_eqs}
\end{equation}
where we have identified the speed of sound as:
\begin{equation}
 c_s = \left(\frac{\partial P_{\eta}}{\partial \rho} + 
 \frac{P_{\kappa}}{n \rho} \frac{\partial P_{\eta}}{\partial T}
 + \frac{n k^2 \sigma}{m}\right)^{1/2}.\label{eq:sound_cs}
\end{equation}
Setting $\nu_0 = 0$ yields the thermal mode $\nu_t = \tau \nu_1$, 
while the cases with $\nu_0 \neq 0$ correspond to the acoustic modes
$\nu_\pm = \nu_a \pm i\nu_s$:
\begin{equation}
 \nu_t = \frac{k^2 \kappa_{T;0}}{\rho_0 c_{s;0}^2} \left[
 \left(\frac{\partial P_{\eta}}{\partial n}\right)_0 + n_0 k^2 \sigma\right], \qquad 
 \nu_a = \frac{k^2(\eta_0 + m \kappa_{T;0})}{2\rho_0} - \frac{\nu_t}{2}, \qquad
 \nu_s = k c_{s;0}.\label{eq:long_coeff}
\end{equation}

In the case of the ideal gas, we have $P_{\eta;0} = P_{\kappa;0} = p^i_0 = n_0 T_0$, 
hence
\begin{equation}
 c_{s;{\rm ideal}} = \sqrt{\frac{2 T}{m}}, \qquad 
 \nu_{t;{\rm ideal}} = \nu_{a;{\rm ideal}} = \frac{\tau k^2 T_0}{m}.
\end{equation}
The above expressions can be seen to coincide with those deduced in Eq.~(39) of 
Ref.~\cite{shan2018} by setting $\gamma = 2$, ${\rm Pr} = 1$, 
$\kappa = \kappa_T / \rho_0 c_p$, $c_p = 2/m$ and 
$\kappa_T = 2 \tau n_0 T_0 /m$.

The behaviour of the true van der Waals fluid can be recovered by 
setting $P_{\eta;0} = P_{\kappa;0} = p^w$. For the model used in this paper,
only $P_{\eta;0} = p^w$ is fulfilled, while $P_{\kappa;0} = p^i$ is approximated through the 
ideal gas pressure. This latter approximation affects the speed of sound $c_s$, 
as well as the damping coefficients $\nu_a$ and $\nu_t$. 
Figure~\ref{fig:long_coef} presents the dependence of $c_s$, 
$\nu_a$ and $\nu_t$ with respect to $n_0$ for the ideal gas, for the true 
van der Waals fluid and for our model. The most important feature for the study of 
phase separation phenomena is the correct recovery of the spinodal curve, which 
is given by the solutions of $\nu_t = 0$ for various values of $T$. It can 
be seen from Eq.~\eqref{eq:long_coeff} that the spinodal curve is determined by 
solving
\begin{equation}
 \frac{\partial P_\eta}{\partial n} + n_0 \sigma k^2 = 0.
\end{equation}
The surface tension term reduces the breadth of the spinodal region. 
For $k =2\pi / L$ and $\sigma = 10^{-4}$, the 
contribution of this term is negligible. However, there is always a minimal wavelength 
under which spinodal decomposition cannot occur, namely 
$L_{\rm min} = 2\pi/ k_{\rm min}$, where 
\begin{equation}
 k_{\rm min} = \frac{3\sqrt{3}}{2\sqrt{\sigma}} \sqrt{\frac{1 - T^{1/3}}{3 - 2T^{1/3}}}.
\end{equation}
At $\sigma = 10^{-4}$ and $T = 0.8$, $L_{\rm min} \simeq 0.0966$, while 
the interface width predicted through Eq.~\eqref{eq:xiwagner} is
$\xi_w \simeq 0.021$. For large wavelengths, the surface tension 
can be neglected and the equation $\partial_n p^w = 0$ predicts that 
spinodal decomposition can occur for densities between 
$0.521$ and $1.574$. Another interesting feature of the real van der Waals 
fluid is that the speed of sound becomes imaginary for densities 
between $0.771$ and $1.631$ (the values are computed for 
$T_0 = 0.8$ and $\sigma = 10^{-4}$), which lie within the spinodal curve. 
In our model, the speed of sound remains real for all values of $n$.
In the regions outside the spinodal curve, the qualitative behaviour 
of the speed of sound in our model and in the true van der Waals model 
is similar. There are however discrepancies between the value of the speed 
of sound in the vapour and liquid phases for the true van der Waals 
fluid ($c_{s;{\rm vapour}} \simeq 1.0289$ and $c_{s;{\rm liquid}} \simeq 1.479$) 
compared to those occuring in our model 
($c_{s;{\rm vapour}} \simeq 1.130$ and $c_{s;{\rm liquid}} \simeq 2.056$).

% For the van der Waals fluid, the sound speed can be computed by substituting the 
% correct equation of state. In the case when $P_{\eta; 0} = P_{\kappa; 0} = p^w$, 
% where $p^w$ is the van der Waals pressure \eqref{vdw}, the speeds of sound 
% in vapour and in liquid at $T_0 = 0.8$ and $\sigma = 10^{-4}$ are 
% $c_{s;{\rm vapour}} \simeq 1.029$ and $c_{s;{\rm liquid}} \simeq 1.479$, 
% respectively. The simplified model used in this paper instead uses 
% $P_{\eta;0} = p^w$ and $P_{\kappa;0} = p^i = n_0 T_0$, such that the 
% sound speeds are $c_{s; {\rm vapour}} \simeq 1.130$ and 
% $c_{s;{\rm liquid}} \simeq 2.056$. Regardless of how $P_{\kappa;0}$ is 
% implemented, it is worthwhile noting that $\nu_\lambda$ becomes negative 
% when $\partial_n p^w + n k^2 \sigma < 0$. Neglecting the effect of surface 
% tension, the critical values of $n_0$ for which longitudinal perturbations 
% become linearly unstable can be found by solving the equation 
% \begin{equation}
%  \frac{\partial p^w}{\partial n} = \frac{9}{(3-n)^2}
%  \left[T_0 - \frac{n_0}{4}(3 - n_0)^2\right] = 0.
% \end{equation}
% The above equation defines the spinodal curve inside which spontaneous 
% phase separation (also known as spinodal decomposition) can occur.
% In the case when $T_0 =0.8$, the above equation gives the values 
% $0.521$, $1.574$ and $3.905$, where the latter value is unattainable 
% since $0 \le n \le 3$ for the van der Waals fluid. 

Let us now write the solution of the longitudinal wave problem. 
Considering the coefficients $\delta u_t$ and $\delta u_\pm$ 
of the $\nu_t$ and $\nu_\pm$ modes as independent variables, 
the solutions $\widetilde{\delta u}$, $\widetilde{\delta n}$ 
and $\widetilde{\delta T}$ can be written as:
\begin{align}
 \widetilde{\delta u} =& e^{-\nu_t t} \delta u_t + 
 e^{-\nu_a t} \left[\delta u_c \cos (c_{s;0} k t) + 
 \delta u_s \sin(c_{s;0} k t)\right], \nonumber\\
 \widetilde{\delta n} =& e^{-\nu_t t} \delta n_t + 
 e^{-\nu_a t} \left[\delta n_c \cos (c_{s;0} k t) + 
 \delta n_s \sin(c_{s;0} k t)\right], \nonumber\\
 \widetilde{\delta T} =& e^{-\nu_t t} \delta T_t + 
 e^{-\nu_a t} \left[\delta T_c \cos (c_{s;0} k t) + 
 \delta T_s \sin(c_{s;0} k t)\right],
\end{align}
where Eq.~\eqref{eq:sound_macro_nu} can be used to obtain:
\begin{gather}
 \delta u_c = \frac{1}{2}(\delta u_+ + \delta u_-), \qquad 
 \delta u_s = \frac{1}{2i}(\delta u_+ - \delta u_-), \qquad 
 \delta n_t = \frac{k n_0}{\nu_t} \delta u_t, \qquad 
 \delta T_t = \frac{k P_{\kappa;0}}{n_0 \nu_t - \kappa_{T;0} k^2} \delta u_t, \nonumber\\
 \delta n_c = \frac{k n_0 (\nu_a \delta u_c + \nu_s \delta u_s)}{\nu_a^2 + \nu_s^2}, \qquad
 \delta n_s = \frac{k n_0 (\nu_a \delta u_s - \nu_s \delta u_c)}{\nu_a^2 + \nu_s^2}, \nonumber\\
 \delta T_c = \frac{k P_{\kappa;0} [(n_0 \nu_a - \kappa_{T;0} k^2) \delta u_c + n_0 \nu_s \delta u_s)}
 {(n_0 \nu_a - \kappa_{T;0} k^2)^2 + n_0^2 \nu_s^2}, \qquad
 \delta T_s = \frac{k P_{\kappa;0} [(n_0 \nu_a - \kappa_{T;0} k^2) \delta u_s - n_0 \nu_s \delta u_c)}
 {(n_0 \nu_a - \kappa_{T;0} k^2)^2 + n_0^2 \nu_s^2}.
\end{gather}
In the case when the density and temperature perturbations vanish at initial time
(i.e., $\delta n_0 = \delta T_0 = 0$), the constants 
$\delta u_c$, $\delta u_t$ and $\delta u_s$ are given up to $\tau^2$ through:
\begin{equation}
 \delta u_c \simeq \delta u_0,\qquad 
 \delta u_t \simeq 0, \qquad \delta u_s \simeq 
 \frac{(k^2 \kappa_{T;0} - n_0 \nu_a)^2 - n_0^2 \nu_a \nu_T}
 {n_0 \nu_s[k^2 \kappa_{T;0} - n_0 (\nu_a -\nu_t)]} \delta u_0.
\end{equation}
In this case, $\delta n_t \simeq 0$ and $\delta P_t \simeq 0$, 
such that the contribution of the purely evanescent mode $\nu_t$ is
negligible. The full solution for the velocity amplitude 
can be written up to $O(\tau^2)$ as:
\begin{equation}
 \widetilde{\delta u} \simeq e^{-\nu_a t} \delta u_0 \left[ 
 \cos (c_{s;0} k t) + 
 \frac{(k^2 \kappa_{T;0} - n_0 \nu_a)^2 - n_0^2 \nu_a \nu_t}
 {n_0 k c_{s;0} [k^2 \kappa_{T;0} - n_0 (\nu_a -\nu_t)]} \sin (c_{s;0} k t)\right].\label{eq:long_usol}
%  \widetilde{\delta n} \simeq& -\frac{n_0 \delta u_0}{c_s} e^{-\nu_d t} \left[ 
%  \sin (c_s k t) - 
%  \frac{k \kappa_T(k^2 \kappa_T - n_0 \nu_d)}
%  {n_0 c_s [k^2 \kappa_T - n_0 (\nu_d -\nu_\lambda)]} \cos(c_s k t)\right],\nonumber\\
%  \widetilde{\delta P} \simeq& -\frac{P_{\kappa;0} \delta u_0}{n_0 c_s} e^{-\nu_d t} 
%  \left[\sin(c_s k t) + \frac{k \kappa_T \nu_\lambda}
%  {c_s(k^2 \kappa_T - n_0(\nu_d - \nu_\lambda)} \cos(c_s k t)\right].
\end{equation}

% In order to highlight the $\nu_\lambda$ mode, the system can be initialised 
% such that $\delta u_c = \delta u_s = 0$ and $\delta u_\lambda = \delta u_0$. 
% In this case, $\delta n_c = \delta n_s = \delta T_c = \delta T_s = 0$
% and there is no contribution from the $\nu_\pm$ sector. The solution in this 
% case is:
% \begin{equation}
%  \widetilde{\delta u} \simeq \delta u_0 e^{-\nu_\lambda t}, \qquad 
%  \widetilde{\delta n} \simeq \frac{k n_0}{\nu_\lambda} \delta u_0 e^{-\nu_\lambda t}, \qquad 
%  \widetilde{\delta T} \simeq \frac{k P_{\kappa;0} \delta u_0}{n_0 \nu_\lambda - \kappa_T k^2} 
%  e^{-\nu_\lambda t}.
% \end{equation}

We now present our simulation results. As in Subsec.~\ref{sec:galilei:shear}, 
we perform three batches of simulations: 
the first is for the ideal gas 
with $n_0 = n_{\rm ideal} = 1$; the second and third are for the vapour 
($n_0 = n_g = 0.2396669$) and liquid ($n_0 = n_l = 1.932703$) 
phases of the van der Waals fluid with $\sigma = 10^{-4}$, respectively.
All simulations are performed at $T = 0.8$ and we consider the wavelength 
fixed at $L = 1$ ($k = 2\pi / L$). The relaxation time is fixed at 
$\tau = 5 \times 10^{-4}$ while the time step is $\delta t = 2 \times 10^{-4}$.
The initial wave amplitudes are $\delta u_0 = 10^{-3}$, $\delta n_0 = \delta T_0 = 0$ 
and the evolution of the velocity amplitude $\widetilde{\delta u}$ is 
given in Eq.~\eqref{eq:long_usol}. In each simulation batch, we use 
between $N_x = 20$ and $N_x = 60$ grid points and the background 
velocity along the $x$ axis is varied between $V = 0$ and $V = 2$.

In all cases, we perform $N_t = 50000$ iterations, up to $t_f = 10$. 
The values of the amplitude of the velocity are stored at intervals 
$\Delta t = 100\delta t$ and 
are labelled $\widetilde{\delta u}_s$ ($0 \le s \le S = 500$). The procedure 
for computing $\widetilde{\delta u}_s$ is:
\begin{equation}
 \widetilde{u}_s = \frac{2}{\mathfrak{N}_x} \sum_{i = 1}^{\mathfrak{N}_x} 
 (u_{x;s;i} - V) \sin [k (x_i - V \hat{t}_s)],
\end{equation}
where $\hat{t}_s = s \times \Delta t = 100s \times \delta t$.
The quantitative analysis is performed at the level of 
three quantities, namely: the acoustic damping coefficient $\nu_a$; the 
sound speed $c_s$; and the $L_2$ norm. In order to extract $\nu_a$ and 
$c_s$ from the numerical data, Eq.~\eqref{eq:long_usol} is written as:
\begin{equation}
 \widetilde{u}(t) = \delta u_0 e^{-\nu_a t} \left[
 \cos (c_s k t) + \mathcal{S} \sin(c_s k t)\right].
 \label{eq:long_usol_2}
\end{equation}
The parameters $\nu_a$, $c_s$ and $\mathcal{S}$ are obtained 
by performing a three-parameter fit of Eq.~\eqref{eq:long_usol_2} for the 
case of the van der Waals fluid, while in the case of the ideal gas, $\mathcal{S}$ 
is set to $0$ and the fit is performed using only two free parameters. The other parameters are 
$\delta u_0 = 10^{-3}$ and $k = 2\pi$. 
The $L_2$ norm is computed as follows:
\begin{equation}
 L_2 = \left\{\int_{0}^{t_{\rm f}} \frac{dt}{t_{\rm f}} 
 \left[\frac{\widetilde{\delta u}_{\rm num}(t)}{\widetilde{\delta u}_{\rm lin}(t)} - 1 
 \right]^2\right\}^{1/2}
 \simeq \left\{\frac{1}{S} \sum_{s = 0}^{S} \mathfrak{f}_s
 \left[\frac{\widetilde{\delta u}_s}{\widetilde{\delta u}_{\rm lin}(\hat{t}_s)} - 
 1\right]^2\right\}^{1/2},
 \label{eq:long_L2} 
\end{equation}
where $\widetilde{\delta u}_{\rm lin}$ is the solution of the linearised hydrodynamic 
equation derived in Eq.~\eqref{eq:long_usol}, evaluated at 
$t = \hat{t}_s$.

Figure~\ref{fig:long_evol} shows the typical evolution of the amplitude 
$\widetilde{u}(t)$, as obtained using our numerical method, compared to the 
analytic solution \eqref{eq:long_usol} of the linearised hydrodynamics equations,
for the cases of the ideal gas and of the vapour and liquid phases of the 
van der Waals fluid at $T = 0.8$. It can be seen that our numerical results are 
well overlapped with the analytic solution even when $\mathfrak{N}_x = 20$ and $V = 2$, 
thus demonstrating the capabilities of the numerical scheme and the 
degree of Galilean invariance of our implementation. Figures~\ref{fig:long_errnx} 
and \ref{fig:long_errux} describe the typical procedure that we employed 
for the quantitative analyses discussed below. 

In Fig.~\ref{fig:long_errnx}, the relative error of the numerically obtained 
values for $\nu_a$ and $c_s$, computed with respect to their analytic expectations
in Eq.~\eqref{eq:long_coeff}, as well as $L_2$, are represented with respect to 
$\delta x = 1/\mathfrak{N}_x$ for the case when $V = 0$. It can be seen that the errors decrease 
with $\delta x$ only for $\delta x \gtrsim 0.04$. Further decreasing $\delta x$ 
allows these quantities to stabilize at values which have a relative difference 
compared to Eq.~\eqref{eq:long_coeff} of about $10^{-3}$. This difference is comparable 
to both $\delta u_0$ and $\tau$, while Eqs.~\eqref{eq:long_coeff} and 
\eqref{eq:long_usol} are valid only at linear order in these quantities. 
Thus, the deviatation of the numerical results from the results obtained 
in the linearised regime is consistent with the assumptions employed in 
deriving the analytic solution. Thus, in order to extract the order of the 
numerical scheme, a numerical fit of the function $a (\delta s)^\gamma$
is performed on the relative errors, but only for $\delta s \ge 1/30$,
thus avoiding the effects of the plateau region which appears at 
smaller values of $\delta s$. The exponents of the above mentioned fits 
are given in the legend. Further analysis was performed by considering  
a selection of values for $V$, between $V =0$ and $V = 2$ and the 
results are summarised in Tab.~\ref{tab:long_errux}. The values of the 
exponents corresponding to $\nu_d$ and $L_2$ are generally confined 
between $4$ and $5$. The exponent corresponding to $c_s$ presents 
wider variations, having values larger than $5$ for $V \lesssim 1$,
and decreasing below $2$ at $V = 2$. It should be noted that 
the relative error in obtaining $c_s$ is several orders of magnitude below 
$(\nu_{a;{\rm app}}/\nu_{a;{\rm lin}} - 1)$ and $L_2$.
We conclude that our numerical results generally support that our numerical 
scheme is at least of order $4$ for small and moderate values of $V$.

Figure~\ref{fig:long_errux} summarises the procedure that we used in 
order to determine the order at which changing the background longitudinal 
velocity $V$ affects our numerical results. Thus, we considered the relative 
errors of the differences of the numerically determined quantities 
$\nu_a$, $c_s$ and $L_2$ for a given value of $V$ and their values 
obtained when $V = 0$. It can be seen that for sufficiently small values
of $V$, a power law can be observed, the exponent of which indicates an
almost quadratic dependence of these differences on $V$. The results 
presented in Fig.~\ref{fig:long_errux} are restricted to the 
case when $\mathfrak{N}_x = 60$. Further results for various values of $\mathfrak{N}_x$ 
between $20$ and $60$ are summarised in Tab.~\ref{tab:long_errnx}. 
These results are generally supportive of the nearly quadratic 
dependence of the relative errors on $V$.

\subsection{Laplace pressure test of a moving bubble}\label{sec:galilei:bubble}

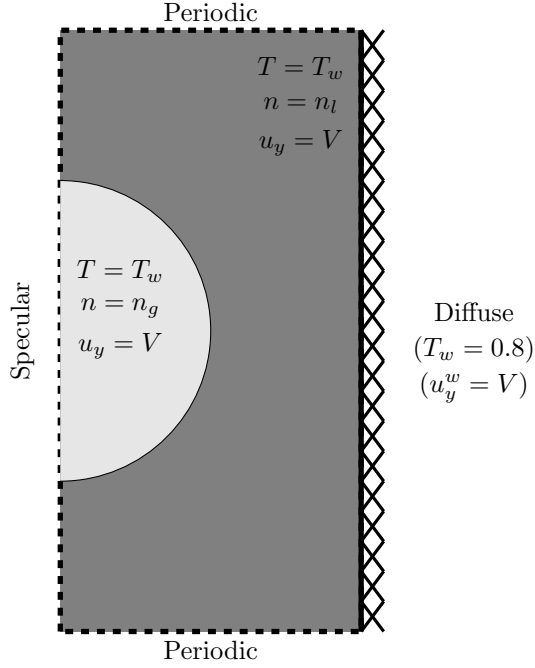
\begin{figure}[t]
\centering
\begin{tikzpicture}
\fill [black!50!white] (0,-40mm) rectangle (40mm,40mm);
\draw [line width=2,dashed] (0,-40mm) rectangle (40mm,40mm);
\draw [line width=2] (40mm,-40mm) -- (40mm,40mm);

\foreach \y in {-9,...,10} {
    \draw [line width=1.2] (40mm,\y*4mm-4mm) -- (43mm,\y*4mm);
}
\foreach \y in {-9,...,10} {
    \draw [line width=1.2] (43mm,\y*4mm-4mm) -- (40mm,\y*4mm);
}

\draw [fill=black!10!white] (0.mm,-20mm) arc[radius = 20mm, start angle= -90, end angle= 90];

\node at (20mm,-42.5mm) {Periodic};
\node [rotate=90] at (-5mm,0mm) {Specular};
\node at (20mm,42.5mm) {Periodic};

% \fill[black!20!white] (0mm,0mm) rectangle (80mm,40mm);

\node at (55mm,2.5mm) {Diffuse};
\node at (55mm,-2.5mm) {$(T_w=0.8)$};
\node at (55mm,-7.5mm) {$(u_y^w=V)$};

\node at (32mm,35mm) {$T=T_w$};
\node at (32mm,30mm) {$n=n_l$};
\node at (32mm,25mm) {$u_y=V$};

\node at (8mm,8mm) {$T=T_w$};
\node at (8mm,3mm) {$n=n_g$};
\node at (8mm,-2mm) {$u_y=V$};

%\draw [<->,dashed] (0,20mm)--(20mm,20mm);
%\node at (10mm,22mm) {$R=R_0$};

\end{tikzpicture}
\caption{Galilean invariance bubble test setup. The system is initialized 
as described in Sec.~\ref{sec:galilei:bubble}.}
\label{fig:galilei:bubble_setup}
\end{figure}

In this subsection we present a study of the evolution of a
gas bubble enclosed between parallel walls kept at constant temperature for various values of the background
velocity parallel to the system walls. We initialised the system with a bubble centered on $x=0,y=0$ using 
formula \eqref{eq:tanh}, temperature $T=T_w=0.8$, fluid velocity $u_y=V$ and wall velocity $u_y^w=V$.
We test the Galilean invariance of our model by tracking the value $\gamma(V,t)$ of the surface tension evaluated using the Laplace law \eqref{eq:Laplace},
as well as of the vertical coordinate $y_c(V,t)$ of the bubble center, evaluated using formula \eqref{eq:tanh}, for background velocities $V=0.01,0.02,0.05$ and $0.1$.

In Fig.\ref{fig:galilei_bubble} (a) we present the evolution of the relative error $\epsilon_\gamma(V,t)=\vert\gamma(V,t)/\gamma(0,t)-1\vert$, 
where $\gamma(0,t)$ is the value of the surface tension for the stationary bubble.
It can be observed that this error is well below $1\%$ for all the velocities 
considered.
The numerical fit of the average values $\overline{\epsilon}_\gamma(V)$, over the interval $20\leq t\leq 30$, against 
the function $a V^b$, as seen in Fig. \ref{fig:galilei_bubble}(b), yields a value of 
$b$ very close to $2$.
In Fig. \ref{fig:galilei_bubble}(c) we plot the evolution of the relative deviation $\epsilon_y(V,t)=[y_c(V,t)/Vt]-1$ 
of the bubble center. In Fig. \ref{fig:galilei_bubble}(d) we plot the deviation $\Delta y_c(V,t)=y_c(V,t)-V t$ 
with respect to background velocity at $t=30$. A numerical fit gives a nearly linear dependence of the deviation $\Delta y_c(V,30)$ on the background velocity.

In order to better illustrate the deviations of the bubble center in time, we present in Fig. \ref{fig:bubble_contour} the density isocontour corresponding to $n = 1.0$ for $V = 0.1$ 
after one ($t = 10$), two ($t = 20$) and three ($t = 30$) cycles.
It can be seen that the bubble constantly lags behind the background flow, 
such that the contours corresponding to successive cycles do not overlap. We
attribute this effect to the spurious currents which are always present at the interface.

\begin{figure}
 \centering
 \begin{tabular}{cc}
 \includegraphics[width=0.45\linewidth]{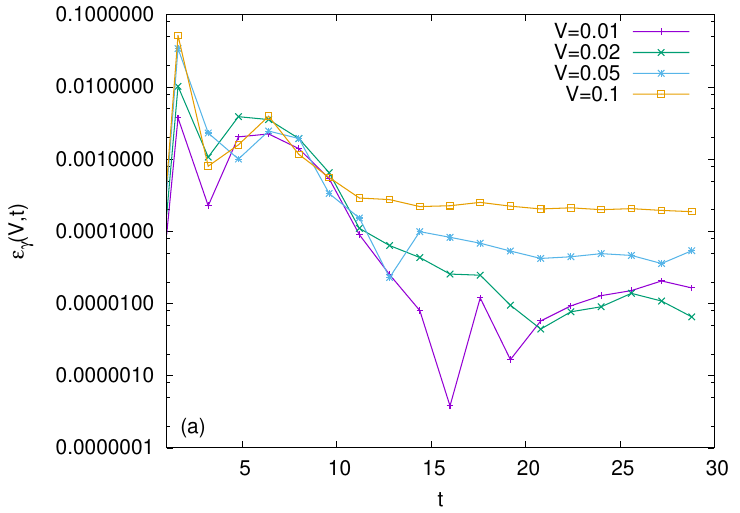}&
 \includegraphics[width=0.45\linewidth]{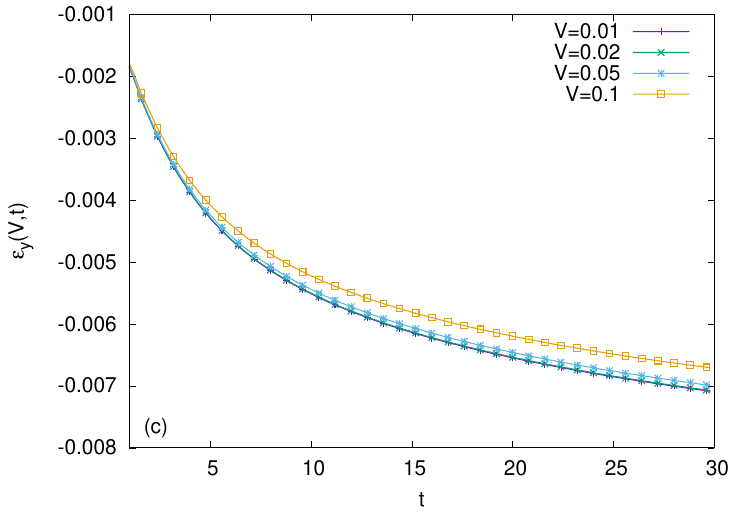}\\
 \includegraphics[width=0.45\linewidth]{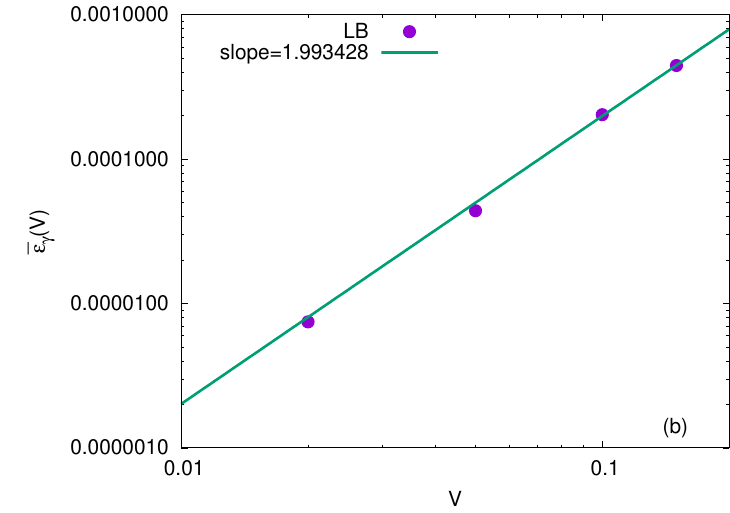}&
 \includegraphics[width=0.45\linewidth]{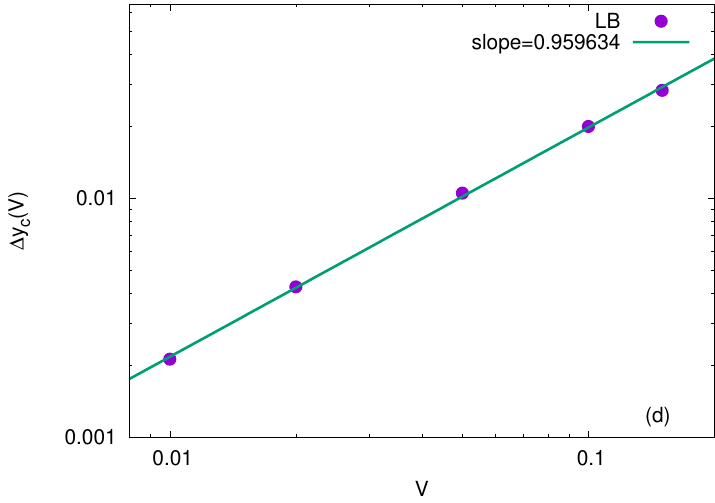}
 \end{tabular}
\caption{(a) evolution of the relative error $\epsilon_\gamma(V,t)$, 
(b) average values $\overline{\epsilon}_\gamma(V)$ with respect to the background velocity $V$,
(c) evolution of the relative deviation $\epsilon_y(V,t)$ of the bubble center and 
(d) deviation $\Delta y_c(V,t)$ with respect to the background velocity $V$.}
\label{fig:galilei_bubble}
\end{figure}

\begin{figure}
\centering
 \includegraphics[width=0.7\linewidth]{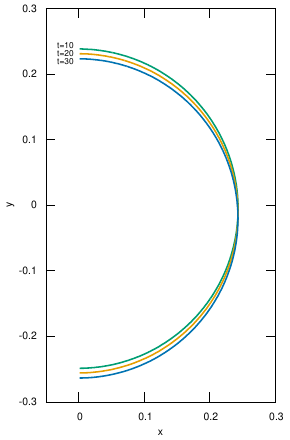}
\caption{Density contour plots($n=1$) at $t=10,\,20,\,30$ for the bubble 
moving with constant background velocity $V=0.1$.}
\label{fig:bubble_contour}
\end{figure}

\section{Phase separation dynamics between heat extracting parallel plates}
\label{sec:pspp}

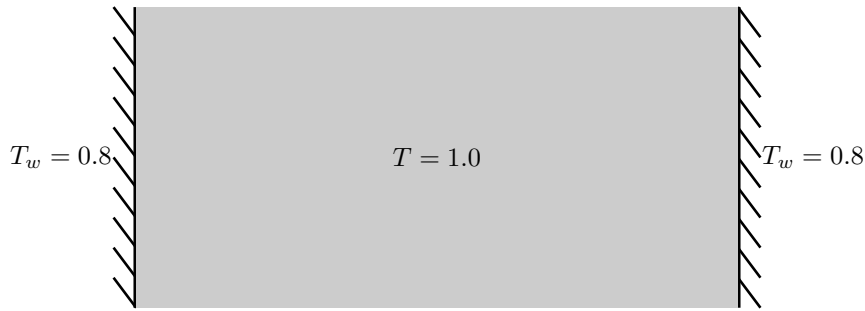
\begin{figure}[t]
\centering
\begin{tikzpicture}
\draw [line width=2] (0mm,0mm) -- (0mm,40mm);
\draw [line width=2] (80mm,0mm) -- (80mm,40mm);
\draw [line width=1] (0mm,0mm)--(-3mm,4mm);
\draw [line width=1] (0mm,4mm)--(-3mm,8mm);
\draw [line width=1] (0mm,8mm)--(-3mm,12mm);
\draw [line width=1] (0mm,12mm)--(-3mm,16mm);
\draw [line width=1] (0mm,16mm)--(-3mm,20mm);
\draw [line width=1] (0mm,20mm)--(-3mm,24mm);
\draw [line width=1] (0mm,24mm)--(-3mm,28mm);
\draw [line width=1] (0mm,28mm)--(-3mm,32mm);
\draw [line width=1] (0mm,32mm)--(-3mm,36mm);
\draw [line width=1] (0mm,36mm)--(-3mm,40mm);

\draw [line width=1] (83mm,0mm)--(80mm,4mm);
\draw [line width=1] (83mm,4mm)--(80mm,8mm);
\draw [line width=1] (83mm,8mm)--(80mm,12mm);
\draw [line width=1] (83mm,12mm)--(80mm,16mm);
\draw [line width=1] (83mm,16mm)--(80mm,20mm);
\draw [line width=1] (83mm,20mm)--(80mm,24mm);
\draw [line width=1] (83mm,24mm)--(80mm,28mm);
\draw [line width=1] (83mm,28mm)--(80mm,32mm);
\draw [line width=1] (83mm,32mm)--(80mm,36mm);
\draw [line width=1] (83mm,36mm)--(80mm,40mm);

\fill[black!20!white] (0mm,0mm) rectangle (80mm,40mm);

\node at (90mm,20mm) {$T_w=0.8$};
\node at (-10mm,20mm) {$T_w=0.8$};
\node at (40mm,20mm) {$T=1.0$};

\end{tikzpicture}
\caption{Initial setup for the phase separation dynamics between heat extracting parallel plates problem considered in Sec.~\ref{sec:pspp}}.
\label{pspp}
\end{figure}

We now consider the phase separation in a van der Waals
fluid placed between two parallel walls. 
The fluid between the walls is in isothermal 
conditions at the critical temperature $T = 1$ and the density 
field is initialized with small fluctuations $n_{i,j} = 1 + \delta n_{i,j}$,
where the indices $i$ and $j$ identify the grid node, while
$\delta n_{i,j}$ is a random point-dependent number satisfying 
$-0.01 \le \delta n_{i,j} \le 0.01$.
At initial time, the temperature of the walls is suddenly decreased to $T_w < 1$ 
(in this section, we only consider $T_w = 0.8$), thus inducing the phase 
separation process by cooling the system via the gradual extraction of heat 
through the diffuse reflecting walls. Figure \ref{pspp} shows the
initial setup of the problem.  The temperature of the walls 
is kept constant throughout the simulation. 
The simulation domain is comprised of $2 \mathfrak{N} \times \mathfrak{N}$ nodes.
The lattice spacing is set to $\delta s = 1/\mathfrak{N}$, where $\mathfrak{N} = 320$,
such that the left and right walls are located at $x = -1$ and $x = 1$, respectively.
The domain has unit vertical span. Diffuse reflection boundary conditions are 
implemented along the walls, which are located at $i = \frac{1}{2}$ and 
$i = 2\mathfrak{N} + \frac{1}{2}$ ($1 \le j \le \mathfrak{N}$). Periodic boundary conditions 
apply along the top and bottom domain boundaries, where $j = \frac{1}{2}$ and 
$j = \mathfrak{N} + \frac{1}{2}$ ($1 \le i \le 2\mathfrak{N}$).

% \begin{figure}[t]
% \begin{center}
% \begin{tabular}{rcc}
% t=2.5
% &
%  \includegraphics[width=160pt,angle=270]{t2p5-eps-converted-to.pdf}&
%  \includegraphics[width=120pt,angle=270]{t2p5_temp-eps-converted-to.pdf}\\
% t=5
% &
%  \includegraphics[width=160pt,angle=270]{t5-eps-converted-to.pdf}&
%  \includegraphics[width=120pt,angle=270]{t5_temp-eps-converted-to.pdf}\\
% t=10
% &
%  \includegraphics[width=160pt,angle=270]{t10-eps-converted-to.pdf}&
%  \includegraphics[width=120pt,angle=270]{t10_temp-eps-converted-to.pdf}\\
%  t=12.5
% &
%  \includegraphics[width=160pt,angle=270]{t12p5-eps-converted-to.pdf}&
%  \includegraphics[width=120pt,angle=270]{t12p5_temp-eps-converted-to.pdf}\\
%  t=25
% &
%  \includegraphics[width=160pt,angle=270]{t25-eps-converted-to.pdf}&
%  \includegraphics[width=120pt,angle=270]{t25_temp-eps-converted-to.pdf}\\
%  t=550
% &
%  \includegraphics[width=160pt]{cold_plates_55000000-eps-converted-to.pdf}&
%  \includegraphics[width=120pt]{cold_plates_55000000_temp_profile-eps-converted-to.pdf}
%  \end{tabular}
% \end{center}
% \caption{Snapshots of the phase separation between parallel walls
% (left column)  and temperature profiles taken along the line
% $j={\mathfrak{N}}_{2}/2$ (right column),
% at $t=10,25,31,40,60$ and $550$ (from top to bottom).
% \label{largesep}}
% \end{figure}

\begin{figure}[t]
\begin{center}
\begin{tabular}{cc}
\vspace*{-16mm}
 \includegraphics[angle=-90,origin=c,width=180pt]{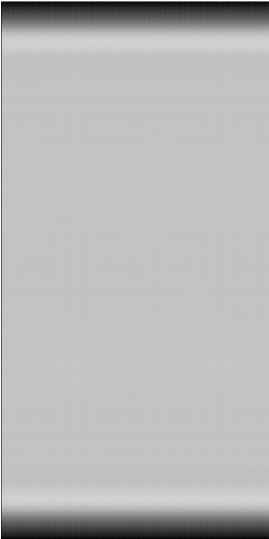}&
 \includegraphics[angle=-90,origin=c,width=180pt]{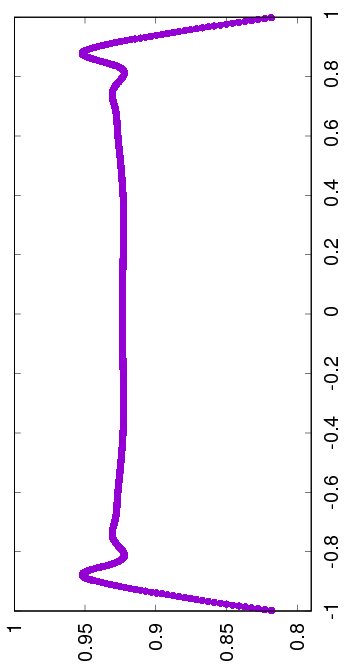}\\
\vspace*{-16mm}
 \includegraphics[angle=-90,origin=c,width=180pt]{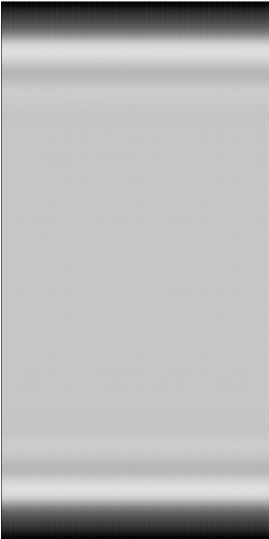}&
 \includegraphics[angle=-90,origin=c,width=180pt]{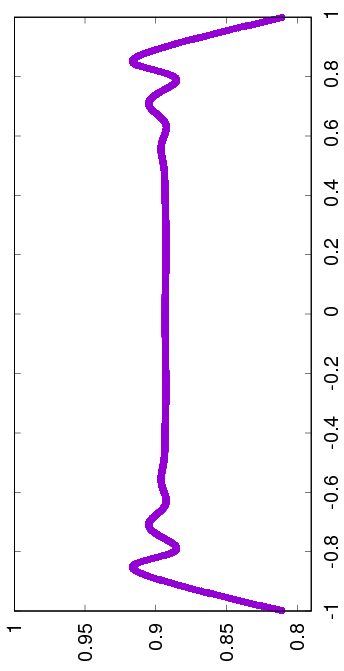}\\
\vspace*{-16mm}
 \includegraphics[angle=-90,origin=c,width=180pt]{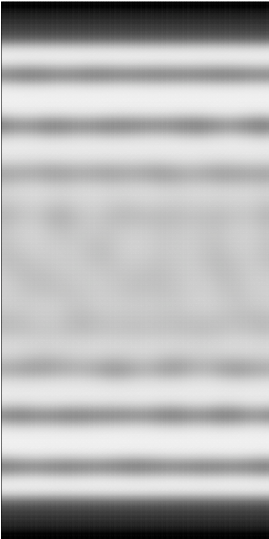}&
 \includegraphics[angle=-90,origin=c,width=180pt]{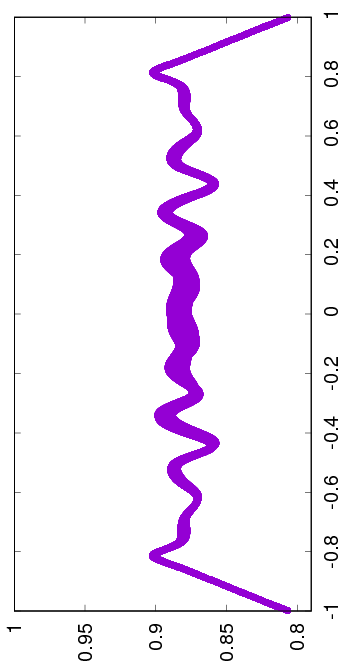}\\
\vspace*{-16mm}
 \includegraphics[angle=-90,origin=c,width=180pt]{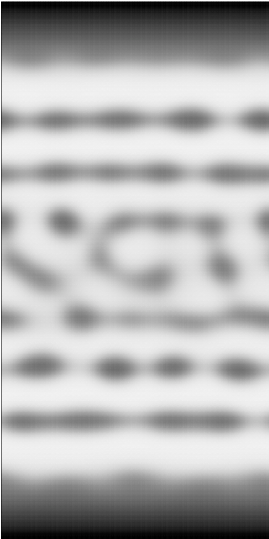}&
 \includegraphics[angle=-90,origin=c,width=180pt]{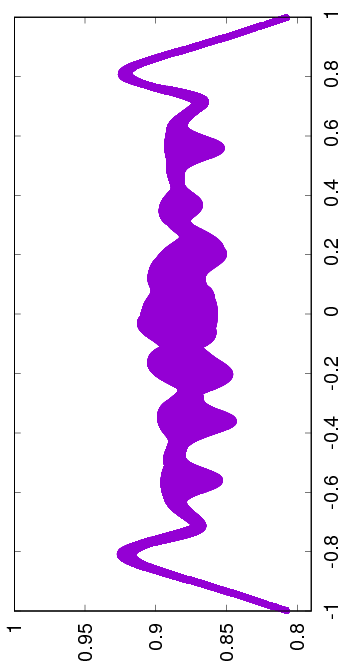}\\
\vspace*{-16mm}
 \includegraphics[angle=-90,origin=c,width=180pt]{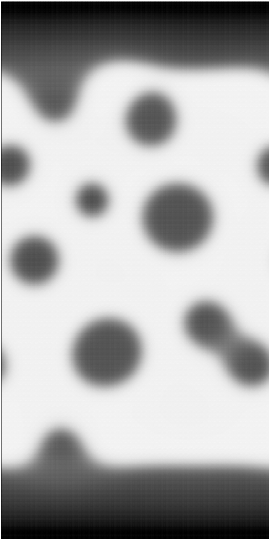}&
 \includegraphics[angle=-90,origin=c,width=180pt]{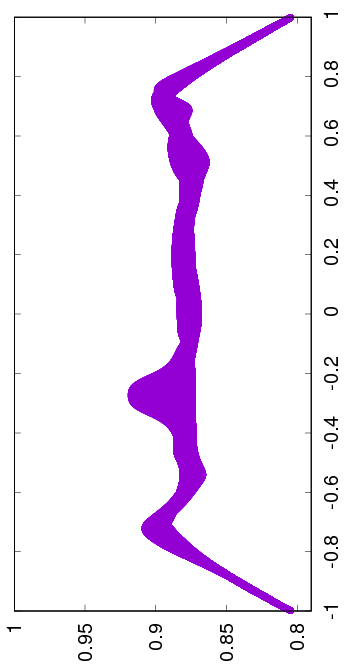}\\
 \vspace*{-16mm}
 \includegraphics[angle=-90,origin=c,width=180pt]{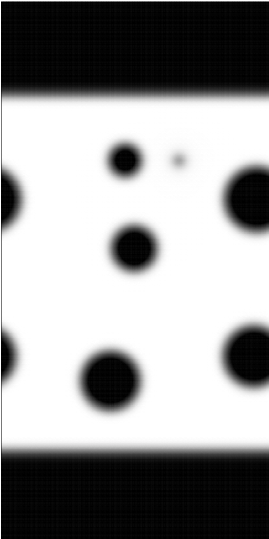}&
 \includegraphics[angle=-90,origin=c,width=180pt]{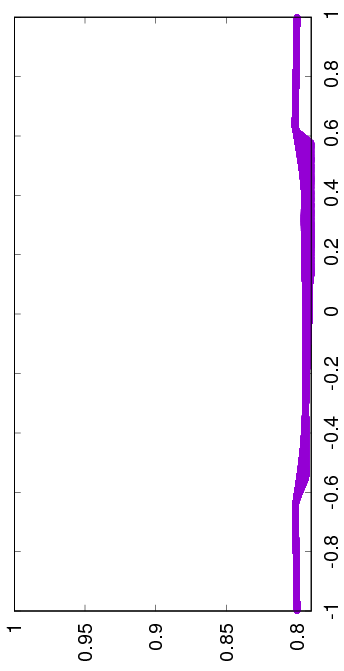}
 \end{tabular}
\end{center}
\vspace*{+1mm}

\caption{Snapshots of the phase separation between parallel walls
(left column)  and temperature profiles taken along the x axis superimposed,
at $t=2.5, 5, 10, 12.5, 25$ and $240$ (from top to bottom).
\label{largesep}}
\end{figure}

Figure \ref{largesep} shows the evolution of the liquid - vapor separation
process on a $640 \times 320$  lattice.
At $t=0$, the fluid temperature was $T=T_{c}=1.0$ and the wall temperature 
was set to $T_w=0.8$. The simulation was conducted with the
time step and lattice spacing $\delta t =2\times 10^{-4}$ and 
$\delta s=1/320$, respectively, while $\sigma = 10^{-4}$ and 
$\tau = 5 \times 10^{-3}$.
At early times, one observes the liquid deposition on the cold
walls. As the bulk temperature decreases, further parallel bands of low and 
high density appear near the walls.
Afterwards, these bands break into individual droplets due to the action of 
surface tension.
The formation of liquid droplets in the central region of
the channel is observed at later stages of the simulation. This happens because the temperature
in the center of the channel decreases during the heat extraction through the walls, 
but always remains higher than the wall temperature,
as seen in the right column of Figure \ref{largesep}\,. This feature
was observed also when investigating the liquid-vapour
phase separation with a different thermal LB model \cite{gonnellaPRE2007}.
Moreover, the right column of Figure \ref{largesep} revealed that the local 
maxima in the temperature profiles are
located in the liquid-vapour interface 
regions. In these regions, where large density 
gradients are present, there is an unphysical heat generation process, due
to the spurious velocity, a numerical effect that plagues the LB models
\cite{liPECS2016,zarghamiPRE2015,hejranfarPRE2015,chenIJHMT2014,
khajeporPRE2015,khajeporPRE2016,ikedaCAF2014,ganCTP2012,
ganFP2012,ganPRE2011,ganIJMPC2014}.
This numerical effect has been succesfully reduced in this paper 
by using the fifth order 
weighted essentially non-oscillatory (WENO-5) numerical scheme.

\section{Conclusion}\label{sec:conc}

A single particle distribution function thermal lattice Boltzmann model based on the full-range
Gauss-Hermite quadrature of order $Q=5$ was tested by simulating
the liquid-vapour phase separation in a van der Waals fluid bounded
by two parallel walls. The Van der Waals force term was implement using 49-point stencils.
We validated our thermal model by considering the 
plane interface problem, the Laplace pressure test and 
by comparing the phase separation results 
against the Maxwell construction results. Good agreement was obtained 
for temperatures as low as $0.72\,T_c$.
We also present a discussion on transport coefficients, sound speed and Galilean invariance.

Starting from an initial state in which the fluid is at the critical
temperature $T_c$, with random density fluctuations of at most $1\%$ 
around the critical density, we investigated the phase separation between 
two walls kept at a constant  temperature $T_w < T_c$. Our simulations 
show that the condensation starts in the vicinity of the walls. As the bulk temperature decreases,
liquid droplets develop further in the channel.
We observed that in the stationary state, spurious currents persist, which 
affect the temperature field through spurious heating at the vapour-liquid interface.
To avoid the large relative errors induced by the numerical effects, it is necessary to use
 high order schemes such as the fifth order weighted essentially non-oscillatory (WENO-5)
 scheme with sufficiently small values of the lattice spacing $\delta s$ and the time step 
 $\delta t$, e.g. $(\delta s,\delta t)=(1/160,2\times 10^{-4})$.
With these parameters, the non-dimensionalised magnitude of the spurious velocity is below $6\times 10^{-5}$, while 
the fluid temperature profile is within the range of $1.0\%$ 
above the wall temperature $T_w$, even when $T_w = 0.72\, T_c$.

\section*{Acknowledgments}
This work is supported by a grant from the Romanian National Authority
for Scientific Research, CNCS-UEFISCDI, project number
PN-II-ID-PCE-2011-3-0516.
The authors are indebted to Adrian Horga for 
invaluable insight regarding the development of our CUDA code.

\section*{References}

%\bibliography{mybibfile}
\bibliography{sergiuICMMES}

\end{document}